\pdfoutput=1
\documentclass[aps, prd, 10 pt, notitlepage, nofootinbib, superscriptaddress, preprintnumbers, longbibliography, floatfix, preprint]{revtex4-2}
\usepackage{amsfonts, amsmath, amssymb, mathtools, graphicx, color, hyperref}
\usepackage{tabularx}
\usepackage[caption=false]{subfig}

\definecolor{nicered}{rgb}{.7,.1,.1}
\definecolor{nicegreen}{rgb}{.1,.5,.1}
\definecolor{darkblue}{rgb}{0,0,.5}
\definecolor{asparagus}{rgb}{0.53, 0.66, 0.42}
\definecolor{blue-violet}{rgb}{0.54, 0.17, 0.89}
\hypersetup{colorlinks, citecolor=nicegreen, linkcolor=nicered, urlcolor=darkblue}

\newcommand{\beq}{\begin{eqnarray}}
\newcommand{\eeq}{\end{eqnarray}}
\newcommand{\be}{\begin{equation}}
\newcommand{\ee}{\end{equation}}
\newcommand{\bea}{\begin{align}}
\newcommand{\eea}{\end{align}}
\newcommand{\rhotiL}{\tilde\rho_L}
\newcommand{\RtiL}{\tilde R_L}
\newcommand{\rhotiC}{\tilde\rho_C}
\newcommand{\RtiC}{\tilde R_C}
\newcommand{\fmsq}{f_{m^2}}
\newcommand{\feta}{f_\eta}
\newcommand{\epsal}{\varepsilon_\alpha}

\newcommand{\nn}{\nonumber \\}
 
\newcommand{\mueps}{\mu_\epsilon}
\newcommand{\FV}{\text{FV}}

\newcommand{\di}{\text{d}}
\newcommand{\hatpsi}{\hat{\psi}_\FV}
\newcommand{\ddV}{V^{(2)}}

\begin{document}
\count\footins = 1000

\title{False Vacuum Decay Rate From Thin To Thick Walls}

\author{Marco Matteini
\href{https://orcid.org/0000-0001-6481-3025}{\includegraphics[scale=0.3]{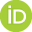}}}
\email{marco.matteini@ijs.si}
\affiliation{Jo\v{z}ef Stefan Institute, Jamova 39, 1000 Ljubljana, Slovenia}
\affiliation{Faculty of Mathematics and Physics, University of Ljubljana, Jadranska 19, 1000 Ljubljana, Slovenia}

\author{Miha Nemev\v{s}ek
\href{https://orcid.org/0000-0003-1110-342X}{\includegraphics[scale=0.3]{figures/orcid_32x32.png}}}
\email{miha.nemevsek@ijs.si}
\affiliation{Jo\v{z}ef Stefan Institute, Jamova 39, 1000 Ljubljana, Slovenia}
\affiliation{Faculty of Mathematics and Physics, University of Ljubljana, Jadranska 19, 1000 Ljubljana, Slovenia}

\author{Yutaro Shoji
\href{https://orcid.org/0000-0003-3932-9199}{\includegraphics[scale=0.3]{figures/orcid_32x32.png}}}
\email{yutaro.shoji@ijs.si}
\affiliation{Jo\v{z}ef Stefan Institute, Jamova 39, 1000 Ljubljana, Slovenia}

\author{Lorenzo Ubaldi
\href{https://orcid.org/0000-0002-9567-9719}{\includegraphics[scale=0.3]{figures/orcid_32x32.png}}}
\email{lorenzo.ubaldi@ijs.si}
\affiliation{Jo\v{z}ef Stefan Institute, Jamova 39, 1000 Ljubljana, Slovenia}

\date{\today}

\begin{abstract}
We consider a single real scalar field in flat spacetime with a polynomial potential up to $\phi^4$,
that has a local minimum, the false vacuum, and a deeper global minimum, the true vacuum.
When the vacua are almost degenerate we are in the thin wall regime, while as their difference in 
potential energy increases, we approach the thick wall regime.
We give explicit simple formulae for the decay rate of the false vacuum in 3 and 4 spacetime dimensions. 
Our results include a careful treatment both of the bounce action, which enters at the exponent 
of the decay rate, and of the functional determinant at one loop, which determines the prefactor.
The bounce action is computed analytically as an expansion in the thin wall parameter in generic $D$ 
dimensions.
We find that truncating such an expansion at second order we obtain a remarkably accurate bounce 
action also deep into thick wall regimes. 
We calculate the functional determinant numerically in 3 and 4 dimensions and fit the results 
with simple polynomials of the same thin wall parameter. 
This allows us to write the complete one-loop decay rate as a compact expression, which 
works accurately from thin to thick wall regimes.
\end{abstract}
\maketitle
\newpage

\tableofcontents

\section{Introduction}

The interest in phase transitions and metastable states has a long history~\cite{Langer:1967ax} and
often appears in the context of quantum~\cite{Kobzarev:1974cp, Coleman:1977py} and thermal field 
theories~\cite{Linde:1981zj, Affleck:1980ac}.
Apart from the theoretical and conceptual appeal, the applications of such phenomena are found in particle 
physics~\cite{Isidori:2001bm, Chigusa:2017dux, Chigusa:2018uuj, Andreassen:2017rzq, Baratella:2024hju}, 
cosmology and condensed matter (see~\cite{Devoto:2022qen} for a pedagogical introduction).

A widely used framework for computing false vacuum decay rates was defined in the seminal
paper~\cite{Coleman:1977py} that introduced a semi-classical approach with the hyper-spherical
bounce solution\footnote{The proof of a spherical symmetric bounce dominating the rate presented
in~\cite{Coleman:1977th} was recently extended to multi-fields~\cite{Blum:2016ipp} and to (classically)
scale invariant theories~\cite{Oshita:2023pwr}.} in Euclidean spacetime.
It showed that the decay rate per unit volume takes the form $\Gamma / V = A \, e^{-B}$, where $B$ 
is the Euclidean action calculated on the classical bounce field configuration. 
The prefactor $A$ is obtained from the quantum fluctuations around the bounce~\cite{Callan:1977pt}, 
and is typically more difficult to compute than $B$. 
Given that $A$ is dimensionful, one might guess it by na\"ive dimensional analysis.
This fact, combined with the exponential sensitivity of the rate on $B$, led to many studies that 
focused on the calculation of $B$.

When multiple scalar fields are present, a situation of interest in contexts beyond the 
Standard Model, computing $B$ becomes quite challenging. 
One option~\cite{Wainwright:2011kj} is to extend the method~\cite{Coleman:1977py} of 
undershoot/overshoot, which is especially hard because one wants to find the saddle point of the 
action in a multi-dimensional field space~\cite{Masoumi:2017trx}.
Recent approaches tackled the bounce issue in a number ways by using gradient 
flow~\cite{Chigusa:2019wxb, Sato:2019axv, Sato:2019wpo, Hong:2023dan}, machine 
learning~\cite{Piscopo:2019txs}, shooting + perturbative linearization~\cite{Athron:2019nbd} and 
optimisation algorithms~\cite{Bardsley:2021lmq}.
Another way is to find a useful closed form solution based on linear segments~\cite{Duncan:1992ai}, 
forming a triangular potential (solved for $D=4$ in~\cite{Duncan:1992ai}).
This leads exactly to the thin wall (TW) action when the two minima are degenerate, but also fails quickly 
outside of it~\cite{Aravind:2014pva}, even though it seems to be useful for multi-field 
estimates~\cite{Masoumi:2017trx}.
Extending the number of segments leads to the polygonal bounce program~\cite{Guada:2018jek}, which 
was worked out for multi-fields in general $D$ and implemented in the~\texttt{FindBounce} 
package~\cite{Guada:2020xnz}.
One method that avoids the bounce altogether is based on the tunneling potential~\cite{Espinosa:2018szu,
Espinosa:2018hue}.

In this paper we focus on a single real scalar field and consider a potential up
to $\phi^4$, for which the calculation of the bounce action
was initially studied in~\cite{Coleman:1977py} at the leading order in the 
thin wall limit, where the two minima are 
nearly degenerate.
Departing from this limit, the solution of~\cite{Coleman:1977py} quickly becomes inaccurate. 
A simple way to improve it is to perform a simple fit to the numerically calculated 
action~\cite{Dine:1992wr, Adams:1993zs}. 
Such fits are useful, as they accurately cover potentials ranging from thin wall to thick wall
configurations~\cite{Sarid:1998sn}, including phenomenologically relevant 
situations~\cite{Levi:2022bzt}.

One of the main aims of this work is to compute the bounce action as a series in terms of the 
thin wall expansion parameter and assess its accuracy away from the thin wall regime. 
While the result in~\cite{Coleman:1977py} only had the leading order of the expansion, we manage 
to compute the series {\em analytically} up to the fourth order. 
Our calculation builds on the work done in Ref.~\cite{Ivanov:2022osf}, which in turn
built on Ref.~\cite{Munster:1999hr}.
We find that, truncating our expansion at the second order, our solution remains 
very accurate in a much wider range of parameter
space away from the thin wall limit, thus providing a significant improvement to the solution of~\cite{Coleman:1977py}. 
We perform several checks of the accuracy of our result by testing it against
numerical evaluations performed using the~\texttt{FindBounce} package~\cite{Guada:2020xnz},
together with the gradient 
flow method~\cite{Chigusa:2019wxb, Sato:2019axv, Sato:2019wpo, Hong:2023dan}.
We find good agreement with the numerics and the fits~\cite{Sarid:1998sn,Levi:2022bzt},
in a large range of parameters, between thin and thick wall configurations.
We believe that our analytic result brings a deeper understanding of the bounce
action in such a range, compared to the numerical fits.
For other recent works, complementary to the current one, that try to get a better
analytic understanding of the bounce action in thick wall regimes 
with different methods, see Refs.~\cite{Mukhanov:2022abt,Espinosa:2023tuj}.

A full calculation of the decay rate must also deal with the prefactor $A$ instead of leaving 
it to guesswork. 
The prescription for handling it was first developed in~\cite{Callan:1977pt} and involves 
computing a functional determinant, which is often challenging. 
Progress in computing $A$ has been made in some directions, including the calculation of the SM 
rate~\cite{Isidori:2001bm, Chigusa:2017dux, Andreassen:2017rzq, Baratella:2024hju} and in gauge 
theories~\cite{Chigusa:2018uuj,Chigusa:2020jbn}.
These works mostly used the Feynman diagrammatic techniques and an explicit Fourier transform 
of the bounce solution to regularize the determinant.
An alternative treatment of the determinant using the WKB/$\zeta$ function formalism was developed
in~\cite{Dunne:2005rt}, where the minimal subtraction in large multipoles is used to regulate 
the finite sum over the multipoles.
This approach, together with the application of the Gelfand-Yaglom theorem, led to the recent
complete calculation of $A$ in the thin wall limit for any $D$ in a simple closed
form~\cite{Ivanov:2022osf}.
There, the determinant was obtained at the leading order in the thin wall expansion, together with the 
zero removal, which is essential to obtain the correct dimension of the prefactor~\footnote{
Incidentally, an exact solution for $A$ and $B$ can be found for the quartic-quartic potential 
for any value of parameters of the potential~\cite{Guada:2020xnz}.}.

Whereas for $B$ we managed to compute analytically higher order corrections, the calculation of $A$ is much more involved.
In this paper we use a numerical approach~\cite{Isidori:2001bm} to calculate $A$, including the 
zero removal procedure. 
We do so in the full range between thin and thick wall,
then provide a simple fit function to the result. 
We run the numerics with the recently released tool \texttt{BubbleDet}~\cite{Ekstedt:2023sqc}, 
which computes the determinant for generic potentials,
and cross check the results with other methods implemented with our own code.
All methods agree with the analytic result of Ref.~\cite{Ivanov:2022osf} in the thin
wall limit.

The paper is organized in the following way.
We first set the notation and conventions for parametrizing the scalar potential and the Euclidean 
action in~\S\ref{sec:LinCub}.
In~\S\ref{sec:SExpansion} we introduce the thin wall expansion of the bounce action.
We calculate for the first time the terms of the series analytically up to the fourth order 
for any spacetime dimension $D$, and numerically for even higher orders when $D = 3, 4$.
In~\S\ref{sec:thick} we show that truncating the thin wall bounce action expansion at second order 
results in an excellent approximation also deep into thick wall regimes.
This is perhaps the most remarkable result of this work.
In~\S\ref{sec:fluctuations} we calculate the functional determinant at one loop,
explaining the zero modes removal procedure and the regularization of the UV
divergences. 
We conclude in~\S\ref{sec:summary} by putting together the full result for the false vacuum decay rate 
in a simple formula, which can be readily used for phenomenological applications.
We leave several technical details and cross checks of the calculations in the Appendices. 

%
%
\section{Two parametrizations, one physics} \label{sec:LinCub}

We consider the theory of a single real scalar field in $D$ space-time dimensions, with a polynomial 
potential up to $\phi^4$, featuring two minima at the tree level.
One is only local and thus is the false vacuum (FV), the other is the absolute minimum\footnote{
When we study thick walls in section~\ref{sec:thick} we also consider limiting cases which, instead
of two minima, feature (i) an absolute minimum and an inflection point, and (ii) a local minimum in an 
unbounded potential.}, the true vacuum (TV).
If the field starts at the FV, it will eventually tunnel to the TV. 
We are interested in computing the decay rate of the FV, relying on the methods introduced in 
Refs.~\cite{Coleman:1977py, Callan:1977pt}.
We want to find the bounce, the field configuration which extremizes the Euclidean action, and compute 
the quantum fluctuations around it, the functional determinant.

As the bounce solution is $O(D)$ symmetric~\cite{Coleman:1977th}, the starting point is 
the action
\begin{align} \label{eq:SDef}
  S &= \Omega \int_0^\infty {\rm d}\rho \ \rho^{D-1} \left( \frac{1}{2} \left( 
  \frac{{\rm d}\phi}{{\rm d}\rho} \right)^2 + V(\phi) - V_{\rm FV} \right) \, ,
  &
  \Omega &= \frac{2 \pi^{D/2}}{\Gamma \left(D/2 \right)} \, ,
\end{align}
where $\Omega$ is the solid angle in $D$ dimensions, $\rho$ the Euclidean radius and $V_\text{FV}$
denotes the value of the potential at the FV. 
Extremizing $S$ to find the bounce corresponds to solving
\begin{align} \label{bounceq}
  \ddot \phi + \frac{D-1}{\rho} \dot \phi &= \frac{{\rm d}V}{{\rm d}\phi} \, , 
  & 
  \phi \left(\rho = \infty \right) &= \phi_{\rm FV} \, , 
  & 
  \dot \phi \left(\rho = 0 \right) &= 0 \, .
\end{align}
Here, the dot denotes a derivative with respect to $\rho$.

As mentioned above, we are going to study polynomial potentials up to $\phi^4$ with a FV and a TV
at the tree level (not radiatively induced). 
In general, a polynomial of up to the fourth power has five parameters.
The constant term is irrelevant, because we are subtracting the $V_\text{FV}$ in the calculation of 
the action, moreover we restrict our analysis to flat spacetime, without including gravity.
One of the remaining four parameters can also be removed by shifting the scalar field by a constant.
We then consider the following two parametrizations: 
\begin{align}
  V_L(\phi_L) &= \frac{\lambda}{8} \left( \phi_L^2 - v^2 \right)^2 + \lambda \Delta v^3 \left( \phi_L - v \right) \, , 
  & \text{linear parametrization} \, , \label{linpar} 
  \\
  V_C(\phi_C) &= \frac{1}{2} m^2 \phi_C^2 + \eta \phi_C^3 + \frac{1}{8} \lambda_C \phi_C^4 \, , 
  & \text{cubic parametrization} \, . \label{cubpar}
\end{align}
The relation between the two is given by
\begin{align}
  \phi_C & = \phi_L - \phi_L^{\rm FV}  \, , 
  &
  \phi_L^{\rm FV} & = v \ \frac{\delta^2 + 3^{1/3} }{3^{2/3} \delta } \, ,
   \label{phishift} \\ 
  \delta &=\left[ 9 \left( \sqrt{\Delta^2 - \Delta_{\max}^2}  - \Delta \right) \right]^{1/3} \, , & \Delta_{\rm max} & = \frac{1}{3\sqrt{3}} \, ,
\end{align}
where $\phi_L^{\rm FV}$ is the value of the field at the FV.
We choose our parameters in the following ranges for linear and cubic:
\begin{align}
  \lambda 	& \geq 0 \, ,	& v    &> 0 \, , & 0 &< \Delta < \Delta_{\max}  \, , 
\nonumber  \\
  m^2      	& \geq 0 \, , 	& \eta &> 0 \, , & \lambda_C &< \frac{4\eta^2}{m^2} \, .  
   \label{paranges}
\end{align}
Taking $\Delta > 0$ places $\phi_L^{\rm FV} > 0$ and $\phi_L^{\rm TV} < 0$, while $\Delta < 0$ would 
switch the roles of FV and TV.

At $\Delta = \Delta_{\rm max}$ the FV in potential \eqref{linpar} turns into an inflection point; 
for larger $\Delta$ there is only one minimum, the TV, and there is no tunneling to discuss any longer.
With $\lambda \geq 0$, the potential considered here is bounded from below, and tunneling proceeds
from positive to negative field values.
In the cubic parametrization, the choice $\lambda_C < 4\eta^2/m^2$ is to fix the FV at $\phi_C^{\rm FV} = 0$, 
and the deeper minimum at $\phi_C^{\rm TV} < 0$.
When $\lambda_C > 0$, the potential is bounded from below. 
The bounce still exists for $\lambda_C \leq 0$, when the potential is unbounded.
Note that for $\lambda_C \geq 0$ tunneling proceeds from $\phi_C^{\rm FV}=0$ towards negative field values.
For strictly negative $\lambda_C$ the potential has a higher barrier at positive $\phi_C$ and a lower
one at negative $\phi_C$. 
This implies that tunneling can proceed in both directions, with a higher probability towards negative field values. 
Indeed, in that direction the bounce action has no discontinuity as one dials $\lambda_C$ from positive to 
negative~\cite{Sarid:1998sn}.
The inflection point at $\Delta = \Delta_{\rm max}$ corresponds to taking $m^2 = 0$ in 
the cubic parametrization.

Deciding which parametrization to use is a matter of convenience for the question to answer 
or the calculation to perform. 
Physical results do not depend on such a choice, as long as the action is calculated exactly.
In the rest of the paper we will see examples of how it can be useful to switch between the two 
parametrizations.

In both cases it is convenient to introduce variables that are dimensionless in any space-time dimensions $D$. 
For the linear case, we define
\begin{align}
\label{dimlessL}
  \varphi_L &\equiv \frac{\phi_L}{v} \, , 
  &
  \rhotiL &\equiv \sqrt{\lambda v^2} \ \rho \, ,
\end{align}
in terms of which the action becomes
\begin{equation} \label{Slin}
  \begin{split}
    S &= \Omega \frac{v^{4-D}}{\lambda^{D/2 - 1}} S_L(\Delta) 
    = \Omega \frac{v^{4-D}}{\lambda^{D/2 - 1}} \int_0^\infty {\rm d}\rhotiL \ \rhotiL^{D-1} \left( 
    \frac{1}{2} \left( \frac{{\rm d}\varphi_L}{{\rm d}\rhotiL} \right)^2 + \tilde{V}_L(\varphi_L) - 
    \tilde{V}_L(\varphi_L^{\rm FV}) \right) \, ,
  \end{split}    
\end{equation}
with the rescaled dimensionless linear potential
\begin{equation} \label{VtL}
  \tilde{V}_L(\varphi_L) = \frac{1}{8} \left( \varphi_L^2 - 1 \right)^2 + \Delta 
  \left( \varphi_L - 1 \right) \, .
\end{equation}
The action dependence on $\lambda$ and $v$ goes into the factorized $v^{4-D}/(\lambda^{D/2 - 1})$ and 
the remaining $S_L(\Delta)$ is solely a function of $\Delta$.

In the cubic case \eqref{cubpar}, we define
\begin{align} \label{dimlessC}
  \varphi_C  \equiv \frac{2\eta}{m^2} \phi_C \, , 
  \qquad 
  \rhotiC \equiv m \rho \, ,
  \qquad
  \epsal \equiv 1 - \lambda_C \frac{m^2}{4\eta^2} \, ,
  \qquad
  0 <  \epsal  \leq 1 \, .
\end{align}
Then the action in the cubic parametrization is given by
\begin{align} \label{Scub}
    S &= \Omega \frac{m^{6-D}}{4\eta^2} S_C(\epsal) 
= \Omega \frac{m^{6-D}}{4\eta^2}  \int_0^\infty {\rm d}\rhotiC \ \rhotiC^{D-1}  
  \left( \frac{1}{2} \left( \frac{{\rm d}\varphi_C}{{\rm d}\rhotiC} \right)^2 + 
  \tilde V_C(\varphi_C) \right) \\
 & = \Omega \frac{m^{4-D}}{\lambda_C}(1-\epsal) S_C(\epsal) \, ,
\end{align}
with the rescaled dimensionless cubic potential
\begin{equation} \label{VtC}
   \tilde V_C(\varphi_C) = \frac{1}{2} \varphi_C^2 + \frac{1}{2} \varphi_C^3 + 
   \frac{1-\epsal}{8} \varphi_C^4 \, ,
\end{equation}
and with $S_C(\epsal)$ only a function of $\epsal$.
In the rest of the paper we refer to the two sets of variables and parameters as
\begin{align}
  & \{ \varphi_L \, , \ \rhotiL \, , \ \lambda \, , \ v\, , \ \Delta \}, & \text{linear parametrization} \, ,  \label{linparam} 
  \\
  & \{ \varphi_C \, , \ \rhotiC \, , \ m \, , \ \eta \, , \ \epsal \}, & \text{cubic parametrization} \, . \label{cubparam}
\end{align}
An exact invertible map between the two sets is given in Appendix~\ref{sec:map}.

Now we want to compute the bounce and the corresponding action. This can be done, and has been done,
numerically for any choice of the parameters.
These numerical calculations can be performed in either parametrization and the results will match trivially after 
one translates between one and the other.
We would like to understand if it is possible to gain some analytic insight into these calculations.
The potentials above do not have an exact analytic bounce solution. However, they do have
an approximate analytic solution in the thin wall limit.
Our strategy is to first define perturbative expansions in powers of $\Delta$ and $\epsal$ 
around the thin wall limit, then study whether these expansions give accurate results when used
away from that limit.
We will see that, remarkably, they work all the way up to thick wall regimes. 

%
%
\section{Thin wall expansion with high order corrections} \label{sec:SExpansion}

In the linear and cubic parametrizations, the thin wall limit, where
FV and TV are almost degenerate and one can compute the tunneling rate analytically,
corresponds to $\Delta\to0$ in \eqref{VtL} 
and $\epsal\to0$ in \eqref{VtC}. 
In this limit, the potentials \eqref{VtL} and \eqref{VtC} have the same shape: 
one is shifted compared to the other along the horizontal $\varphi$ axis, which does 
not affect the physics. 
This implies that results obtained in the linear parametrization as an expansion in $\Delta$ must be 
in a one-to-one correspondence to the analogous result in the cubic parametrization at the same order 
in $\epsal$.
Some of the current authors performed the tunneling rate calculation in previous work~\cite{Ivanov:2022osf}, 
using the linear parametrization.
In this section we summarize and extend some of the results of Ref.~\cite{Ivanov:2022osf}. 
The goal here is to compute the bounce action in the thin wall limit, including high order corrections, 
in the expansions in $\Delta$ and in $\epsal$.

%
\subsection{Linear parametrization}

To compute the bounce it is useful to introduce first the dimensionless variable 
$z_L \equiv \rhotiL - r_L$ to describe the shape of the bubble.
Here $r_L$ is a dimensionless constant that corresponds to the size of the instanton, the bubble radius.
We can use the following thin wall expansions,
\begin{align}
  \varphi_L(z_L) &= \sum_{n \geq 0} \Delta^n \varphi_{Ln}(z_L) \, , 
  &
  r_L &= \frac{1}{\Delta} \sum_{n \geq 0} \Delta^n r_{Ln} \, ,\label{phi_r_expansion_linear}
\end{align}
plug them into the bounce equation~\eqref{bounceq}, and solve it order by order in $\Delta$.
The procedure is described in detail in~\cite{Ivanov:2022osf}.
Such a double expansion is useful for analytic calculations, but it is redundant:
the corrections to $r_L$ can be understood as the resummation of a series of corrections to $\varphi_L$ via derivative expansions.
In Appendix~\ref{app:bounce} we repackage the result into a single expansion.

With this setup, one can get the bounce action as a series in even powers of $\Delta$, 
\begin{equation} \label{SLexpa}
  S_L^{(N)}(\Delta) = S_L^{(0)} \left( 1 + \sum_{n = 1}^{N/2} \Delta^{2n} s^L_{2n}  \right) \, ,
\end{equation}
that truncates at a given order $N$. 
The leading order is given by
\begin{equation}
  S_L^{(0)} = \frac{1}{\Delta^{D-1}} \left( \frac{D-1}{3} \right)^{D-1} \frac{2}{3D} \, .
\end{equation}
The coefficients $s^L_2$ and $s^L_4$ can be computed analytically,
\begin{equation}
\begin{split}
  s^L_2 & = \frac{-8D^2 + (25 - 3 \pi^2)D +1}{2(D-1)} \, ,  \\
  s^L_4 & = \frac{1}{40 (D-1)^3}  \biggl( 320 D^5+80 D^4 \left(3 \pi ^2-49\right) 
  -3 D^3 \left(550 \pi ^2 + 3 \pi ^4 - 6185 \right)   \\
  & \quad \qquad \qquad \qquad + 5 D^2  \left( 426 \pi ^2+45 \pi^4 - 648 \zeta (3) - 7843 \right)  \\
  & \quad \qquad \qquad \qquad + D \left(3240 \zeta(3)+30635+360 \pi ^2-414 \pi ^4\right)+105 \biggr) \, .
\end{split}
\end{equation}
$s^L_2$ was computed in~\cite{Ivanov:2022osf}, while the calculation of $s_L^4$ is in Appendix~\ref{sec:sl4}.
The higher orders can be computed numerically, as we describe
in Appendix~\ref{sec:semianal}.
The functions $\varphi_{Ln}(z)$ have odd (even) $z$-parity for even (odd) $n$, while the radius 
coefficients $r_{Ln}$ vanish for odd $n$. We compute up to $s^L_{16}$ in $D = 3,4$, 
and report our results in Table~\ref{tab:slcoeff}.

\begin{table}
\centering
\begin{tabular}{ | l || *{8}{ c |}}
  \hline 
            & $s^L_2$ & $s^L_4$ & $s^L_6$   & $s^L_8$           & $s^L_{10}$        & $s^L_{12}$        
  & $s^L_{14}$          & $s^L_{16}$  
  \\ \hline
  $D = 3$   & $-21.2$ & $-57.6$ & $-977$    & $-2.01\cdot 10^4$ & $-4.73\cdot 10^5$ & $-1.24 \cdot 10^7$ 
  & $-3.65 \cdot 10^8$  & $-1.23 \cdot 10^{10}$     
  \\ \hline
  $D = 4$   & $-24.2$ & $7.53$  & $-266$    & $-5.86\cdot 10^3$ & $-1.21\cdot 10^5$ & $-2.50 \cdot 10^6$ 
  & $-5.08 \cdot 10^7$  & $-9.34 \cdot 10^8$     
  \\ \hline
\end{tabular}
  \caption{Coefficients $s^L_{2n}$ that enter \eqref{SLexpa}, evaluated numerically in $D = 3$ 
  and $4$.}
  \label{tab:slcoeff}
\end{table}

%
\subsection{Cubic parametrization}

We can apply the same strategy to the cubic parametrization and compute the bounce in the TW limit.
We define $z_C \equiv \rhotiC - r_C$ and the following expansions
\begin{align} \label{bounceandrc}
  \varphi_C(z_C) &= \sum_{n \geq 0} \epsal^n \varphi_{Cn}(z_C) \, , 
  & 
  r_C &= \frac{1}{\epsal} \sum_{n \geq 0} \epsal^n r_{Cn} \, .
\end{align}
%
We can solve them analytically up to $\varphi_{C2}$ and $r_{C2}$, see Appendix~\ref{app:cubic}.
The bounce action is 
\begin{equation} \label{SCexpa}
  S_C^{(N)}(\epsal) =  S_C^{(0)} \left( 1 + \sum_{n = 1}^N \epsal^n s^C_n \right) \, ,
\end{equation}
with
\begin{equation} \label{SC0}
  S_C^{(0)} = \frac{1}{\epsal^{D-1}} \left( \frac{D-1}{3} \right)^{D-1} \frac{2}{3D} \, .
\end{equation}
Whereas in the linear parametrization we only had even powers of $\Delta$, in the cubic we have 
both even and odd powers of the expansion parameter $\epsal$ in the action. 
With the knowledge of the bounce up to second order \eqref{bounceandrc},
we can find $s^C_1$ and $s^C_2$ via an explicit computation. 
However, there is no need to do it from scratch: after all the work done in the linear parametrization, 
we can simply translate those results into the cubic parametrization. 
The actions defined in~\eqref{Slin} and~\eqref{Scub} must be equal, which implies, using the map
in Appendix~\ref{sec:map},
\begin{align} \label{SLtoC}
  S_C(\epsal) = \frac{4\eta^2}{m^{6-D}} \frac{v^{4-D}}{\lambda^{D/2 - 1}} 
  S_L(\Delta(\epsal)) =  \frac{(1+ 2\epsal)^{2-D/2}}{(1-\epsal)^{3-D/2}} S_L(\Delta(\epsal)) \, .
\end{align}
Now we can replace $S_L$ on the right hand side with the expansion in \eqref{SLexpa}, translate from
$\Delta$ to $\epsal$, expand everything for small $\epsal$,
and read off the coefficients defined in~\eqref{SCexpa}. 
The leading order term matches with~\eqref{SC0} and beyond that we get
\begin{align}
  s^C_1 & =  \frac{3D}{2} + 4 \, , 
  \\
  s^C_2 & = \frac{9 D^3 -11D^2 +(138 -12\pi^2)D -64}{8 (D-1)} \, ,
  \\
  s^C_3 & = \frac{9 D^4-87 D^3 +(510 -36\pi^2)D^2 + (48 \pi ^2 -248)D   -256}{16 (D-1)} \, , 
  \\
  s^C_4 & = \frac{1}{640 (D-1)^3}   \left(135 D^7-3465 D^6+5 \left(7153-216 \pi ^2\right) D^5  \right. \nn
  & \qquad\qquad\qquad  + 5 \left(2208 \pi^2-34627\right) D^4-8 \left(-64250+5715 \pi ^2+18 \pi ^4\right) D^3   \nn
  & \qquad\qquad\qquad  \left. 
  +\left(  720 \pi ^2 \left(77+5 \pi ^2\right) - 848420  - 51840 \zeta(3) \right)D^2  \right. \nn
  & \qquad\qquad\qquad \left. + \left(51840\zeta(3)  -6624 \pi ^4-2400 \pi^2+589040  \right) D -10240 \right) \, .
\end{align}
Since the function $S_L(\Delta)$ is known analytically up to the fourth order in $\Delta$ expansion, we 
get $S_C(\epsal)$ analytically up to the same order. 
We can obtain up to order sixteen numerically, translating again the results we got in the linear parametrization, 
using~\eqref{SLtoC} and expanding in $\epsal$. 
The numerical coefficients $s^C_n$ are shown in Table~\ref{tab:sccoeff}.

\begin{table}[t!]
\centering
\begin{tabular}{ |  l  || *{8}{ c |}}
\hline 
  & $s^C_1$ & $s^C_2$ & $s^C_3$ & $s^C_4$ & $s^C_{5}$ & $s^C_{6}$ & $s^C_{7}$ & $s^C_{8}$  
  \\
  \hline
  $D = 3$ & $8.50$ & $8.67$ & $6.05$ & $-37.4$ & $217$ & $-1.63 \cdot 10^3$ & $1.12 \cdot 10^4$ & $-8.08 \cdot 10^{4}$     
  \\
  \hline
  $D = 4$ & $10.0$ & $17.3$ & $-2.96$ & $7.59$ & $5.00$ & $-215$ & $2.05 \cdot 10^3$ & $-1.67 \cdot 10^4$  
  \\
  \hline 
  \hline
  & $s^C_9$ & $s^C_{10}$ & $s^C_{11}$ & $s^C_{12}$ & $s^C_{13}$ & $s^C_{14}$ & $s^C_{15}$ & $s^C_{16}$  
  \\
  \hline
  $D = 3$ & $5.87 \cdot 10^5$ & $-4.35 \cdot 10^6$ & $3.27 \cdot 10^7$ & $-2.49 \cdot 10^8$ & $1.92 \cdot 10^9$ & 
  $-1.50 \cdot 10^{10}$ &   $1.19 \cdot 10^{11}$ & $-9.54 \cdot 10^{11}$     
  \\
  \hline
  $D = 4$ & $1.28 \cdot 10^5$ & $-9.53 \cdot 10^5$ & $7.04 \cdot 10^6$ & $-5.18 \cdot 10^7$ & $3.80 \cdot 10^8$ & 
  $-2.79 \cdot 10^9$ & $2.04 \cdot 10^{10}$ & $-1.49 \cdot 10^{11}$     
  \\
  \hline
\end{tabular}
\caption{Coefficients $s^C_{n}$ that enter \eqref{SCexpa} evaluated numerically in $D = 3$ and $4$.}
\label{tab:sccoeff}
\end{table}

%
%
\section{Approaching the thick wall} \label{sec:thick}

In the previous section we obtained the bounce action as a series expansion in the small parameters
$\Delta$ and $\epsal$. 
In this section we explore what happens when we apply those results away from that limit. 
There are different ways to depart from the thin wall, depending on which parameters we vary
and which we hold fixed. We saw that in the action \eqref{Slin} the parameters $\lambda$ and $v$
factor out in a fixed combination; the same is true for the parameters $m$ and $\eta$ in \eqref{Scub}.
Hence, it is trivial to get the bounce action for different values of those parameters.  
What is less trivial is to vary the dimensionless parameters $\Delta$ and $\epsal$. 
These considerations suggest to define the following thick wall limits:
\begin{figure}[t]
    \subfloat{{\includegraphics[width=.40\textwidth]{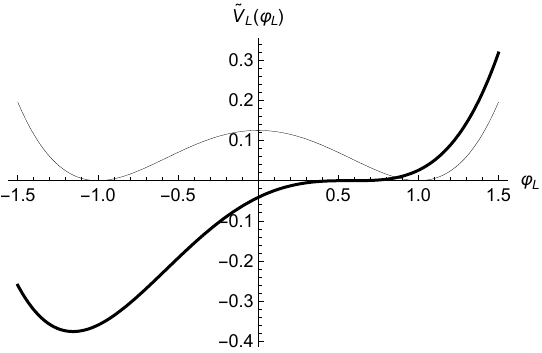} }}%
    \qquad \qquad
    \subfloat{{\includegraphics[width=.40\textwidth]{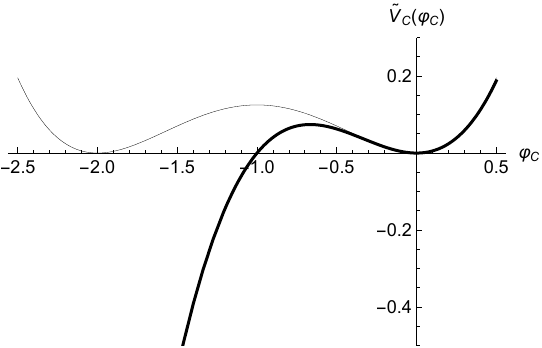} }}%
    \caption{On the left we show $\tilde V_L$ from Eq.~\eqref{VtL}. 
    The thin line corresponds to $\Delta = 0$, the thin wall limit; the thick line corresponds to 
    $\Delta = \Delta_{\rm max}$ and defines the inflection point thick wall. 
    On the right we have $\tilde V_C$ of Eq.~\eqref{VtC}. 
    Here the thin line corresponds to $\epsal = 0$, while the thick line corresponds to $\epsal = 1$ 
    and defines the vanishing quartic thick wall.}
    \label{fig:thick}
\end{figure}

\begin{enumerate}

\item{\em Vanishing quartic thick wall.}
We keep $m$ and $\eta$ fixed in the cubic parametrization and vary $\epsal$. 
In this case, $\epsal = 1$ corresponds to setting the quartic interaction to zero and the potential 
becomes unbounded from below as $\varphi_C \to - \infty$, see FIG.~\ref{fig:thick}. 
In this limit the bounce and its corresponding action stay finite.

\item {\em Inflection point thick wall.}
The potential in the linear parametrization is deformed by changing $\Delta$ and keeping $\lambda$ 
and $v$ fixed. 
At $\Delta = \Delta_{\rm max} = 1/\sqrt{27}$, the FV disappears and becomes an inflection point, 
see FIG.~\ref{fig:thick}. 
As we approach $\Delta_{\rm max}$ the bounce shrinks to a point in field space and the bounce 
action vanishes.
\end{enumerate}

In what follows we first examine whether the expansions in $\epsal$ and $\Delta$
work well up to these thick wall limits. 
In the ranges $0 \leq \epsal \leq 1$ and $0 \leq \Delta \leq \Delta_{\rm max}$ we can use the 
exact map of Appendix~\ref{sec:map} to translate between the two parametrizations; we will take
advantage of such a translation.

In the linear parametrization, keeping $\lambda$ and $v$ fixed and positive
and dialing $\Delta$ above $\Delta_{\rm max}$, we lose the false vacuum,
we are left only with an absolute minimum, and there is no longer tunneling.
On the other hand, in the cubic parametrization, we can keep $m$ and $\eta$ fixed and positive,
and dial $\epsal$ above 1, which corresponds to dialing $\lambda_C$ to negative values. In this
case the potential still has a local minimum at the origin, one high barrier to its right, one lower 
barrier to its left, and it is unbounded for $\varphi_C \to \pm \infty$. The probability of tunneling
to the left is higher, and the bounce solution for this tunneling direction has no discontinuity
as we go from $\epsal < 1$ to $\epsal > 1$. 
It is quite natural then to extend our study in the cubic parametrization to the region
$\epsal > 1$. We will do so in section~\ref{sec:unbounded}. 

%
\subsection{Vanishing quartic}

The vanishing quartic thick wall is defined in the cubic parametrization, fixing $m$ and $\eta$, 
and taking $\epsal \to 1$ (corresponding to $\lambda_C \to 0$).
In this limit the potential becomes unbounded from below as $\phi_C \to - \infty$, but there is still a false 
vacuum at the origin and a barrier to its left. 
As a consequence, there exists a finite bounce with the corresponding finite bounce action. 
It is straightforward to compute them numerically in the range $0 < \epsal \leq 1$, going 
from thin to thick wall. 
The numerical results can then be fitted by simple functions. 
Such a fit was performed in Ref.~\cite{Sarid:1998sn} for $D = 4$ and more recently in Ref.~\cite{Levi:2022bzt}, 
both in $D = 3$ and $D = 4$. 
We also ran our numerics and found agreement with both references. 
The fits provided in Ref.~\cite{Levi:2022bzt}, translated into our notation, are
\begin{align} \label{fit3} 
  S_C^{\rm fit}(\epsal) & = S_C^{(0)} \left( 
  1 + 8.50 \epsal + (8.21 + 1.35 \sqrt{1-\epsal}) \epsal^2 - 2.51 \epsal^3  \right) \, , & D = 3 \, , 
  \\ \label{fit4}
  S_C^{\rm fit}(\epsal) & = S_C^{(0)} \left( 1 + 10.0 \epsal + 17.0 \epsal^2 - 0.43 \epsal^3 \right) \, , & D = 4 \, , 
\end{align}
with $S_C^{(0)}$ given by~\eqref{SC0}.
Plugging $S_C^{\rm fit}(\epsal)$ into \eqref{Scub} one obtains a very accurate estimate of the bounce action 
in the whole range $0 < \epsal \leq 1$.
These fits bear a very close resemblance to our thin wall expansions \eqref{SCexpa} (apart from the term with $\sqrt{1-\epsal}$ 
in $D=3$, which is anyway small compared to its companion 8.21 inside the parentheses for the range of $\epsal$ 
under consideration), which up to $N = 5$ are
\begin{align} \label{thin3} 
  S_C^{(5)}(\epsal) & = S_C^{(0)} \left( 1 + 8.50 \epsal + 8.67  \epsal^2 + 6.05 \epsal^3 -37.4 \epsal^4 + 217 \epsal^5  \right) \, , & D = 3 \, , 
  \\ \label{thin4}
  S_C^{(5)}(\epsal) & = S_C^{(0)} \left( 1 + 10.0 \epsal + 17.3 \epsal^2 - 2.96 \epsal^3 +7.59 \epsal^4 + 5.00 \epsal^5 \right)  \, , & D = 4 \, . 
\end{align}

Up to $\epsal^2$ the coefficients of our expansion are in very good agreement with those obtained via the numerical fit.
This is remarkable, as we found them analytically with an ab-initio 
calculation starting from the thin-wall limit. 
However, our expansion, defined in \eqref{SCexpa} and derived for $\epsal \ll 1$, does not converge as $\epsal$ approaches 1. 
This is obvious from Table~\ref{tab:sccoeff}, where we see that the coefficients grow very large at higher orders.
Also, this should be expected, due to the following argument. 
The $S_C^{(N)}(\epsal)$ expansion can be constructed starting from $S_L(\Delta)$, see \eqref{SLtoC},
 and using the map \eqref{eg:CubToLin} 
to translate from $\Delta$ to $\epsal$. The function $\Delta(\epsal)$ has a singularity at $\epsal = -1/2$. This implies that,
when we expand $S_L(\Delta(\epsal))$ with respect to $\epsal$, the series will not converge for $|\epsal| > 1/2$.

 \begin{figure}[!t]
  \includegraphics[width=0.8\textwidth]{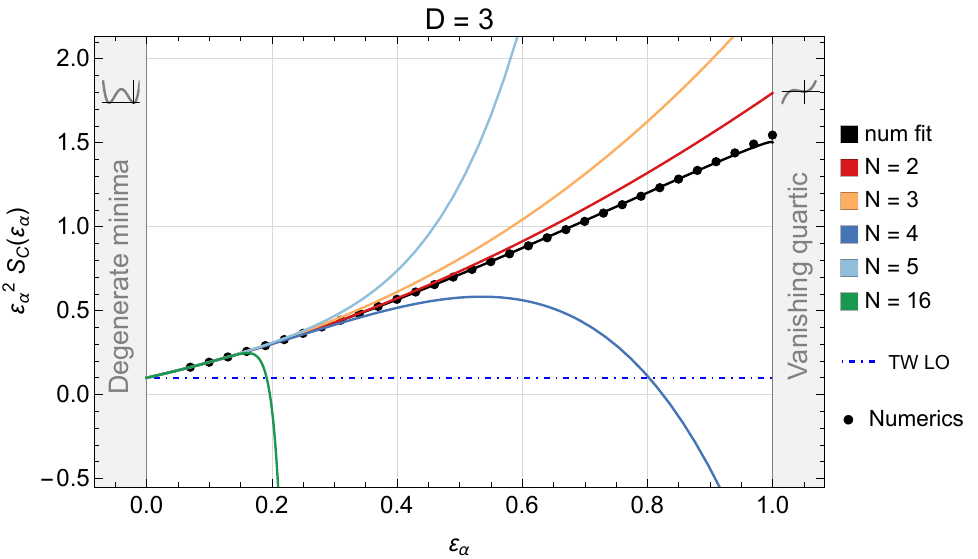}
  \caption{We plot $\epsal^2 S_C(\epsal)$ as a function of $\epsal$ in $D = 3$.
  The dots are our numerical results, obtained with a simple ``shooting'' method,
  as described in~\cite{Coleman:1977py}. 
  The black line is from the fit \eqref{fit3}, in perfect agreement with the numerics. 
  The dot-dashed blue line is the leading order thin wall result \eqref{SC0}.
  The colored lines are from our
  (semi)analytic thin-wall expansion of the bounce action \eqref{SCexpa},
  truncated at the order indicated in the legend.
  We see that at $N = 2$ we get the best approximation.
  Including up to $N = 16$ the bounce action diverges around $\epsal \simeq 0.2$.
}
   \label{fig:VanQuart3}
\end{figure}
\begin{figure}[!t]
  \includegraphics[width=0.8\textwidth]{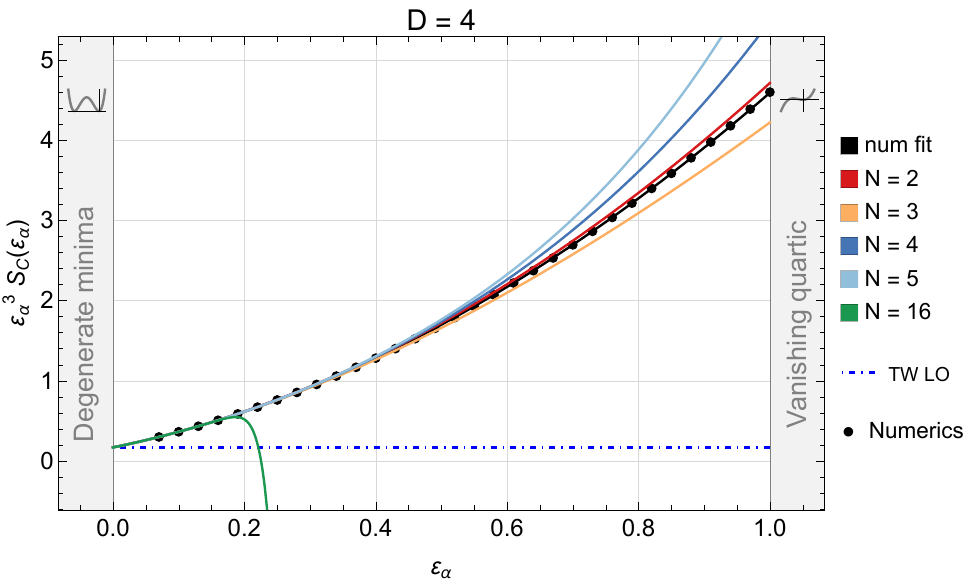}
 \caption{
 We plot $\epsal^3 S_C(\epsal)$ as a function of $\epsal$ in $D = 4$.
 The color code and the comments are the same as in FIG.~\ref{fig:VanQuart3}.
 Compared to $D = 3$, the expansion truncated at second order is 
 an even better approximation in the whole range from thin ($\epsal \to 0$) to
 thick ($\epsal \to 1$) wall.
}
  \label{fig:VanQuart4}
\end{figure}

Given these considerations, and comparing \eqref{thin3} and \eqref{thin4} to \eqref{fit3} and \eqref{fit4},
 it seems that simply truncating our analytic expansion at second order will give a good approximation
to the bounce action in the whole range $0<\epsal \leq 1$, from thin to thick wall. We can check this visually with FIGs.~\ref{fig:VanQuart3}
and~\ref{fig:VanQuart4}.
In FIG.~\ref{fig:VanQuart3} we plot the $D = 3$ case. We see that the 
value of the bounce action computed numerically quickly departs from the
leading order thin wall result~\cite{Coleman:1977py} given in \eqref{SC0} 
as we move away from $\epsal = 0$. 
Including higher order thin-wall corrections, as in \eqref{SCexpa}, we get a 
significant improvement. In particular, it is clear that truncating the expansion
at $N = 2$ gives the best approximation of the bounce action in the whole range
between thin ($\epsal \to 0$) and vanishing-quartic thick wall ($\epsal \to 1$).

In FIG.~\ref{fig:VanQuart4} we plot the $D = 4$ case. We have a picture very similar
to the $D = 3$ case. Truncating at order $N = 2$ the approximation is even better than in $D = 3$.
Both in $D = 3$ and $D = 4$ we find that including higher order in the expansions
makes the fit worse.

One can investigate further the nature of these expansions. 
In Appendix~\ref{app:Borel} we show that, including orders up to $n=40$, the coefficients grow factorially 
at large $n$.
This indicates an asymptotic series, so one could do a Borel analysis. 
This is beyond the scope of this paper and we leave a detailed such study to future work. 
We report some preliminary considerations in this direction in Appendix~\ref{app:Borel}.

%
\subsection{Inflection point}

\begin{figure}
  \includegraphics[width=0.8\textwidth]{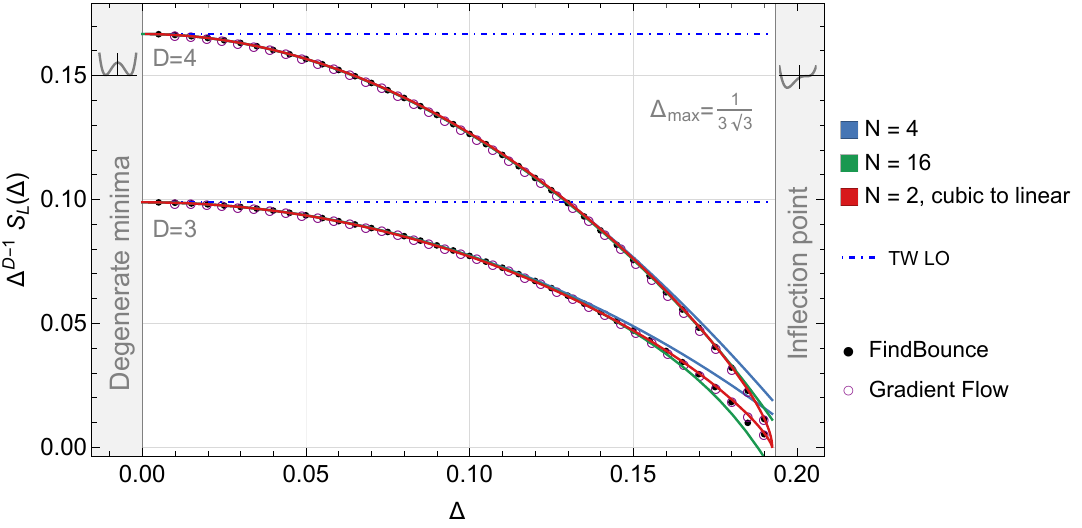}
  \caption{
  Euclidean bounce action times $\Delta^{D-1}$ in the linear parametrization for $D =3, (4)$ in lower (upper) lines, going from thin ($\Delta \simeq 0$) 
  to thick wall ($\Delta \simeq \Delta_{\max}$).
  Dot-dashed lines are the leading order thin-wall approximations, the solid ones include higher order corrections.
  The blue lines show the linear analytic result \eqref{SLexpa} up to $N=4$, the gray ones the 
  semi-analytic up to $N=16$. 
  The red lines are obtained from the action in the cubic parametrization, truncated
  at $N=2$ and translated into the linear parametrization, see \eqref{Scubtolin}.
  The filled  and empty dots are the numerical results from~\texttt{FindBounce} and gradient flow, respectively.
  }
  \label{fig:SD}
\end{figure}

Consider the bounce action in the linear parametrization, keeping $\lambda$ and $v$ fixed
and changing $\Delta$ from 0 to $\Delta_{\rm max} = 1/\sqrt{27}$. As we have seen, 
$\Delta \to 0$ corresponds to the thin wall regime, with the two vacua almost 
degenerate; as $\Delta$ approaches $\Delta_{\rm max}$ the false vacuum merges
with the maximum into an inflection point, and the potential difference with
the true vacuum becomes large. We refer to this configuration as the inflection
point thick wall.

In FIG.~\ref{fig:SD} we plot the bounce action $S_L(\Delta)$
multiplied by $\Delta^{D-1}$ in the range $0 < \Delta < \Delta_{\rm max}$. 
The filled and emtpy dots are the product of numerical evaluations, with two different
methods.
The plain lines are the result of our analytic calculation up to $N=4$, and semi-analytic up to 
$N=16$.
The dot-dashed blue line is the leading order result in the thin wall.
We see that the latter quickly ceases to provide a good approximation of the bounce
action as we move away from $\Delta = 0$. 

On the other hand, and quite remarkably, keeping higher order thin-wall corrections appears to give 
an excellent approximation in a wide range of $\Delta$.
By eye we see that close to $\Delta_{\rm max}$ we start having some discrepancies compared to 
the numerical results. 
%
%

\begin{figure}
  \includegraphics[width=0.8\textwidth]{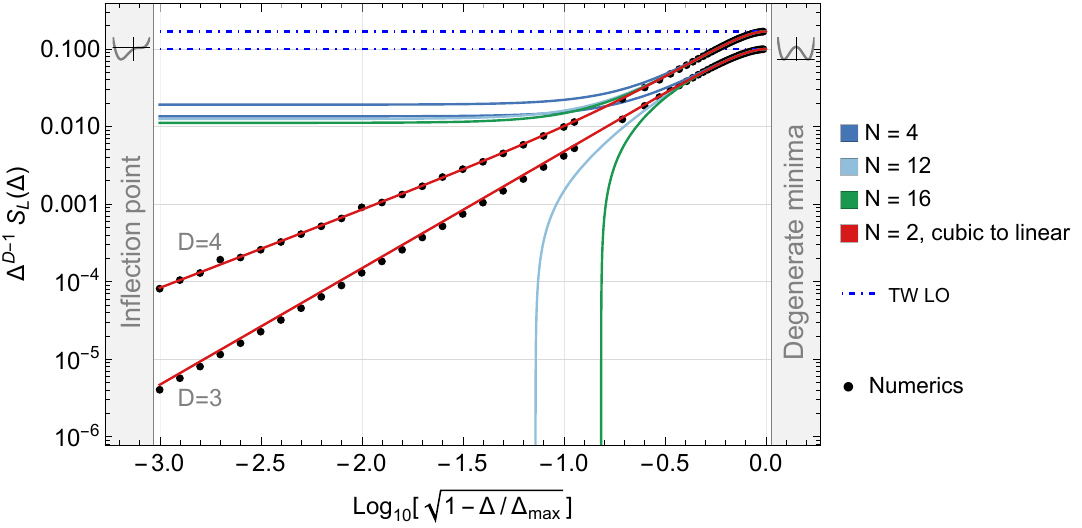}
  \caption{
  This is the logarithmic version of FIG.~\ref{fig:SD}; we have changed the horizontal axis such that now 
  we approach the inflection point as we move to the left. 
  Compared to FIG.~\ref{fig:SD} we have added the green line which corresponds to \eqref{SLexpa} 
  truncated at $N=12$. 
  We see that, no matter where we truncate, the expansion \eqref{SLexpa} does not work well as we 
  get close to the inflection point. 
  Instead the red line, obtained from the cubic expansion \eqref{SCexpa} truncated at second order 
  and translated into the linear parametrization, see \eqref{Scubtolin}, provides an excellent approximation 
  in the whole range.
  }
  \label{fig:Inflection}
\end{figure}

Let us check more explicitly by plotting $\Delta^{D-1} S_L(\Delta)$ on a logarithmic scale as function of 
${\rm Log}_{10}[\sqrt{1 - \Delta/\Delta_{\rm max}}]$ in FIG.~\ref{fig:Inflection},
in order to focus on values of $\Delta$ closer to $\Delta_{\rm max}$.
We see that both in $D = 3$ and $D = 4$ it turns out that truncating at $N = 12$ is slightly better 
than $N=16$. 
However, as we get closer to the inflection point, none of these expansions provides a good approximation.

It is time to talk about the red line, which we show both in FIGs.~\ref{fig:SD} and~\ref{fig:Inflection}, 
and clearly provides the best approximation.
It is calculated as follows. 
By equating \eqref{Slin} and \eqref{Scub} we have
\begin{equation} \label{Scubtolin}
  S_L(\Delta) = \frac{\lambda^{D/2-1}}{v^{4-D}} \frac{m^{6-D}}{4\eta^2}S_C(\epsal) 
  \simeq \frac{f_{m^2}^{3-D/2}(\Delta)}{f_\eta^2(\Delta)} S_C^{(2)}(\epsal(\Delta)) \, .
\end{equation}
In the last equality we substituted $S_C(\epsal)$ with its approximate version~\eqref{SCexpa} truncated at 
second order, then translated from $\epsal$ to $\Delta$ using the map~\eqref{epsaldelta}.
We also used~\eqref{metatoL}, \eqref{fmsqdef}, \eqref{fetadef} to get the factor in front of 
$S_C^{(2)}(\epsal(\Delta))$.
Note that $f_{m^2}(\Delta \to \Delta_{\rm max}) \to 0$, while $f_\eta(\Delta \to \Delta_{\rm max})$ stays 
finite. 
It follows that at the inflection point the bounce action vanishes, $S_L(\Delta \to \Delta_{\rm max}) \to 0$. 
This is the expected behavior: at $\Delta \to \Delta_{\rm max}$ there is no longer a barrier in the 
potential and no field space to tunnel through.
Indeed, the bounce field configuration shrinks to a single point, $\phi_L(\rho = 0) = 
\phi_L(\rho = \infty) = \phi_{L{\rm inflection}} = v/\sqrt{3}$, and the bounce action vanishes.
While the expansion in the linear parametrization does not capture this limiting behavior,
the construction in the cubic parametrization naturally does. 
In the latter, the inflection point limit corresponds to taking $m^2 \to 0$. 
Then the bounce field configuration $\phi_C = m^2/(2\eta) \, \varphi_C$ shrinks to zero,
even if its dimensionless counterpart $\varphi_C$ remains finite\footnote{
The dimensionless bounce field $\varphi_C$ is computed following the thin-wall construction described
in~\cite{Ivanov:2022osf}, and in Appendix~\ref{app:cubic} for the cubic parametrization.
The construction is such that $\varphi_C$ connects true (at $\rho = 0$) and false vacuum (at $\rho = \infty$)
at every order in the expansion $\epsal$, hence it always has a finite extension.
}. 
Likewise, the action \eqref{Scub},
proportional to $m^{6-D}$, goes to zero.
For these reasons, starting from the cubic parametrization to get $S_L(\Delta)$
as in \eqref{Scubtolin} produces an excellent approximation of the bounce action in
the whole range of $\Delta$ between 0 and $\Delta_{\rm max}$, as clearly shown by the
red lines in FIGs.~\ref{fig:SD} and \ref{fig:Inflection}.

\subsection{Unbounded potential} \label{sec:unbounded}
\begin{figure}[t]
  \includegraphics[width=0.8\textwidth]{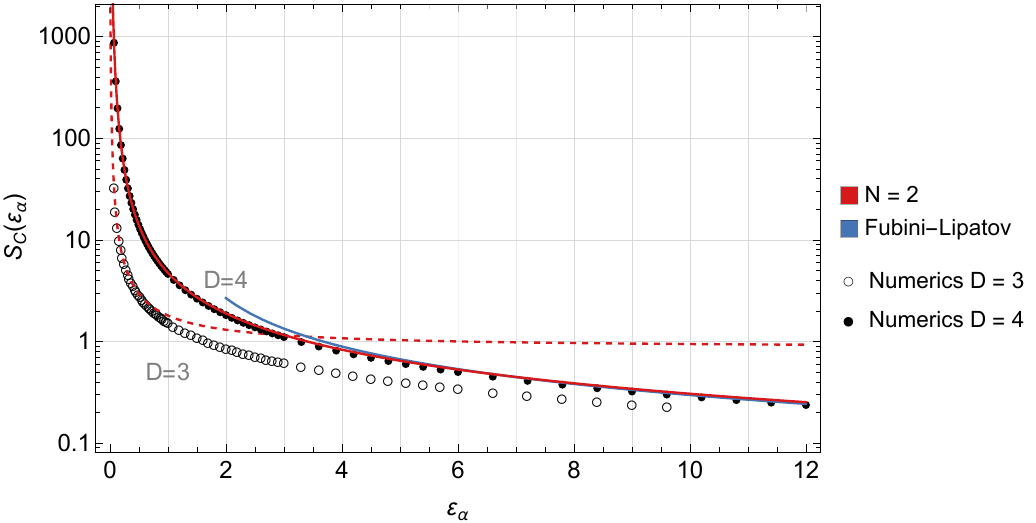}
 \caption{We show $S_C(\epsal)$ as a function of $\epsal$. 
 The red lines correspond to \eqref{SCexpa} truncated at $N=2$ in $D=3$ (dashed) and
 $D=4$ (plain). The blue line corresponds to the Fubini-Lipatov bounce action in $D=4$,
 see \eqref{SCFL}.
 The filled and empty dots are the results of numerical evaluations using the shooting method.
}
  \label{fig:unbounded}
\end{figure}

Let us consider again the cubic parametrization with $m$ and $\eta$ fixed. Now we want to push $\epsal$ to values
larger than 1, which correspond to a negative $\lambda_C$. The potential then becomes unbounded, but there 
is still a bounce solution for tunneling from the false vacuum toward negative field values\footnote{
There is also a solution for tunneling toward positive field values, but it is suppressed as the barrier is higher
in that direction. We will only deal with tunneling toward negative field values.
}. 
Such a solution is readily evaluated numerically, and the results are shown in 
FIG.~\ref{fig:unbounded} for $D=3$ (empty circles) and $D=4$ (black dots). We see,
as expected, that there is no discontinuity in the bounce action as we go over $\epsal =1$.

Now let us make a bold move and use our bounce action \eqref{SCexpa}, 
calculated analytically in the thin wall regime and truncated at $N=2$, 
also in the range $\epsal > 1$, which is clearly far away from thin wall. 
The result is shown by the red lines in FIG.~\ref{fig:unbounded}. 
In $D=3$ (red-dashed) the approximation for $\epsal > 1$ is decent, but not
great; the line deviates from the numerical result, but the scale on the 
vertical axis is logarithmic, so the discrepancy is not so large.
What is striking is the remarkable agreement in $D=4$ (red-plain) of our
truncated thin-wall expansion with the numerics all the way to large $\epsal$.

How can it be that the result obtained in the thin wall approximation works 
so well so far away from the regime of its derivation?
In $D=4$ we can rewrite the action as
\begin{equation} \label{SCD4}
S = 2\pi^2 \frac{m^2}{4 \eta^2} S_C(\epsal) = 2\pi^2 \frac{1-\epsal}{\lambda_C}S_C(\epsal) \, ,
\end{equation}
where we used $1- \epsal = \lambda_C m^2/(4\eta^2)$. Now we can think 
of keeping the ratio $m^2/(4\eta^2)$ fixed and send $\lambda_C$ to large
negative values. 
In this limit, the original potential is approximated by
$V \approx \frac{\lambda_C}{8}\phi_C^4$, which is scale invariant in $D=4$.
This admits the Fubini-Lipatov~\cite{Fubini:1976jm,Lipatov:1976ny} 
bounce solution with the action
\begin{align} \label{SFL}
S_{\rm FL} &= \frac{16\pi^2}{3(-\lambda_C)} \, , & \lambda_C &< 0 \, .
\end{align}
Comparing \eqref{SCD4} and \eqref{SFL} we get
\begin{align} \label{SCFL}
S_C^{\rm FL}(\epsal) &= \frac{8}{3}\frac{1}{\epsal -1} \approx \frac{8}{3 \epsal} \, , & {\rm for} \ \epsal &\gg 1 \, ,
\end{align}
which we plot as the blue line in FIG.~\ref{fig:unbounded}.
We see that the Fubini-Lipatov solution is indeed an excellent approximation at 
large $\epsal$.
Our thin wall expansion truncated at second order is
\begin{equation} \label{SC4tr2}
S_C^{(2)}(\epsal) = \frac{1}{6 \epsal^3} \left(1 + 10 \epsal + (37 - 2\pi^2)\epsal^2 \right) \ 
\xrightarrow[\epsal \gg 1]{} \ \frac{1}{6\epsal}(37 - 2\pi^2) \, .
\end{equation}
We get the same form as \eqref{SCFL} with a slightly different coefficient. 
It is surprising that the two coefficients are actually so close, given that 
the two analytic results are derived in opposite limits. 

It is crucial to truncate the thin wall expansion at second order,
or we would not match the behavior $S_C(\epsal) \propto \epsal^{-1}$ at 
large $\epsal$. It is remarkable that the same truncation gives an excellent
approximation in such a wide range of $\epsal$.

\subsection{Thick wall summary}
We have analyzed different thick wall limits for a real scalar 
with a polynomial potential up to $\phi^4$. 
Starting with a bounce action written as an expansion in the thin-wall parameter, we have explored whether it 
provided a good fit also in thick wall regimes. 
We have found that, in all the cases considered, the bounce action
calculated in the cubic parametrization, with the thin-wall expansion parameter
$\epsal$ defined in \eqref{dimlessC}, truncated at second order in $\epsal$,
gives an excellent approximation of the true bounce action. 

Concretely, starting with the potential
\begin{equation}
V = \frac{1}{2}m^2 \phi^2 + \eta \phi^3 + \frac{\lambda}{8} \phi^4 \, ,    
\end{equation}
and defining
\begin{equation}
\epsal = 1 - \lambda \frac{m^2}{4\eta^2} \, ,    
\end{equation}
one could use the expressions
\begin{align}
S & = 4\pi \frac{m^3}{4\eta^2} \frac{1}{\epsal^2}\left(\frac{2}{3}\right)^2 \frac{2}{9}
\left[1 + \frac{17}{2} \epsal + \left(\frac{247}{8} - \frac{9\pi^2}{4} \right) \epsal^2 \right] & D = 3 \, , \label{SsumD3} \\
S & = 2\pi^2 \frac{m^2}{4\eta^2} \frac{1}{6 \epsal^3} 
\left[1 + 10 \epsal + \left(37 - 2\pi^2 \right) \epsal^2 \right] & D = 4 \, ,
\label{SsumD4}
\end{align}
to evaluate accurately the bounce action from thin ($\epsal \ll 1$)
to thick wall configurations (any larger $\epsal$). 
In $D=3$ this approximation deteriorates a bit for
$\epsal > 1$ (see FIG.~\ref{fig:unbounded}), while in $D=4$ this approximation is 
excellent for any $\epsal > 0$.
In the range $0 < \epsal \leq 1$ the expressions \eqref{SsumD3} and \eqref{SsumD4}
can easily be translated into the linear parametrization using the exact map
in Appendix~\ref{sec:map}, and used to obtain very accurately the bounce action for $\lambda > 0$,
$v>0$ and $0< \Delta < \Delta_{\max}$.

We stress that our expansion for the bounce action up to second order was computed
analytically, and the fact that it works so well even far away from the regime of its
derivation is remarkable. 
Until now the common lore in the literature was that the thin wall approximation
does not provide a good description as soon as one departs from the thin wall limit.
Here we have shown that improving the original result of~\cite{Coleman:1977py}, by
including the next two corrections in the
expansion, one gets a result that describes accurately the bounce
action also deep into thick wall regimes.

%
%
\section{Functional determinants} \label{sec:fluctuations}
The vacuum decay rate per volume is given by~\cite{Callan:1977pt}
\begin{align}
\frac{\Gamma}{{\cal V}} & =  \left(\frac{S}{2\pi \hbar}\right)^{D/2} e^{-\frac{S_R}{\hbar} - S_{\rm ct}-\frac{1}{2}\ln  \left\vert \frac{{\rm det}'{\cal O}}{{\rm det}{\cal O}_{\rm FV}} \right\vert} (1 + O(\hbar)) \, .
\label{GamovV}
\end{align}
Here, $S_R$ is the renormalized bounce action and $S_{\rm ct}$ its one-loop counterterm. $S_R$ is obtained from the bounce actions $S$ we studied in the previous section,
upon computing the one-loop running of the couplings. Recall that the action takes the form
$S = \Omega \frac{v^{4-D}}{\lambda^{D/2-1}} S_L(\Delta)$ in the linear parametrization, or $S =\Omega \frac{m^{4-D}}{\lambda}(1-\epsal) S_C(\epsal)$ in the cubic, in units of $\hbar = 1$.
Here $S_L(\Delta)$ and $S_C(\epsal)$ are functions only of the dimensionless parameters $\Delta$ and $\epsal$.
In order to restore the $\hbar$ power counting, one has to perform the following rescalings: $\lambda \to \hbar \lambda$, $v \to \hbar^{-1/2} v$, $m \to m$, $\eta \to \hbar^{1/2} \eta$.
Note that $\Delta$ and $\epsal$ do not rescale with $\hbar$. Then from \eqref{GamovV}, where the powers of $\hbar$ are explicit, it is clear that $S_{\rm ct}$ and
$\ln \left\vert \frac{{\rm det}'{\cal O}}{{\rm det}{\cal O}_{\rm FV}} \right\vert$ must be proportional to $\frac{\lambda^{D/2-1}}{v^{4-D}} S_R$ (linear parametrization)
or to $\frac{\lambda}{m^{4-D}} S_R$ (cubic parametrization). In other words, both $S_{\rm ct}$ and the ln of the ratio of determinants are functions only of the dimensionless
parameters $\Delta$ or $\epsal$. 
Via explicit calculations we find that $\Delta$ and $\epsal$ do not run at one loop.
The recipe \eqref{GamovV}, with the explicit small $\hbar$ expansion, makes it clear that the calculation is done at one loop, 
ignoring higher loop contributions.
It follows that our final result for the decay rate will include the first order (one loop) quantum corrections
and thin-wall corrections in $\Delta$ or $\epsal$ up to higher orders.  

In this section we discuss the calculation of $\ln \left\vert \frac{{\rm det}'{\cal O}}{{\rm det}{\cal O}_{\rm FV}} \right\vert$, where ${\cal O} = -\partial^2 + \frac{{\rm d}^2V}{{\rm d}\phi^2}$,
with the second derivative of the potential evaluated at the bounce field configuration. In ${\cal O}_{\rm FV}$ the second derivative of the potential is evaluated at the constant
false vacuum field value.
 The prime on the determinant at the numerator
denotes that the zero modes, related to translational invariance, are removed.
In Ref.~\cite{Ivanov:2022osf}, the functional determinant was computed analytically at the leading order in the thin wall expansion parameter $\Delta$. It was found
that $\ln \left\vert \frac{{\rm det}'{\cal O}}{{\rm det}{\cal O}_{\rm FV}} \right\vert$ is proportional to $\Delta^{1-D}$, like the bounce action.
In the spirit of the first part of the paper, we would like to compute corrections to this result as a series expansion in $\Delta$.
However, the analytic calculation of the functional determinant already proved to be formidably challenging
at the leading order, and we do not think it is possible to get higher orders
analytically.
Thus we follow a different route.
We focus on the cubic parametrization and 
compute the functional determinant numerically in the range $0 < \epsal \leq 1$,
 then we fit it with a polynomial function.
In Appendix \ref{app:FunDet} we provide several checks of the calculation, 
using the linear parametrization in the range $0 < \Delta < \Delta_{\rm max}$. 
In the rest of this section we describe the setup and the technical points of the calculation.

Given the $O(D)$ symmetry of the problem, the radial part of the functional determinant can be separated 
and the angular part expanded in hyper-spherical multipoles denoted by $l$
\begin{equation}
  \ln\left(\frac{\det \mathcal O}{\det \mathcal O_\text{FV}}\right)  = 
   \sum_{l=0}^\infty\ln \left(\frac{\det{\cal O}_l}{\det{\cal O}_{l{\text{FV}}}}\right) \, ,
\end{equation}
with
\begin{align}
  {\mathcal O}_l &= -\frac{\mathrm{d}^2}{\mathrm{d} \rho^2} - \frac{D-1}{\rho} \frac{\mathrm{d}}{\mathrm{d} \rho} + 
  \frac{l \left( l + D - 2 \right)}{\rho^2} + V^{(2)} \, ,
\end{align}
where $V^{(2)} = \mathrm{d}^2 V/\mathrm{d} \phi^2$ is evaluated on the bounce. 
We have dropped for now the prime from the determinant at the numerator, we will get back to 
the zero removal shortly.
Using the Gelfand-Yaglom method~\cite{Gelfand:1959nq}, the ratio of determinants is recast as
\begin{equation} \label{Rldef}
  \frac{\det{\cal O}_l}{\det{\cal O}_{l{\text{FV}}}} = \lim_{\rho \to \infty} R_l(\rho)^{d_l} \, ,
\end{equation}
where the degeneracy factor is
\begin{align} \label{eqDegL}
  d_l &= \frac{(2l + D-2)(l+D-3)!}{l! (D-2)!} \, .
\end{align}
The quantity $R_l(\rho)$ solves the following differential equation
\begin{align} \label{Rlequ}
  & \frac{1}{R_\nu} \frac{\mathrm{d}^2 R_\nu}{\mathrm{d}\rho^2} + 2 \frac{1}{R_\nu} \frac{\mathrm{d} R_\nu}{\mathrm{d}\rho} 
  \left( \frac{\mathrm{d} \psi_{\nu\text{FV}} / \mathrm{d}\rho}{\psi_{\nu\text{FV}}} \right) -
  \left(\ddV - \ddV_\text{FV} \right)  = 0  \, , 
\end{align}
where we traded the multipole label $l$ for its better $D$-dimensional version
\begin{equation}
\nu = l - 1 + D/2 \, , 
\end{equation}
and $\psi_{\nu{\rm FV}}$ satisfies
\begin{align}   
  \left( -\frac{\mathrm{d}^2}{\mathrm{d} \rho^2} + \frac{\nu^2 - \frac{1}{4}}{\rho^2} + 
  \ddV_\text{FV} \right) \psi_{\nu \text{FV}} = 0 \, ,
\end{align}
with boundary conditions
\begin{align}
  \psi_{\nu \text{FV}} (\rho \to 0) &\sim \rho^{\nu + \frac{1}{2}} \, , 
  & R_\nu(\rho = 0) &= 1 \, , & \frac{\mathrm{d} R_\nu}{\mathrm{d}\rho}(\rho = 0) &= 0 \, .
\end{align}
The sum over the multipoles can be recast as
\begin{equation} \label{logdetRnumain} 
  \ln \left(\frac{\det \mathcal O}{\det \mathcal O_\text{FV}} \right) = 
  \sum_{\nu = D/2-1}^\infty d_\nu \ln R_\nu \, ,
\end{equation}
where it is understood that $R_\nu = \lim_{\rho \to \infty} R_\nu(\rho)$.

\subsection{Removal of zero modes}\label{sec:removal}
The functional determinant contains zero eigenvalues related to the translational invariance of 
the bounce, which must be removed. This is crucial in order to get the correct 
dimensions~\cite{Callan:1977pt} of the vacuum decay rate per volume \eqref{GamovV}.
Such zero modes correspond to the $l=1$ multipole, and imply that ${\rm lim}_{\rho\to\infty}R_{l=1}(\rho) = 0$, 
with $R_l$ defined in \eqref{Rldef}.

In the Gelfand-Yaglom method the zero removal procedure can be implemented as follows.
We start with
\begin{align} \label{Rlequoff}
  &\left[ \frac{\mathrm{d}^2}{\mathrm{d}\rho^2} + 2  
  \left( \frac{\mathrm{d} \psi_{1\text{FV}} / \mathrm{d}\rho}{\psi_{1\text{FV}}} \right) \frac{\mathrm{d} }{\mathrm{d}\rho} -
  \left(\ddV - \ddV_\text{FV} \right) -\mueps^2 \right] R_1^\epsilon(\rho) = 0  \, .
\end{align}
This is the operator of \eqref{Rlequ} written for $l=1$ (that is $\nu = D/2$), to which we have added a small offset
$\mueps^2$
We have also defined
\begin{align} \label{delR1def}
  R_1^\epsilon(\rho) =R_1(\rho) +\mueps^2 \, \delta R_1(\rho) \, .
\end{align}
To satisfy the boundary conditions we must have
\begin{align}
  & R_1^\epsilon(0)=R_1(0)=1 \, ,& \delta R_1(0)=0 \, ,
  \\
  & \dot{R_1^\epsilon}(0)=\dot{R_1}(0)=0 \, , & \dot{\delta R_1}(0)=0 \, ,
\end{align}
where the dot denotes a derivative with respect to $\rho$.
We then need to compute 
\begin{equation} \label{R1prime}
\lim_{\mueps^2 \to 0} \frac{1}{\mueps^2} R^\epsilon_1(\infty) = \delta R_1(\infty) \equiv  \frac{R^\prime_1}{m^2} \, .
\end{equation}
Note that $\delta R_1(\infty)$ has dimensions of an inverse squared mass, while with this definition\footnote{
In Ref.~\cite{Ivanov:2022osf} a dimensionful definition of $R_1^\prime$ was used. The definition in this work 
is more convenient.
}
$R_1^\prime$ is dimensionless, in any $D$. 
$R^\prime_1$ replaces $R_1 = 0$ in the functional determinant:
\begin{equation}
\left\vert \frac{{\rm det}^\prime {\cal O}}{{\rm det} {\cal O}_{\rm FV}} \right\vert^{-1/2} = \left[ |R_0|^{d_0} \left(\frac{R_1^\prime}{m^2}\right)^{d_1} \prod_{l=2}^\infty R_l^{d_l} \right]^{-1/2} 
= m^D \left[ \prod_{l=0}^\infty R_l^{d_l} \right]^{-1/2}  \, .
\end{equation}
Here, $R_0 < 0$ corresponds to the negative eigenvalue at $l=0$ with $d_0 = 1$; at $l=1$ we have $d_1 = D$ [see \eqref{eqDegL}]. In the last equality the factor $m^D$ makes it evident that the decay rate has the 
correct dimensions; in the final product it is understood that we must take the absolute value for $R_0$, 
and we must use $R_1^\prime$ in place of $R_1$.

\begin{figure}[t]
    \centering
    \includegraphics[width=0.9\textwidth]{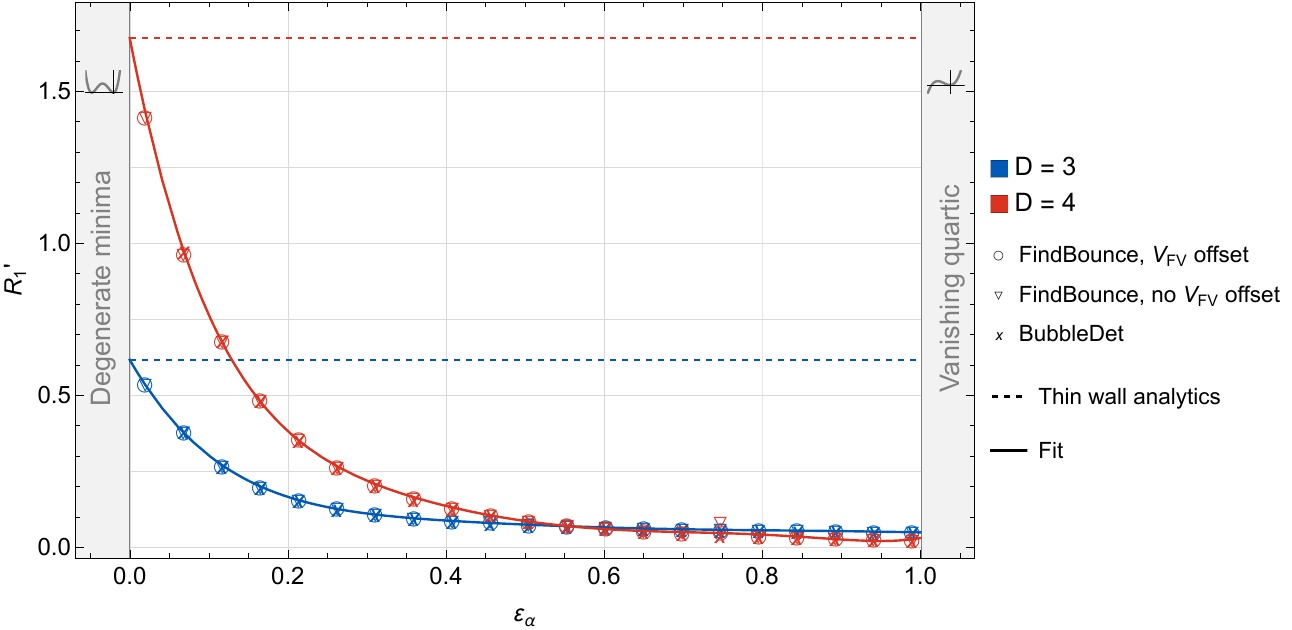}
    \caption{\footnotesize{
   We plot $R_1^\prime$, defined in \eqref{R1prime}, as a function of $\epsal$.
   We use a few different methods for the numerical evaluation: (i) we compute
   the bounce using \texttt{FindBounce}, then plug it into \eqref{delR1eq} [no $V_{\rm FV}$ offset];
    (ii) with the same bounce we use the method ``$V_{\rm FV}$ offset'' described in Appendix~\ref{app:FunDet};
     (iii) we use the package \texttt{BubbleDet} \cite{Ekstedt:2023sqc}. All these methods are in good agreement. We also
   show the fits \eqref{fitzero3} and \eqref{fitzero4} as solid lines, and 
   the analytic result \eqref{R1primethin} obtained in \cite{Ivanov:2022osf} 
   in the thin-wall limit ($\epsal \to 0$) as dashed lines.
   }
    }
    \label{fig:zero_removal}
\end{figure}

In order to compute $\delta R_1(\infty)$ we plug \eqref{delR1def} into \eqref{Rlequoff} and collect terms
of order $\mueps^2$:
\begin{align} \label{delR1eq}
  \left[ \frac{\mathrm{d}^2}{\mathrm{d}\rho^2} + 2  
  \left( \frac{\mathrm{d} \psi_{1\text{FV}} / \mathrm{d}\rho}{\psi_{1\text{FV}}} \right) \frac{\mathrm{d} }{\mathrm{d}\rho} -
  \left(\ddV - \ddV_\text{FV} \right) \right] \delta R_1(\rho) = R_1(\rho) \, .
\end{align}
On the right hand side, $R_1(\rho)$ is the solution to \eqref{Rlequ} with $l=1$. 
Note that, as $R_1(\rho)$ is dimensionless, switching to dimensionless variables 
 \eqref{dimlessC}
on the left hand side of \eqref{delR1eq} makes it clear that $\delta R_1$ is proportional to $(m^2)^{-1}$ times a function of $\epsal$.
It follows that $R_1^\prime$ is a function of $\epsal$ only. 
We solve \eqref{delR1eq} numerically for different 
values of $\epsal$, then extract ${\rm lim}_{\rho\to\infty} \delta R_1(\rho)$, and obtain the result shown in FIG.~\ref{fig:zero_removal}.

The analytic thin-wall result at the leading order in $\epsal$ found in Ref.~\cite{Ivanov:2022osf} corresponds to
\begin{align} \label{R1primethin}
 R^\prime_1 = \frac{e^{D-1}}{12} \, .
\end{align}
Our numerics are in agreement with this as $\epsal$ approaches zero. 
The factor $e^{D-1}$ 
appeared in a non trivial way in \cite{Ivanov:2022osf}, and we are not aware of other thin-wall
calculations which reproduce it.

The numerical results obtained by solving \eqref{delR1eq} and shown in FIG.~\ref{fig:zero_removal} are well fitted by the functions
\begin{align}
\!\!\! \!\!\! \!\! R_1^\prime(\epsal) & = \frac{e^2}{12} \left(1 - 7.32 \epsal + 27.06 \epsal^2 - 54.82 \epsal^3 + 61.96 \epsal^4 - 36.56 \epsal^5 + 8.76 \epsal^6 \right) \, , & D = 3 \, , \label{fitzero3} \\
\!\!\! \!\!\! \!\! R_1^\prime(\epsal) & = \frac{e^3}{12} \left(1 - 8.00 \epsal + 32.10 \epsal^2 - 73.94 \epsal^3 + 97.01 \epsal^4 - 66.78 \epsal^5 + 18.63 \epsal^6 \right)  \, , & D = 4 \, . \label{fitzero4}
\end{align}

%
\subsection{Regularized sums}
Next we want to compute 
\begin{equation}
\ln \left\vert \frac{{\rm det}^\prime {\cal O}}{{\rm det}{\cal O}_{\rm FV}} \right \vert 
 = \sum_{\nu = D/2-1}^\infty d_\nu \ln R_\nu \, ,
\end{equation}
with the $R_\nu$ component corresponding to $l=0$ replaced by its absolute value, and the one
corresponding to $l=1$ replaced by $R_1^\prime$ evaluated in the previous section.

The sum, which relates to a one-loop calculation, is UV divergent and must be regularized.
The degree of divergence depends on the number of spacetime dimensions, and is seen at large multipoles: 
%
\begin{equation}
  \sum_{\nu} d_\nu \ln R_\nu \xrightarrow[\nu \to \infty]{} \sum_\nu \frac{2}{\Gamma(D-1)} \nu^{D-2} \left( \frac{c_{\nu 1}}{\nu} + \frac{c_{\nu 3}}{\nu^3} + O(\nu^{-5})  \right) \, .
\end{equation}
Here $c_{\nu 1}$ and $c_{\nu 3}$ are numerical coefficients. 
We see, as expected, that in $D = 3$ we have a linear divergence, in $D=4$ we have a quadratic and a logarithmic divergence, and so on. 
We compute the sum numerically with \texttt{BubbleDet}~\cite{Ekstedt:2023sqc}, which 
uses a regularization scheme equivalent to the one in~\cite{Dunne:2006ct}. 
In Appendix~\ref{app:FunDet} we perform several checks of the calculation, 
using the linear parametrization, different numerical methods, and also another
regularization scheme.

\begin{figure}[t]
    \centering
    \includegraphics[width=0.94\textwidth]{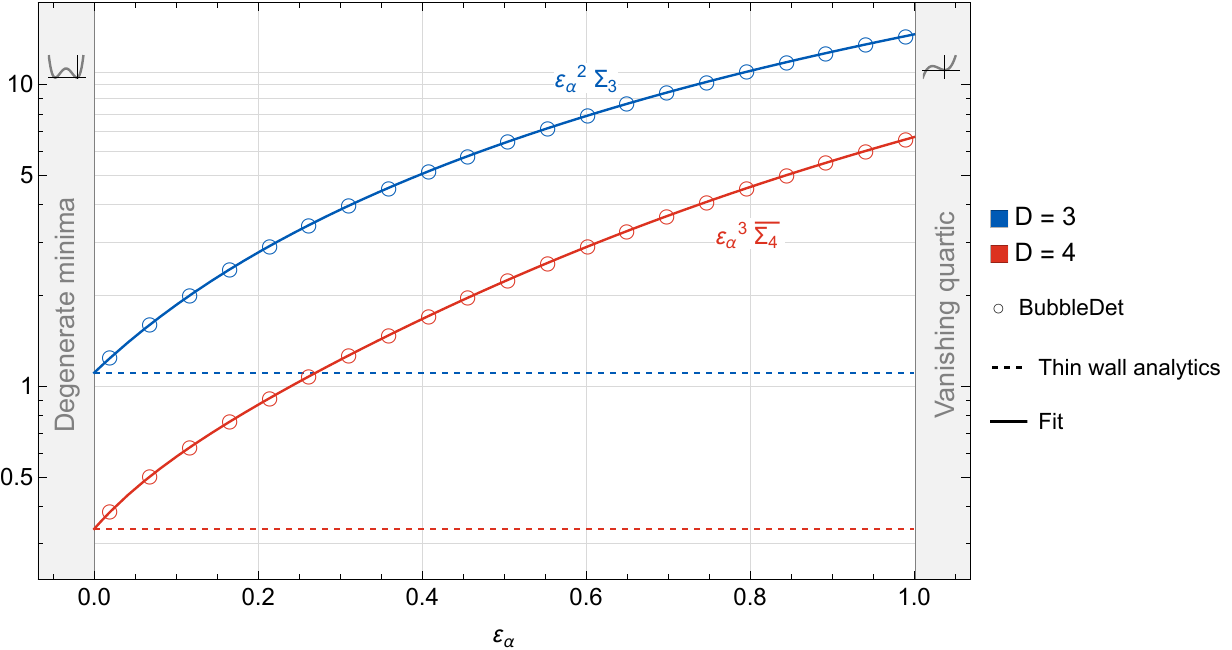}
    \caption{\footnotesize{ 
 We plot the rescaled regularized sums $\epsal^2 \Sigma_3$ and $\epsal^3 \overline\Sigma_4$ as functions of $\epsal$. Here, $\Sigma_3$ is given
 by \eqref{rensum3}, while $\overline\Sigma_4$ by \eqref{rensumD4}.
 The empty circles are the numerical values obtained with \texttt{BubbleDet}. 
 The horizontal dashed lines correspond to the analytic results $\epsal^2\Sigma_3 = \frac{20 + 9\ln 3}{27}$ and 
 $\epsal^3\overline \Sigma_4 = \frac{27 - 2\pi \sqrt{3}}{48}$ computed in~\cite{Ivanov:2022osf} in the thin wall limit ($\epsal \to 0$). 
 The solid lines correspond to the fits \eqref{fitC3} and \eqref{fitC4}.  
 Here we have set the scale $\mu$, that appears in $\overline\Sigma_4$ through \eqref{I2tilde}, to $\mu = m$.
     }}
    \label{fig:RenSum_cubic}
\end{figure}

In the $\zeta$-function scheme of~\cite{Dunne:2006ct}, 
the sums are are regularized as follows: 
\begin{align} \label{rensum3}
  \Sigma_3 = 
  \sum_{\nu=1/2} 2\nu \left(\ln R_\nu - \frac{1}{2 \nu} I_1 \right) \, ,
\end{align}  
in $D = 3$, and\footnote{
The notation $\Sigma_D$ for the sum over multipoles was introduced in~\cite{Ivanov:2022osf}. Here, in $D=4$ we use
$\overline \Sigma_4 = \Sigma_4 - \frac{1}{8} \tilde I_2$.
}
\begin{align} \label{rensumD4}
\begin{split}
 \overline\Sigma_4= 
   \sum_{\nu=1} \nu^2 \left(\ln R_\nu - \frac{1}{2 \nu} I_1 + \frac{1}{8 \nu^3} I_2 \right) - \frac{1}{8} \tilde I_2 \, ,
\end{split}
\end{align}
in $D = 4$.
The subtractions are given by the integrals
\begin{align} \label{Isubtract}
  I_m &= \int_0^\infty \mathrm{d} \rho \, \rho^{2 m - 1} \left( V^{(2) m} - V^{(2) m}_\text{FV} \right) =   \int_0^\infty \mathrm{d} \rhotiC \, \rhotiC^{2 m - 1} \left( \tilde V_C^{(2) m} - \tilde V^{(2) m}_{C\text{FV}} \right)\, .
\end{align}
In the last equality we have the dimensionless $\rhotiC$ introduced in \eqref{dimlessC}, and $\tilde V^{(2)}_C \equiv \frac{{\rm d}^2 \tilde V_C}{{\rm d}\varphi_C^2}$, with 
$\tilde V_C$ given in \eqref{VtC}, and the second derivative evaluated on the bounce.
This makes it clear that $I_m$ is a function only of the parameter $\epsal$, through $\tilde V^{(2)}_C(\epsal)$. 
Analogously, in the linear parametrization $I_m$ is a function only of $\Delta$.
In $D = 4$, outside the sum over multipoles, we have to add back
\begin{align} 
  \tilde I_2 &=  \int_0^\infty \mathrm{d} \rho \, \rho^3 \left( V^{(2) 2} - V^{(2) 2}_\text{FV} \right) \left[1 + \gamma_E + \ln\left( \frac{\mu \rho}{2} \right) \right] \nonumber \\
&=  \int_0^\infty \mathrm{d} \rhotiC \, \rhotiC^3 \left( \tilde V^{(2) 2}_C - \tilde V^{(2) 2}_{C\text{FV}} \right) \left[ 1 + \gamma_E + \ln\left( \frac{\rhotiC}{2} \right) +  \ln\left( \frac{\mu}{m} \right) \right]  \, .
  \label{I2tilde}
\end{align}
To go from the first to the second line, we rescaled again to the dimensionless variables  \eqref{dimlessC}.
In the final result for the decay rate \eqref{GamovV}, the dependence on the arbitrary 
scale $\mu$ will cancel exactly against the contribution from the renormalized 
Euclidean action. As in this section we are computing the functional determinant
only, we have to pick a value of $\mu$ to run the numerics. We choose $\mu = m$.

The results of the numerical evaluations of the regularized sums $\Sigma_3$ and $\overline\Sigma_4$ are shown in FIG.~\ref{fig:RenSum_cubic}. 
We see that in the thin wall limit, $\epsal \to 0$, the numerics are in good agreement with the analytic calculations of~\cite{Ivanov:2022osf}.
It is useful to provide fits to these numerical results in the range of $\epsal$ between 0 and 1:
\begin{align}
\Sigma_3(\epsal) & = \frac{20 + 9 \ln 3}{27} \frac{1}{\epsal^2}
\left(1 + 6.0 \epsal + 8.0 \epsal^2 - 1.8 \epsal^3 \right) \, , \label{fitC3} 
\\
\overline\Sigma_4(\epsal) & = \frac{27 - 2\pi \sqrt{3}}{48} \frac{1}{\epsal^3} \left(1 + 7.2 \epsal - 0.6 \epsal^2 + 24 \epsal^3 -15 \epsal^4 + 3.5 \epsal^5 \right) \, . \label{fitC4}
\end{align}

%
%
\section{Summary} \label{sec:summary}

We have studied quantum tunneling in a real-scalar field theory with a polynomial
potential up to $\phi^4$, in flat spacetime. We have computed the bounce action,
in generic $D$ spacetime dimensions, as an expansion in the thin wall parameter, 
and showed that, upon truncating the expansion at second order, 
it provides an excellent approximation also in thick wall regimes. 
Out of two possible parametrizations of the scalar potential,~\eqref{linpar} and~\eqref{cubpar}, 
we have found that
\begin{equation} \label{quarticp}
  V = \frac{1}{2}m^2 \phi^2 + \eta \phi^3 + \frac{\lambda}{8}\phi^4 \, , \qquad \epsal \equiv  1 - \lambda \frac{m^2}{4\eta^2} \, ,
\end{equation}
is a more convenient choice for departing from the thin wall limit and describing various thick wall regimes. 
It is useful to introduce the parameter $\epsal$, dimensionless in any $D$, to express the final results.
For $\epsal \to 0$ we are in the thin wall limit, and we approach thick wall regimes as $\epsal$ increases. 
For $\epsal > 1$ the potential becomes unbounded. 
We can still compute the bounce action in that regime, and we have found that our truncated expansion still 
works remarkably well.
Our results for the bounce action are given in generic $D$.

We have also computed the functional determinant numerically in the range of
$\epsal$ between 0 and 1, that is for the bounded potential case, in $D=3$ and 
$D = 4$. 
As we are considering a scalar theory, we only have scalar fluctuations in
the determinant.
For the unbounded case one should include fermions and/or gauge bosons in the 
fluctuations, in order to stabilize the potential with quantum corrections,
as is the case in the Standard Model~\cite{Isidori:2001bm, Chigusa:2017dux, Andreassen:2017rzq, Baratella:2024hju}, 
for instance. 
Considering spin $1/2$ and spin $1$ is beyond the scope of this work, so we restrict the functional 
determinant result to the case of the bounded potential for the scalar.

Given \eqref{quarticp}, our results for the decay rate of the false
vacuum per unit volume, including the prefactor calculated at one loop,
are summarized as follows,
\begin{equation} \label{finalGamma}
\frac{\Gamma}{{\cal V}} = \left( \frac{S_D}{2\pi} \right)^{D/2} m^D e^{-S_D - \frac{1}{2} \Sigma_D} \, ,    
\end{equation}
where
\begin{align}
  S_3 & = \frac{32\pi}{81} \frac{m^3}{4\eta^2} \frac{1}{\epsal^2} 
  \left(1 + \frac{17}{2} \epsal + \left(\frac{247}{8} - \frac{9\pi^2}{4} \right) \epsal^2 \right) \, ,  
  \\
  S_4 & = \frac{\pi^2}{3} \frac{m^2}{4\eta^2} \frac{1}{\epsal^3}  
  \left(1 + 10 \epsal + \left(37 - 2\pi^2 \right) \epsal^2 \right) \, , 
  \\
  \Sigma_3 & = \frac{20 + 9 \ln 3}{27} \frac{1}{\epsal^2} \left(1 + 6.0 \epsal + 8.0 \epsal^2 - 1.8 \epsal^3 \right) \, , 
  \\
  \Sigma_4 & = \frac{27 - 2\pi \sqrt{3}}{48} \frac{1}{\epsal^3} 
  \left(1 + 7.2 \epsal - 0.6 \epsal^2 + 24 \epsal^3 -15 \epsal^4 + 3.5 \epsal^5 \right) \, .
\end{align}
Note the factor $m^D$ in \eqref{finalGamma}, which gives the correct dimensions
of the rate, and was obtained from the procedure of removing the zero modes from
the determinant. Expression \eqref{finalGamma} provides a very accurate 
evaluation of the decay rate in $D=3$ and $D=4$ in the range $0 < \epsal \leq 1$,
from thin ($\epsal \to 0$) to thick ($\epsal \to 1$) wall.
While the parameter $\epsal$ does not run at one loop,
the couplings $m^2$ and $\eta$ that appear in the formula are understood as
renormalized, $m^2 \equiv m^2(\mu_0)$ and $\eta \equiv \eta(\mu_0)$, where
we have fixed the renormalization condition at the scale $\mu_0 = m$.
To arrive at the result \eqref{finalGamma} we have adopted a scheme to deal 
with the UV divergences, that appear in intermediate steps of the calculation,
which is equivalent to dimensional regularization and $\overline{\rm MS}$.

If one is interested in evaluating the decay rate in cases that are more intuitively	
parametrized by the potential \eqref{linpar}, where the cubic $\eta \phi^3$ is traded
for a term linear in $\phi$, then \eqref{finalGamma} must be 
modified as follows. The factor $m^D$ is replaced by $(\sqrt{\lambda}v)^D$;
the bounce action $S_D$ is translated from the expressions above
into the linear parametrization using the exact map given in Appendix~\ref{sec:map};
$\Sigma_3$ and $\Sigma_4$ must be translated with some care, as we explain in 
Appendix~\ref{app:FunDet}.

\smallskip
{\bf Note added:} During the proofing stage of this work, a question regarding Coleman's
original thin wall action was brought to our attention thanks to Alonso Rodrigo and Adam Pluciennik.
We found a couple of typos in~\cite{Coleman:1977py}: equation (4.12) is missing a factor
of 2 and should read $S_1 = 2 \mu^3/(3 \lambda)$.
Equation (4.15) is missing a factor of 2 in the second term and should read 
$S_E = -1/2 \ \pi^2 R^4 \epsilon + 2 \pi^2 R^3 S_1$.
Equation (4.19) is correct and using the correct $S_1 = 2 \mu^3/(3 \lambda)$, one arrives at 
$B = 2^4 \, (\pi^2 \mu^{12})/(6 \epsilon^3 \lambda^4)$, a bounce action 
in the thin-wall limit, which is larger by a factor of 16 compared to the one given in equation (4.21)
of~\cite{Coleman:1977py}.
This agrees with our leading order result for the action in $D=4$ when using the translation
$\Delta = \epsilon/(2 \lambda v^4)$ and $\mu^2 = \lambda v^2$.

%
%
\begin{acknowledgments}
  We thank Sašo Grozdanov for an interesting discussion on convergence of series. 
  This work is supported by the Slovenian Research Agency under the research core funding No. P1-0035 and in 
  part by the research grants N1-0253, J1-4389 and J1-60026.
  MM is supported by the Slovenian Research Agency's young researcher program under grant No. PR-11241.
  The work of LU and MN is partially supported by the grant J1-60026.
\end{acknowledgments}

\appendix
%
%
\section{Map between linear and cubic parametrizations} \label{sec:map}

The relation between the parameters of the linear potential $\{ \lambda, \ v, \ \Delta \}$ and 
those of the cubic $\{ m, \ \eta, \ \epsal \}$ can be translated in both directions.
To go from cubic to linear, we have:
\begin{align} \label{eg:CubToLin}
  \Delta(\epsal) &= \frac{\epsal}{\left(1 + 2 \epsal \right)^{3/2}} \, ,
  &
  \lambda(m,\eta,\epsal) &= \frac{4\eta^2}{m^2} \left( 1 - \epsal \right) \, , 
  &
  v(m,\eta,\epsal) &= \frac{m^2}{2 \eta}\frac{\sqrt{1 + 2 \epsal}}{1 - \epsal} \, . 
\end{align}

Going in the other direction is slightly more involved.
The first equation directly relates $\epsal$ to $\Delta$ without any dependence on the other parameters. Defining
\begin{align}
  \delta &=\left[ 9 \left( \sqrt{\Delta^2 - \Delta_{\max}^2}  - \Delta \right)\right]^{1/3} \, , 
  & \Delta_{\rm max} &= \frac{1}{3\sqrt{3}} \, ,
\end{align}
we have
\begin{equation}\label{epsaldelta}
\epsal(\Delta) = \frac{3^{1/3}\delta^2 - \delta^4 - 3^{2/3}}{2(3^{1/3} + \delta^2)^2} \, ,
\end{equation}
and 
\begin{align} \label{metatoL}
  m^2(\lambda, v, \Delta) &= \lambda v^2 \fmsq(\Delta)  \, , & 
  \eta(\lambda,v,\Delta)  &= \frac{\lambda v}{2} \feta(\Delta) \, ,
\end{align}
with
\begin{align} \label{fmsqdef}
  \fmsq(\Delta) & = \frac{1}{6} \left(3^{2/3} \delta^2
  + \frac{3^{4/3}}{\delta^2}+3\right) \, , 
  \\ \label{fetadef}
  \feta(\Delta) & = \frac{\delta^2 +3^{1/3}}{ 
  3^{2/3}\delta} \, .
\end{align}

Given the definitions of the dimensionless Euclidean radii $\rhotiL \equiv \sqrt{\lambda v^2} \rho$ 
and $\rhotiC \equiv m \rho$, we have $\rhotiL$ as a function of $\rhotiC$ and vice-versa,
\begin{align}
  \rhotiL(\rhotiC) &= \sqrt{\frac{1+2\epsal}{1-\epsal}}  \ \rhotiC \, ,
  &
  \rhotiC(\rhotiL) &= \sqrt{\fmsq(\Delta)} \ \rhotiL \, .
\end{align}
From the relation \eqref{phishift} between $\phi_C$ and $\phi_L$ we obtain the relations between the 
corresponding dimensionless fields.
The cubic $\varphi_C$ is related to the linear $\varphi_L$ as
\begin{equation} \label{phiLtoC}
  \varphi_C(\rhotiC; \epsal) = \frac{\sqrt{1+2\epsal}}{1-\epsal} 
  \left[\varphi_L\left( \sqrt{\frac{1+2\epsal}{1-\epsal}}  \ 
  \rhotiC \right) - \varphi_L^{\rm FV}\left(\Delta(\epsal)\right)  \right] \, ,
\end{equation}
where $\varphi_L^{\rm FV} = \phi_L^{\rm FV}/v$, with $\phi_L^{\rm FV}$ given in \eqref{phishift}.
The linear field is related to the cubic one as
\begin{equation} \label{phiCtoL}
  \varphi_L(\rhotiL; \Delta) = \frac{\fmsq(\Delta)}{\feta(\Delta)} \varphi_C\left( \sqrt{\fmsq(\Delta)} 
  \ \rhotiL \right) + \varphi_L^{\rm FV}(\Delta) \, .
\end{equation}

%
%
\section{Bounce action up to the fourth order analytically} \label{sec:sl4}

Consider the Euclidean action in $D$ dimensions
\begin{align} \label{eqEuclidAction}
  S &= \Omega \int_0^\infty \text{d} \rho \, \rho^{D-1} \left( 
  \frac{1}{2} {\dot \phi}^2  + V - V_\text{FV} \right) \, ,
  &
  V &= \frac{\lambda}{8} \left( \phi^2 - v^2 \right)^2 
  + \lambda \Delta v^3 \left( \phi - v \right) \, .
\end{align}
In this section we work in the linear parametrization, but drop the label $L$ to avoid clutter.
The bounce extremizes the action by solving the Euler-Lagrange equation
\begin{align}
  \ddot \phi + \frac{D-1}{\rho} \dot \phi &= \frac{\text{d}V}{\text{d} \phi} \, ,
  &
  \phi(\rho = 0, \infty) &= (\phi_0, \phi_\text{FV}) ,
  &
  \dot \phi(\rho = 0, \infty) &= (0, 0) \, .
\end{align}
We will construct the bounce solution $\phi(\rho)$ in the TW limit by considering higher orders in the $\Delta$ 
series and compute the action to high precision, including corrections up to $\mathcal O(\Delta^4)$.
We factorize out the dependence on $v$ and $\lambda$ by introducing the dimensionless field $\varphi = \phi/v$ and 
the dimensionless coordinate $z = \sqrt \lambda v \rho - r$, such that the center of the instanton sits at $z=0$.
The constant $r$ measures the distance from the center of the bubble to its wall, and
$\text{d} z = \sqrt \lambda v \, \text{d} \rho$.
This gives
\begin{align} \label{eqDefS}
  S &= \frac{\Omega v^{4-D}}{\lambda^{D/2-1} \Delta^{D-1}} \int_{-r}^\infty
   \text{d} z \, \left( \Delta r + \Delta z \right)^{D-1} \left( 
  \frac{1}{2} {\varphi^\prime}^2  + \tilde V - \tilde V_\text{FV} \right) \, ,
  \\ \label{eqVDef}
  \tilde V &= \frac{V}{\lambda v^4} = 
  \frac{1}{8} \left( \varphi^2 - 1 \right)^2 + \Delta \left( \varphi - 1 \right) \, ,
\end{align}
where $\Omega = 2 \pi^{D/2}/\Gamma(D/2)$. 
Expanding the minima of $V$ in small $\Delta$, we have
\begin{align} \label{eqPhiTVFV}
  \varphi_\text{TV} &= -1 - \Delta + \frac{3}{2} \Delta^2 + \ldots \, , 
  &
  \varphi_\text{FV} &= +1 - \Delta - \frac{3}{2} \Delta^2 + \ldots \, , 
  \\ \label{eqVTVFV}
  \tilde V_\text{TV} &= - 2 \Delta - \frac{1}{2} \left( \Delta^2 + \Delta^3 \right) + \ldots \, ,
  &
  \tilde V_\text{FV} &= -\frac{1}{2}\left( \Delta^2 + \Delta^3 \right) + \ldots \, .
\end{align}
We set up the perturbative ansatz for the field and the Euclidean radius
\begin{align} \label{eqPhiRt}
  \varphi &= \sum \Delta^n \varphi_n \, ,  &  r &= \frac{1}{\Delta} \sum \Delta^n r_n \, .
\end{align}
The peculiar form of the radius expansion follows from the fact that the bounce radius diverges 
$r \to \infty$, as $\Delta \to 0$.
In this TW limit, the two vacua become degenerate and the decay rate vanishes.
It is useful to define
\begin{align} \label{eqStld}
  S &= \frac{\Omega \, v^{4-D}}{\lambda^{D/2-1} \Delta^{D-1}} \, \tilde S \, ,
\end{align}
where $\tilde S$ is a dimensionless integral, a function of $\Delta$ only, that we wish to compute.
  
%
%
{\bf Leading order.} 
The bounce equation at $n = 0$ is given by
\begin{align} \label{eqBEom0}
  \varphi_0^{\prime \prime} &= \frac{1}{2} \varphi_0 \left( \varphi_0^2 -1 \right) \, .
\end{align}
It can be integrated using $\varphi^{\prime \prime} = \text{d} \varphi^\prime/\text{d} z 
= \text{d} \varphi^\prime/\text{d}  \varphi \, \varphi^\prime$ and
\begin{align}
  \int \varphi_0^\prime \text{d} \varphi_0^\prime & 
  = \frac{1}{2} \varphi_0^{\prime 2} = \frac{1}{2}
  \int \varphi_0 \left( \varphi_0^2 -1 \right) \text{d} \varphi_0 \, ,
\end{align}
which gives $\varphi_0^\prime = -1/2 (\varphi_0^2 -1),$ when the appropriate boundary
conditions fix the integration constants. 
Integrating once more, we have
\begin{align} \label{eqBounce0}
  \varphi_0^\prime = \frac{\text{d}  \varphi_0}{\text{d} z} &= -\frac{1}{2} \left(\varphi_0^2 - 1 \right) \, ,
  &
  \int \frac{\text{d} \varphi_0}{1 - \varphi_0^2} &= \text{ath} \varphi_0 = \frac{z}{2}  \, ,
  &
  \varphi_0 &= \tanh \frac{z}{2} \, .
\end{align}
The equation of motion in~\eqref{eqBEom0} is odd under parity $z \to -z$ and so is the final 
solution in~\eqref{eqBounce0}.
The remaining free parameter $r_0$ gets fixed by extremizing the last term of the potential in~\eqref{eqVDef}.
We take the leading term in the $r$ expansion in~\eqref{eqPhiRt}, such that
\begin{align} \label{eqStld0}
  \tilde S_0 &=
  \int_{-r_0/\Delta}^\infty \text{d} z \,
   \left( r_0 + \Delta z \right)^{D-1} \left( \frac{1}{2} {\varphi_0^\prime}^2  + 
  \frac{1}{8} \left( \varphi_0^2 - 1 \right)^2 + \Delta \left( \varphi_0 - 1 \right) \right) \, .
\end{align}
The first two terms are even under parity and vanish exponentially as $z \to \pm \infty$,
because $1/4 (\varphi_0^2 -1)^2 = \varphi_0^{\prime2} \propto e^{\pm 2z}$. 
Thus we can safely extend the lower limit of integration to $-\infty$
\begin{align} \label{eqStld0a}
  \tilde S_0 \ni &\int_{-r_0/\Delta}^\infty \left(r_0 + \Delta z \right)^{D-1} \left( \frac{1}{2} {\varphi_0^\prime}^2  + 
  \frac{1}{8} \left( \varphi_0^2 - 1 \right)^2 \right) 
  \simeq r_0^{D-1} \int_{-\infty}^\infty {\varphi_0^\prime}^2 \text{d} z
  \\
  &
  = r_0^{D-1} \int_{-1}^1 \varphi_0^\prime \, \text{d} \varphi_0
  = r_0^{D-1} \frac{1}{2} \int_{-1}^1 \left( 1 - \varphi_0^2  \right) \, \text{d} \varphi_0
  = \frac{2}{3} r_0^{D-1} \, .
\end{align}
We dropped the subleading terms from the $\left( r_0 + \Delta z \right)^{D-1}$ polynomial and kept $r_0$ only.

The last term of the potential $\propto \Delta \varphi_0$ in~\eqref{eqStld0} is odd under $z$ and goes 
to a finite value when $z \to -r_0/\Delta$. 
Therefore, we cannot simply extend the integration limit to $-\infty$ when $\Delta$ is small.
We can resolve this issue using integration by parts
\begin{align}
  \tilde S_0 \ni &\int_{-r_0/\Delta}^\infty \left( r_0 + \Delta z \right)^{D-1} \left( \varphi_0 - 1 \right) \Delta \text{d} z 
  \\  \label{eqStld0b}
  &= \frac{1}{D} \left( r_0 + \Delta z \right)^D \left( \varphi_0 - 1 \right) \Bigr|_{-r_0/\Delta}^\infty
  - \frac{1}{D} \int_{-\infty}^\infty \left( r_0 + \Delta z \right)^D \varphi_0^\prime \text{d} z
  \\
  &\simeq - \frac{1}{D} r_0^D \int_{-\infty}^\infty \varphi_0^\prime \text{d} z
  = - \frac{1}{D} r_0^D \varphi_0 \bigr|_{-1}^1 = - \frac{2}{D} r_0^D \, ,
\end{align}
where we only kept the leading $r_0^D$ volume term.
Combining the surface and volume terms
\begin{align}
  \tilde S_0 &= \frac{2}{3} r_0^{D-1} - \frac{2}{D} r_0^D \, .
\end{align}
Extremizing over $r_0$ gives us the final result at the leading order
\begin{align}
  \frac{2}{3} \left( D - 1 \right) r_0^{D-2} - 2 r_0^{D-1} &= 0 \, ,
  && \Rightarrow & 
  r_0 &= \frac{D-1}{3} \, ,
  &
  \tilde S_0 &= \frac{2}{3 D} r_0^{D-1} \, .
\end{align}
The final Euclidean action at the leading order is
\begin{align} \label{eqSD0}
  S_0 &= \frac{\Omega \, v^{4-D}}{\lambda^{D/2-1} \Delta^{D-1}} \frac{2}{3 D}
  r_0^{D-1} = \frac{1}{\Delta^{D-1}}
  \begin{cases}
    \frac{2^5 \pi v}{3^4 \sqrt{\lambda}}, & D=3 \, ,
    \\
    \frac{\pi^2}{3 \lambda}, & D=4  \, .
  \end{cases}
\end{align}
It can be split into the kinetic $\mathcal T$ ($1/2$ of the surface term) and the potential 
piece $\mathcal V$ ($1/2$ of the surface + the volume term)
\begin{align}
  S_0 &= \frac{\Omega \, v^{4-D}}{\lambda^{D/2-1} \Delta^{D-1}} \frac{r_0^{D-1}}{3}
  \left( 1 + \frac{D - 6 r_0}{D} \right) \, .
\end{align}
These two are related (at any order in $\Delta$), in agreement with Derrick's theorem 
\begin{align}
  \left( D - 2 \right) \mathcal T = -D \mathcal V \, .
\end{align}

%
%
{\bf Higher orders up to $\Delta^4$.}
Let us continue with the higher order corrections up to $\Delta^4$ with the following notation
\begin{align}
  \tilde S = \tilde S_0 + \Delta^2 \tilde S_2 + \Delta^4 \tilde S_4 + \ldots = \sum_{p=0}^\infty  \Delta^p \tilde S_p \, .
\end{align}
We separate the contributions from the different parts of the bounce, such that  
\begin{align}
  \tilde S_p = \tilde S_p^{(0)} + \tilde S_p^{(1)} + \tilde S_p^{(2)} + \ldots =  \sum_{n=0}^\infty \tilde S_p^{(n)} \, ,
\end{align}
where $\tilde S_p^{(n)}$ comes only from the addition of $\varphi_n$ at the $\Delta^p$ order.

\begin{description}
  %
  %
  \item[Parts from $\varphi_0$] 
  Let us begin with the $\Delta^2$ parts, i.e. by calculating $\tilde S_2$.
  We expand the integrals in~\eqref{eqStld0a} and~\eqref{eqStld0b} to get the parts coming from $\varphi_0$ only
  \begin{align}
    \begin{split}
      \tilde S_2^{(0)} &= r_0^{D-3} \binom{D-1}{2} \int_{-\infty}^\infty \text{d} z \, z^2 {\varphi_0^\prime}^2 
      - \frac{1}{D} r_0^{D-2} \binom{D}{2} \int_{-\infty}^\infty \text{d} z \, z^2 \varphi_0^\prime
    \end{split}
    \\
    &= \frac{3}{2} r_0^{D-1} \left( 3 \frac{D-2}{D-1} \int_{-\infty}^\infty \text{d} z \, z^2 {\varphi_0^\prime}^2 
    - \int_{-\infty}^\infty \text{d} z \, z^2 \varphi_0^\prime \right)
    \\
    &
    = - \left( \frac{\pi^2 + 6(D - 2)}{D-1} \right) r_0^{D-1} \, ,
  \end{align}
  where the relevant integrals are calculated in~\eqref{eqIntPhi0p} and~\eqref{eqIntPhi0pz}.
  The $r_2$ correction does not affect the action at the $\Delta^2$ order because we already extremized it
  around $r_0$ and there the first derivative vanishes, such that
  $S(r_0 + \Delta^2 r_2) \simeq S(r_0) + \text{d}S/\text{d} r_0(r_0) \Delta^2 r_2 = S(r_0)$.

  Further expanding the integrals in~\eqref{eqStld0a} and~\eqref{eqStld0b}, we get the $\Delta^4$ corrections
  from $\varphi_0$, which we separate into even $\tilde S^{(0)\text{e}}_4$ and odd $\tilde S^{(0)\text{o}}_4$ parts.
  The even part comes from multiplying the kinetic part with a binomial expansion in powers of $\Delta z$
  \begin{align}
  \begin{split}
     \tilde S^{(0)\text{e}} &\ni \int_{-\infty}^{\infty} \text{d} z \biggl(
     \left(r_0 + \Delta^2 r_2 + \Delta^4 r_4 \right)^{D-1} + 
     \binom{D-1}{2} \left(r_0 + \Delta^2 r_2 \right)^{D-3} \Delta^2 z^2 
     \\
     & + \binom{D-1}{4} r_0^{D-5} \Delta^4 z^4 \biggr) {\varphi_0^\prime}^2 \, .
  \end{split}    
  \end{align}
  Furthermore, we can safely drop the $r_4$ terms for the same reason as the $r_2$ did not contribute at the
  $\Delta^2$ order, i.e. because $\text{d} S/\text{d} r|_{r_0} = 0$.
  We also dropped the $r_2$ in the last line, which was already $\mathcal O(\Delta^4)$ after the $z$ expansion.
  This gives the following $\Delta^4$ terms from the even part of the potential
  \begin{align}
  \begin{split}
     \tilde S^{(0) \text{e}}_4 &= \binom{D - 1}{2} r_0^{D - 3} r_2^2 \int_{-\infty}^\infty \text{d} z {\varphi_0^\prime}^2 
+ \binom{D - 1}{2} \binom{D - 3}{1} r_0^{D - 4} r_2 \int_{-\infty}^\infty \text{d} z \, z^2 \, {\varphi_0^\prime}^2
    \\
    &+ \binom{D - 1}{4} r_0^{D - 5} \int_{-\infty}^\infty \text{d} z \, z^4 \, {\varphi_0^\prime}^2
  \end{split}  
  \\
  \begin{split}
   &= \binom{D - 1}{2} r_0^{D - 3} r_2^2 \frac{2}{3}
+ \binom{D - 1}{2} (D - 3) r_0^{D - 4} r_2 \frac{2}{9} \left( \pi^2 - 6 \right)
    \\
    &+ \binom{D - 1}{4} r_0^{D - 5} \frac{2}{45} \pi^2 \left( 7 \pi^2 - 60 \right) \, ,
  \end{split}    
  \end{align}
  using integrals in~\eqref{eqIntPhi0p}.
  For the odd part, we expand~\eqref{eqStld0b} (dropping $r_4$)
  \begin{align}
     \tilde S^{(0) \text{o}} &\ni - \frac{1}{D} \int_{-\infty}^\infty \text{d} z \, 
     \left( r_0 + \Delta^2 r_2 + \Delta z \right)^D \, \varphi_0^\prime \, ,
  \end{align}  
  such that we get
  \begin{align}
  \begin{split}
     \tilde S^{(0) \text{o}}_4 &= -\frac{1}{D} \biggl( 
     \binom{D}{2} r_0^{D - 2} r_2^2 \int_{-\infty}^\infty \text{d} z \, \varphi_0^\prime + 
     \binom{D}{2} \binom{D - 2}{1} r_0^{D - 3} r_2 \int_{-\infty}^\infty \text{d} z \, z^2 \, \varphi_0^\prime
    \\
    &+ \binom{D}{4} r_0^{D - 4} \int_{-\infty}^\infty \text{d} z \, z^4 \, \varphi_0^\prime \biggr)
  \end{split}  
  \\
  \begin{split}
     &= -\frac{1}{D} \left( 
     \binom{D}{2} r_0^{D - 2} r_2^2 \, 2 + 
     \binom{D}{2} \binom{D - 2}{1} r_0^{D - 3} r_2 \frac{2 \pi^2}{3} + 
     \binom{D}{4} r_0^{D - 4} \frac{14}{15} \pi^4 \right) \, .
  \end{split}  
  \end{align}
    
  %
  %
  \item[Parts from $\varphi_1$]
  The bounce equation at $n=1$ is given by
  \begin{align} \label{eqBEom1}
    \varphi_1^{\prime \prime} + \frac{1}{2} \left( 1 - 3 \varphi_0^2 \right) \varphi_1
    &= 1 - \frac{D-1}{r_0} \varphi_0^\prime = 1 + \frac{3}{2} \left(\varphi_0^2 - 1 \right) \, .
  \end{align}
  Plugging in $r_0 = (D - 1)/3$ and $\varphi_0^\prime = -1/2(\varphi_0^2 - 1)$, we get the trivial solution
  \begin{align} \label{eqBEom1p}
    \varphi_1^{\prime \prime} &= \frac{1}{2} \left(3 \varphi_0^2 - 1 \right) \left(\varphi_1 + 1 \right) \, ,
    &
    \varphi_1 &= -1 \, .
  \end{align}
  The equation and its solution are even in $z$, in accordance with the boundary conditions.
  Starting with the complete Euclidean action
  \begin{align}
    \tilde S &\simeq \int_{-r}^\infty \text{d} z \,
    \left( \Delta r + \Delta z \right)^{D-1} \biggl( \frac{1}{2} {\varphi^\prime}^2 + \frac{1}{8} \left( 
    \varphi^2 - 1 \right)^2 + \Delta \left(\varphi - 1 \right) + \frac{\Delta^2}{2} + \frac{\Delta^3}{2} \biggr) \, ,
  \end{align}
  we plug in $\varphi = \varphi_0 - \Delta$ and isolate the corrections coming from $\varphi_1$
  \begin{align}
    \tilde S^{(1)} 
    & = \int_{-r}^\infty \text{d} z \,
    \left( \Delta r + \Delta z \right)^{D-1} \left( \frac{3}{4} \Delta^2 \left(\varphi_0^2 - 1 \right)
    - \frac{1}{2} \Delta \varphi_0 \left(\varphi_0^2 - 1 \right)  
    - \frac{1}{2} \Delta^3 \left( \varphi_0 - 1 \right)  \right)  
    \\ \label{eqS1gen}
    &= \int_{-r}^\infty \text{d} z \,
    \left( \Delta r + \Delta z \right)^{D-1} \left( -\frac{3}{2} \Delta^2 \varphi_0^\prime
    + \Delta \varphi_0 \varphi_0^\prime 
    - \frac{1}{2} \Delta^3 \left( \varphi_0 - 1 \right) \right) \, .
  \end{align}
  We expand the integrands and first work out the $\Delta^2$ terms
  \begin{align} \label{eqS21nl}
    \tilde S^{(1)}_2 &= -\frac{3}{2} r_0^{D-1} \int_{-1}^1 \text{d} \varphi_0
    + r_0^{D-2} \left( D - 1 \right) \int_{-\infty}^\infty \text{d} z \, z \varphi_0 \varphi_0^\prime
    + \frac{1}{D} r_0^D
    \\ \label{eqS21eval}
    &= - 3 r_0^{D-1} + 2 (D-1) r_0^{D-2} + \frac{1}{D} r_0^D 
    = r_0^{D-1} \left( 3 + \frac{r_0}{D} \right) \, .
  \end{align}
  The third term in~\eqref{eqS21nl} was evaluated as in~\eqref{eqStld0b}, namely
  \begin{align}
    - \Delta^2 \frac{1}{2} \int_{-r}^\infty \text{d} z \, \left( r_0 + \Delta z \right)^{D-1} \Delta \left( \varphi_0 - 1 \right)
    = \Delta^2 r_0^{D-1} \frac{r_0}{D}  = \Delta^2 \frac{1}{D} r_0^D \, .
  \end{align}
  At this order, the correction is independent of $r_1$, which will turn out to be zero.

  Further expanding $(r_0 + \Delta^2 r_2 + \Delta z)^{D-1}$ in~\eqref{eqS1gen}, we get the following $\Delta^4$ expressions
  \begin{equation}
  \begin{split}
    \tilde S_4^{(1)} &= -\frac{3}{2} r_0^{D-3} \binom{D-1}{2} \int_{-\infty}^\infty \text{d} z \, z^2 \varphi_0^{\prime}
    + r_0^{D-4} \binom{D-1}{3} \int_{-\infty}^\infty \text{d} z \, z^3 \varphi_0 \varphi_0^{\prime}
    \\
    &+ \frac{1}{2 D} \binom{D}{2} r_0^{D - 2} \int_{-\infty}^\infty \text{d} z \, z^2 \varphi_0^{\prime}
    \\
    &- \left(3 (D - 1) r_0 - 2 (D-1)(D-2) - r_0^2 \right) r_0^{D - 3} r_2 \, .
  \end{split}
  \end{equation}  
  The first two lines come from expanding $(r_0 + \Delta z)^p$ in higher powers of $z$, while the last line 
  comes from the expansion of $\Delta^2 r_2$ and can be easily read off of~\eqref{eqS21eval} by substituting
  $r_0 \to r_0 + \Delta^2 r_2$ and expanding in $\Delta$.
  Evaluating the integrals in~\eqref{eqIntPhi0pz} and \eqref{eqIntPhi0Phi0pz}, we end up with
  \begin{equation}
  \begin{split}
    \tilde S_4^{(1)} &= - \pi^2 \binom{D-1}{2} r_0^{D-3}
    + 2 \pi^2 \binom{D-1}{3} r_0^{D-4} + \frac{\pi^2}{3 D} \binom{D}{2} r_0^{D - 2} 
    \\
    & - \left(3 \left(D - 1 \right) r_0 - 2 \left(D - 1 \right) \left(D - 2 \right) - r_0^2 \right) r_0^{D - 3} r_2 \, .
  \end{split}
  \end{equation}    
  
  %
  %
  \item[Parts from $\varphi_2$]
  The bounce equation at $n=2$ is
  \begin{align} \label{eqBEom2}
    \varphi_2^{\prime \prime} + \frac{1}{2} \left( 1 - 3 \varphi_0^2 \right) \varphi_2
    &= \frac{3}{2} \varphi_0 + \frac{9}{D-1} \varphi_0^\prime (z + r_1) \, .
  \end{align}
  The solution to such a differential equation has two homogeneous pieces, one even and one odd, and
  a non-homogeneous part from the Wronskian.
  If $r_1 \neq 0$, the odd part of the non-homogeneous function grows exponentially with $z \to \pm \infty$.
  One limit, e.g. $z\to -\infty$ can be regulated by the odd part of the homogeneous solution.
  However, the other limit $z \to \infty$ cannot be cancelled and the solution cannot reach the FV.
  Therefore, the only consistent solution is to set both $r_1 = 0$ and the odd part of the homogeneous
  solution to zero.

  The remaining free coefficient of the $z-$even part of the homogeneous solution $4/\text{ch}^2(z/2)$, 
  can be set to
  \begin{equation}
    c_{2, H}^{\text{even}} = \frac{\pi^2 + 3( D - 3)}{2 r_0} \, ,
  \end{equation}
  such that the complete field solution at $n=2$ becomes manifestly odd
  \begin{align}
  \begin{split}
    \varphi_2 = \frac{1}{4 r_0 \text{ch}^2(z/2)} \bigl( 
      & \left( 2 - D - 2 \left(4 + \text{ch} z \right) \ln(1 + e^z) \right) \text{sh} z
      \\
      & - z \left(D - e^z \left(4 + \text{sh} z \right) \right) + 3 (\text{Li}_2(-e^z) - \text{Li}_2(-e^{-z}))
    \bigr)  \,.
  \end{split}  
  \end{align}
  The asymptotics of this solution are the TV and FV, such that $\varphi_2(\pm\infty) = \mp 3/2$.

  With $\varphi_2$ at hand we can evaluate the $n = 2$ contribution to the Euclidean action.
  To begin, we will show that $\varphi_2$ does not contribute at the $\Delta^2$ order, i.e. $\tilde S^{(2)}_2 = 0$. 
  To demonstrate this, we look at the terms up to and including $\Delta^3$ and rewrite them as
  \begin{align} \label{eqStld2}
    \tilde S^{(2)} &\ni \Delta^2 \int_{-r}^\infty \text{d} z \, \left( r_0 + \Delta z \right)^{D-1} 
    \left( 
      \varphi_0^\prime \varphi_2^\prime - \varphi_0 \varphi_0^\prime \varphi_2
      + 3 \Delta \varphi_0^\prime \varphi_2 \right) \, .
  \end{align}
  We will add higher orders later on to get all the $\Delta^4$ terms of $\tilde S^{(2)}$.
  To proceed at the current order, we use the following identity
  \begin{align}
    &\int_{-\infty}^\infty \text{d} z \, \frac{\text{d}}{\text{d} z} \left( \varphi_0^\prime \varphi_2 \right)
    = \varphi_0^\prime \varphi_2 \biggr |_{-\infty}^\infty = 0
    = \int_{-\infty}^\infty \text{d} z \, \left( \varphi_0^\prime \varphi_2^\prime + \varphi_0^{\prime \prime} \varphi_2 \right)
    \\ \label{eqIdPhi2}
    &= \int_{-\infty}^\infty \text{d} z \, \left( \varphi_0^\prime \varphi_2^\prime + \frac{1}{2} \varphi_0 \left( \varphi_0^2 - 1 \right) \varphi_2 \right)
    =  \int_{-\infty}^\infty \text{d} z \, \left( \varphi_0^\prime \varphi_2^\prime - \varphi_0 \varphi_0^{\prime} \varphi_2 \right) \, .
  \end{align}
  This shows that the first two terms in~\eqref{eqStld2} cancel away at the $\Delta^2$ order.
  The last term in~\eqref{eqStld2} also vanishes at this order, because $\varphi_0^\prime$ is even and vanishes
  at the boundaries, while $\varphi_2$ is odd, so this is not a volume term and we need a further expansion 
  in $\Delta z$. 
  Thus, $\varphi_2$ does not contribute to the action at the $\Delta^2$ order.

  Before moving on to higher order terms, let us work out the $\Delta^4$ corrections from the terms 
  in~\eqref{eqStld2}, after expanding in $\Delta z$.
  We take advantage of another identity
  \begin{align}
    &\int_{-\infty}^\infty \text{d} z \, \frac{\text{d}}{\text{d} z} \left( z^2 \varphi_0^\prime \varphi_2 \right)
    = z^2 \varphi_0^\prime \varphi_2 \biggr |_{-\infty}^\infty = 0
    \\
    &= 2 \int_{-\infty}^\infty \text{d} z \, z \, \varphi_0^\prime \varphi_2 +
    \int_{-\infty}^\infty \text{d} z \, z^2 \left( \varphi_0^\prime \varphi_2^\prime + \varphi_0^{\prime \prime} \varphi_2 \right) \, ,
  \end{align}
  such that the kinetic term goes into
  \begin{align}
    \int_{-\infty}^\infty \text{d} z \, z^2 \left( \varphi_0^\prime \varphi_2^\prime - \varphi_0 \varphi_0^{\prime} \varphi_2 \right) 
    = -2 \int_{-\infty}^\infty \text{d} z \, z \, \varphi_0^\prime \varphi_2 \, .
  \end{align}
  The last term in~\eqref{eqStld2} is non-zero when we expand in $z$, and when we combine the even (kinetic) and the 
  odd terms, we get
  \begin{align} 
    \tilde S^{(2)}_4 &\ni \left( -2 \binom{D-1}{2} + 3 \left( D - 1 \right) r_0 \right) r_0^{D-3}
    \int_{-\infty}^\infty \text{d} z \, z \, \varphi_0^\prime \varphi_2 
    \\ \label{eqS42vDelta2}
    &= 3 r_0^{D - 2} \int_{-\infty}^\infty \text{d} z \, z \, \varphi_0^\prime \varphi_2  \, ,
  \end{align}
  where we used $r_0 = (D-1)/3$ to get to the last line.

 Now we add the rest of the terms with $\varphi_2$ (dropping $\Delta^2 r_2$) up to the $\Delta^5$ volume terms
  \begin{equation}  \label{eqStld2Del4}
   \begin{split}
    \tilde S^{(2)}_4 &\ni \int_{-r}^\infty \text{d} z \, \left(r_0 + \Delta z \right)^{D-1} 
    \biggl( \frac{1}{2} {\varphi_2^\prime}^2 + \frac{1}{4} \left( 3 \varphi_0^2 - 1 \right) \varphi_2^2
    + \frac{3}{2} \varphi_0 \varphi_2 + \frac{9}{8} 
    \\
    & - \frac{\Delta}{2} \left( \varphi_2 + 3 \varphi_0 \varphi_2^2 - \frac{21}{4} \right) \biggr) \, .
    \end{split}
  \end{equation} 
  We focus separately on the even pieces in the first line and the odd volume parts in the second.
  The even bits can be simplified by using the bounce equation for $\varphi_2$ in~\eqref{eqBEom2}, 
  such that the kinetic terms cancel against some of the other pieces:
  \begin{align}
    \int_{-\infty}^\infty \text{d} z \, \frac{\text{d}}{\text{d} z} \left( \varphi_2 \varphi_2^\prime \right) &= 
    0 = \int_{-\infty}^\infty \text{d} z \, \left( {\varphi_2^\prime}^2 + \varphi_2 \varphi_2 ^{\prime \prime} \right)
    \\
    \varphi_2^{\prime \prime} &= \frac{1}{2} \left( 3 \varphi_0^2 - 1 \right) \varphi_2
    + \frac{3}{2} \varphi_0 + \frac{3}{r_0} z \varphi_0^\prime \, .
  \end{align}
  Thus the even terms give us
  \begin{align} \label{eqS42even}
    \tilde S^{(2)\text{e}}_4 &= 
    \frac{3}{4} r_0^{D-1} \int_{-\infty}^\infty \text{d} z \, \left( \varphi_0 \varphi_2 + \frac{3}{2} \right)
    - \frac{3}{2} r_0^{D-2} \int_{-\infty}^\infty \text{d} z \, z \varphi_0^\prime \varphi_2 \, .
  \end{align}

  For the odd pieces, we integrate by parts and obtain
  \begin{align}
    \tilde S^{(2)\text{o}}_4 &= -\frac{\Delta}{2} \int_{-r_0/\Delta}^\infty \text{d} z \, \left(r_0 + \Delta z \right)^{D-1} 
    \left( \varphi_2 + 3 \varphi_0 \varphi_2^2 - \frac{21}{4} \right) \biggr)
    \\
    &\simeq \frac{r_0^D}{2 D} \int_{-\infty}^\infty \text{d} z \, \left( \varphi_2^\prime + 
    3 \left(\varphi_0 \varphi_2^2 \right)^\prime \right) 
    \\ \label{eqS42odd}
    &= \frac{r_0^D}{2 D} \left( - 3  + 3 \frac{9}{2} \right ) = \frac{21}{4 D} r_0^D \, ,
  \end{align}
  where we used the asymptotic behaviour of $\varphi_2(\pm \infty) = \mp 3/2$ as well as
  \begin{align}
    & \int_{-\infty}^\infty \text{d} z \, \varphi_2^\prime = \varphi_2 \bigr|_{-\infty}^{\infty} = - 3 \, ,
    \\
    & \int_{-\infty}^\infty \text{d} z \, \frac{\text{d}}{\text{d} z} \left( \varphi_0 \varphi_2^2 \right) =
    \left(\varphi_0 \varphi_2^2 \right) \bigr|_{-\infty}^{\infty} = 2 \frac{9}{4} = \frac{9}{2} \, .
  \end{align}
  Combining~\eqref{eqS42vDelta2} with~\eqref{eqS42even} and~\eqref{eqS42odd}, we collect all the terms 
  from $\varphi_2$ that enter the action at the $\Delta^4$ order
  \begin{align}
    \tilde S^{(2)\text{o}}_4 &= 
    \frac{3}{4} r_0^{D-1} \int_{-\infty}^\infty \text{d} z \, \left( \varphi_0 \varphi_2 + \frac{3}{2} \right)
    + \frac{3}{2} r_0^{D-2} \int_{-\infty}^\infty \text{d} z \, z \varphi_0^\prime \varphi_2
    + \frac{21}{4 D} r_0^D 
    \\
    &= \frac{1}{4} \left( 3 D - 2 \pi^2 - 9 \right) r_0^{D-2} 
    - \frac{1}{6} \left( (D-2)(3+\pi^2) + 36 \zeta(3) \right) r_0^{D-3}
    + \frac{21}{4 D} r_0^D \, ,
  \end{align}
  where we took the two integrals from~\eqref{eqIntPhi0Phi2} and~\eqref{eqIntzPhi0PPhi2}.

  %
  %
  \item[Parts from $\varphi_3$]
  After adding $\varphi_3$ to the action, we get the following terms, up to $\mathcal O(\Delta^4)$
  \begin{align}
    \tilde S^{(3)} &= \Delta^3 \int_{-\infty}^\infty \text{d} z \, \left( r_0 + \Delta z \right)^{D-1} 
    \left( \varphi_0^\prime \varphi_3^\prime - \varphi_0^\prime \varphi_0 \varphi_3 + 3 \Delta \varphi_0^\prime \varphi_3 \right) \, .
  \end{align}
  By construction, $\varphi_0$ is odd, while $\varphi_0^\prime$ and $\varphi_3$ are even, which makes the
  first two terms odd in $z$.
  To get a non-zero integral, we need to expand the $\left( r_0 + \Delta z \right)^{D-1} \simeq (D-1) r_0^{D-2} \Delta z
  = 3 r_0^{D-1} \Delta z$, where we used $r_0 = (D-1)/3$.
  This gives us
  \begin{align}
    \tilde S^{(3)}_4 &=  3 r_0^{D - 1} \int_{-\infty}^\infty \text{d} z \, 
    \left( z \left(\varphi_0^\prime \varphi_3^\prime - \varphi_0^\prime \varphi_0 \varphi_3 \right) + \varphi_0^\prime \varphi_3 \right) \, .
  \end{align}
  Now consider the following identity
  \begin{align}
    &\int_{-\infty}^\infty \text{d} z \, \frac{\text{d}}{\text{d} z} \left( z \varphi_0^\prime \varphi_3 \right)
    = z \varphi_0^\prime \varphi_3 \biggr |_{-\infty}^\infty = 0
    =\int_{-\infty}^\infty \text{d} z \, \left( z \left( \varphi_0^\prime \varphi_3^\prime + \varphi_0^{\prime \prime} \varphi_3 \right) +
    \varphi_0^\prime \varphi_3  \right)
    \\
    &=\int_{-\infty}^\infty \text{d} z \, \left( z \left( \varphi_0^\prime \varphi_3^\prime - \varphi_0 \varphi_0^\prime \varphi_3 \right) +
    \varphi_0^\prime \varphi_3  \right) \, ,
  \end{align}    
  which demonstrates that in fact $\tilde S^{(3)}_4 = 0$ and that $\varphi_3$ does not contribute to the action at the
  $\Delta^4$ order.
  %
  %
  \item[Parts from $\varphi_4$]
  The situation is similar and even simpler when we add $\varphi_4$ to the action
  \begin{align}
    \tilde S^{(4)} &=  \Delta^4 \int_{-\infty}^\infty \text{d} z \, \left( r_0 + \Delta z \right)^{D-1} 
    \left( \varphi_0^\prime \varphi_4^\prime - \varphi_0 \varphi_0^\prime \varphi_4 + 3 \Delta \varphi_0^\prime \varphi_4 \right) \, .
  \end{align}
  The last term is odd and needs another insertion of $\Delta z$ and is thus of $\mathcal O(\Delta^6)$.
  For the first two terms we have 
  \begin{align}
     \int_{-\infty}^\infty \text{d} z \, \left( \varphi_0^\prime \varphi_4^\prime - \varphi_0 \varphi_0^{\prime} \varphi_4 \right) = 0 \, ,
  \end{align}
  which follows from the same logic as when we derived~\eqref{eqIdPhi2} but replacing $\varphi_2$ with $\varphi_4$, which
  has similar properties of being odd under $z$ and going to a finite value when $z \to \pm \infty$.
  We have thus shown that $\tilde S^{(4)}_4 = 0$.
\end{description}

{\bf Summary of $\Delta^2$.}
Combining all of the pieces, the action at the $\mathcal O(\Delta^2)$ is given by
\begin{align}
  \tilde S_2 &= \tilde S_2^{(0)} + \tilde S_2^{(1)} + \tilde S_2^{(2)}
  \\
  &= \Delta^2 \left(
  - \left( \frac{\pi^2 + 6(D - 2)}{D-1} \right) + \left( 3 + \frac{r_0}{D} \right) + 0
  \right) r_0^{D-1}
  \\
  &= \Delta^2 \left( \frac{1 + D \left(25 - 8 D - 3 \pi^2 \right)}{3 D (D-1)} \right) r_0^{D-1} \, .
\end{align}
such that the action is
\begin{align}
   S = S_0 \left( 1 + \Delta^2 \left(\frac{1 + D \left(25 - 8 D - 3 \pi^2 \right)}{2 (D-1)} \right) \right) \, .
\end{align}
Derrick also holds at this order.

%
%
{\bf Summary of $\Delta^4$.}
At the $\mathcal O(\Delta^4)$, the $\tilde S_4$ came from 
\begin{align}
  \tilde S_4 = \tilde S_4^{(0)} + \tilde S_4^{(1)} + \tilde S_4^{(2)} + \tilde S_4^{(3)} + \tilde S_4^{(4)} \, ,
\end{align}
where we showed that $\tilde S_4^{(3)} = \tilde S_4^{(4)} = 0$.
The separate terms are summarized here
\begin{align}
  \begin{split}
    \tilde S_4^{(0)} &= \frac{D-1}{540} r_0^{d-5} \biggl((D - 3) (D-2) ((D-4) (7 \pi^2 - 60) - 21 \pi^4 r_0) 
    \\
    &+ 60 (D - 2) r_0 ((D-3) (\pi^2 -6) - 3 \pi^2 r_0) r_2 + 180 (D - 2 - 3 r_0) r_0^2 r_2^2 \biggr)  \, ,
  \end{split}  
   \\
   \begin{split}
   \tilde S_4^{(1)} &= \frac{r_0^{D - 4}}{6} \biggl( (D - 1) \pi^2 (2 (D - 5) D - 3 D r_0 + r_0^2 + 6 (2 + r_0)) 
  \\
  & + 6 r_0 (4 + 2 D^2 - 3 D (2 + r_0) + r_0 (3 + r_0)) r_2 \biggr) \, ,
  \end{split}
  \\
  \tilde S_4^{(2)} &=  \frac{r_0^{D-2}(-21 r_0^2 + D^2 (-6 - 2 \pi^2 + 9 r_0) + 
   D (12 + \pi^2 (4 - 6 r_0) + 3 r_0 (-9 + 7 r_0) - 72 \zeta(3)))}{4 D (D - 1)} \, .
\end{align}
and the $r_2$ is given by
\begin{align}
  r_2 & = \frac{6 \pi^2 - 40 + D (26 - 4 D - 3 \pi^2)}{3(D-1)} \, .
\end{align}

Here are some useful integrals containing $\varphi_0$
\begin{align} \label{eqIntPhi0p}
  \int_{-\infty}^\infty \text{d} z \, {\varphi_0^\prime}^2 &= \frac{2}{3} \, ,
  &
  \int_{-\infty}^\infty \text{d} z \, z^2 \varphi_0^{\prime 2} &= \frac{2}{9} \left( \pi^2 - 6 \right) \, ,
  \\
  \int_{-\infty}^\infty \text{d} z \, z^4 \varphi_0^{\prime 2} &= \frac{2}{45} \pi^2 \left( 7 \pi^2 - 60 \right) \, ,
  \\ \label{eqIntPhi0pz}
  \int_{-\infty}^\infty \text{d} z \, \varphi_0^\prime &= 2\, ,
  &
  \int_{-\infty}^\infty \text{d} z \, z^2 \varphi_0^\prime &= \frac{2}{3} \pi^2 \, ,
  \\
  \int_{-\infty}^\infty \text{d} z \, z^4 \varphi_0^\prime &= \frac{14}{15} \pi^4 \, ,
  \\ \label{eqIntPhi0Phi0pz}
  \int_{-\infty}^\infty \text{d} z \, z \varphi_0 \varphi_0^\prime &= 2 \, ,
  &
  \int_{-\infty}^\infty \text{d} z \, z^3 \varphi_0 \varphi_0^\prime &= 2 \pi^2 \, .
\end{align}
The relevant integrals where $\varphi_0$ and $\varphi_2$ appear are given by
\begin{align} \label{eqIntPhi0Phi2}
  \int_{-\infty}^\infty \text{d} z \, \left( \varphi_0 \varphi_2 + \frac{3}{2} \right) &=  \frac{3 D - 2 \pi^2 - 9}{D-1} \, ,
  \\ 
  \int_{-\infty}^\infty \text{d} z \, \varphi_0^\prime \varphi_2^\prime &=
  \int_{-\infty}^\infty \text{d} z \, \varphi_0 \varphi_0^\prime \varphi_2 = 
  -\frac{2 \pi^2 + 3 (D - 5)}{2 (D - 1)} \, ,
  \\ \label{eqIntzPhi0PPhi2}
  \int_{-\infty}^\infty \text{d} z \, z \varphi_0^\prime \varphi_2 &= -\frac{(D-2)(3+\pi^2) + 36 \zeta(3)}{3 (D - 1)} \, .
\end{align}

%
%
\section{Semi-analytic expansion} \label{sec:semianal}
In this section, we show a systematic expansion of the bounce with respect to $\Delta$ in the linear parametrization.
This allows us to calculate the coefficients of $\Delta^n$ for the bounce and for the action numerically.

Plugging the expansion \eqref{phi_r_expansion_linear} into the bounce equation \eqref{bounceq}, we obtain a 
differential equation for $\varphi_{Ln}$ with $n > 0$ as
\begin{equation}
  \left[\partial_z^2 - \frac12(3\varphi_{L0}^2(z) - 1)\right]\varphi_{Ln}(z) = F_n(z).
\end{equation}
Here, $\varphi_{L0}(z)=\tanh z/2$, $r_{L0} = (D-1)/3$, $\varphi'_{Ln}(\pm\infty)=0$, and 
\begin{equation}
  F_n(z) = \frac{D-1}{r_{L0}^2}r_{L(n-1)} \varphi'_{L0}(z) + G_n(z),
\end{equation}
with $G_n$ defined through
\begin{equation}
    \sum_m\Delta^mG_m(z)=\frac12(\varphi^2(z)-3\varphi_{L0}^2(z))\varphi(z)+\Delta-\frac{D-1}{\Delta r+\Delta z}\Delta\varphi'(z)-\frac{D-1}{r_{L0}^2}\Delta^2\varphi'_{L0}(z)r.
\end{equation}
Notice that $G_n$ only contains $\varphi_{Lm}$ and $r_{L(m-1)}$ with $m<n$.

The general solution to the differential equation can be constructed as
\begin{align}\label{eq_general_solution}
  \varphi_{Ln}(z)&=-\lambda_1(z)\int_0^z \text{d} y\lambda_2(y)F_n(y)+\lambda_2(z)
  \int_{-\infty}^z \text{d} y\lambda_1(y)F_n(y)+C_n\lambda_1(z)+C'_n\lambda_2(z),
\end{align}
where $C_n$ and $C'_n$ are constants and
\begin{align}
  \lambda_1(z)&=\frac{1}{4\cosh^2\frac{z}{2}},
  \\
  \lambda_2(z)&=\frac{1}{4\cosh^2\frac{z}{2}} \left(6 z + 8\sinh z + \sinh 2z \right),
\end{align}
are the solutions of the homogeneous differential equation. From $\lambda'_1(\pm\infty)=0$ and 
$\lambda'_2(\pm\infty)=\infty$, we see that the boundary condition, $\varphi'_{Ln}(\pm\infty)=0$, requires $C'_n=0$ and
\begin{align}
  0 = \int_{-\infty}^\infty \text{d} y \lambda_1(y) F_n(y).
\end{align}
It determines $r_{n-1}$ as
\begin{equation}
    r_{L(n-1)}=-r_{L0}\int_{-\infty}^\infty \text{d} y \lambda_1(y) G_n(y).
\end{equation}
In particular, one gets $r_{L1} = 0$ since $G_2(z)$ is always an odd function.
The undetermined coefficient, $C_n$, affects $r_{L(n+2)}$ through the above equation and 
only a linear combination of these can be determined. 
Since we have expanded both $r$ and $\varphi$, there appears extra freedom to pre-include 
a part of $r_{n+2}$ into $\varphi_{Ln}$. 
Although the intermediate products are different, this does not affect $s_n^L$ since the 
$\Delta$-expansion of the action is unique. 
Notice that one can choose $C_n$ such that $r_{L2} = r_{L3} = \cdots = 0$, which corresponds 
to the expansion of~\cite{Konoplich:1987yd}. 
Another choice of $C_n = 0$ is also useful since it fixes the zero point, $\varphi_{Ln}(0)=0$.

Finally, the expansion coefficients of the action are calculated either from the kinetic part 
or the potential part of the action using
\begin{align}\label{eq_numerical_snL}
  \sum_{n}\Delta^n s_n^L &= \frac{1}{D}\int_{-\infty}^\infty \text{d} z(\Delta r+\Delta z)^{D-1}\varphi'^2(z)\nonumber
  \\
  &= \frac{2}{D(D-2)}\int_{-\infty}^\infty \text{d} z(\Delta r+\Delta z)^{D}\varphi'(z)\nonumber
  \\
  & \hspace{3ex}+\frac{1}{D(D-2)}\frac{1}{\Delta}\int_{-\infty}^\infty \text{d} z(\Delta r+
  \Delta z)^{D}(\varphi^2(z)-1)\varphi(z)\varphi'(z).
\end{align}

%
%
\section{Asymptotic series} \label{app:Borel}

\begin{figure}[t]
    \subfloat{{\includegraphics[width=.45\textwidth]{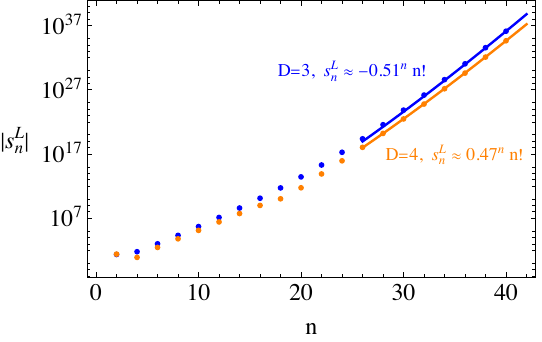} }}%
    \qquad
    \subfloat{{\includegraphics[width=.45\textwidth]{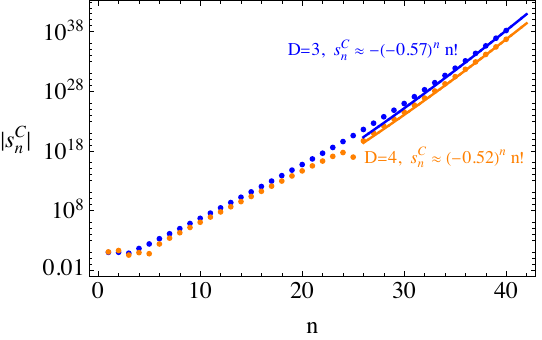} }}%
    \caption{The absolute values of the coefficients $s_n^L$ and $s_n^C$ up to $n=40$. 
    For $n > 4$ the coefficients are computed numerically, using the iterative routine described in Appendix~\ref{sec:semianal}. 
    Both in the linear and in the cubic parametrizations we observe a factorial growth of the coefficients at large $n$, that is roughly $n > 25$.
    The plain lines show fits to the large $n$ points of the form $|s_n| = (1/A)^{n} \ n!$.
    }
    \label{fig:asymptoticoef}
\end{figure}

\begin{figure}[t]
    \subfloat{{\includegraphics[width=.45\textwidth]{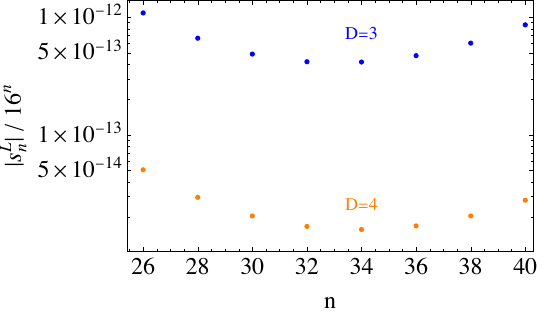} }}%
    \qquad
    \subfloat{{\includegraphics[width=.45\textwidth]{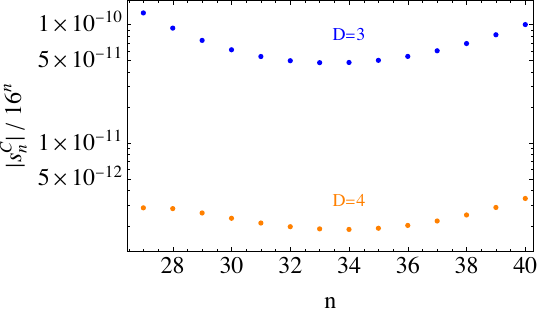} }}%
    \caption{
    The absolute values of the coefficients $s_n^L$ and $s_n^C$, normalized to $16^n$, showing
    the faster than exponential growth at large $n > 25$.
     }
    \label{fig:asymptoticoefnorm}
\end{figure}

In the previous Appendix we described a routine to compute numerically the $s_n^L$ coefficients. 
In the main body of the paper we considered terms up to $n=16$. 
Here we go to higher orders, which involves some additional numerical challenges.
We observe a cancellation among terms in the right hand side of~\eqref{eq_numerical_snL}, where the leading term is 
$\mathcal O(10^4)$ times larger than the sum. 
This creates some challenges to keep track of all the terms and we have performed a number of numerical checks to 
justify the numerical stability and accuracy of our result.
The first check was to compute both the first line and the second line in~\eqref{eq_numerical_snL}, which should give 
the same number due to the Derrick's theorem, meaning the action is extremized precisely enough. 
The second check is to increase the integration precision by increasing the number of subdivisions of the integrand 
and also going from double to quadrupole precision. 
The third check on numerical stability is to change the coefficient of coherent terms, $C_{2n+1}$'s, 
in~\eqref{eq_general_solution} by order one. 
We found that a few highest order points require quadrupole precision, and that at least three significant digits
are stable for all the other checks.

With all the improvements described above, we pushed our numerical results up to $n = 40$ in the linear parametrization. 
On the left of FIG.~\ref{fig:asymptoticoef}, where we plot $|s_n^L|$, we see that for $n > 25$ the growth of the coefficients 
steepens and becomes factorial. 
In $D=3$ all the coefficients with $n$ between 25 and 40 are negative, in $D=4$ they are positive. 
In either case their growth goes as $A^{-n} n!$, with $A = 1.97$ in $D = 3$, and $A = 2.15$ in $D = 4$. 
This result, computed numerically in the linear parametrization, can readily be translated into the cubic expansion, 
using the methods described in the main text. 
On the right of FIG.~\ref{fig:asymptoticoef} we show the absolute value of the coefficients of the cubic expansion 
up to $n=40$; again we observe a factorial growth for $n > 25$. 
In the cubic case the coefficients at large $n$ alternate signs, and are of the form $s_n^C = \pm (-A)^{-n} n!$, with the 
overall minus sign in $D=3$. 
The best fit, taking into account $n \geq 26$, gives $A = 1.75$ in $D=3$, and $A = 1.91$ in $D = 4$.
Note that in $D=4$ we have sort of an outlier at $n=25$. 
This is where the behavior of the coefficients switches, so it is not surprising that the two competing contributions accidentally cancel out to a precision of 1\%.
%
The Derrick's theorem is checked up to six digits at $n=25$, and thus we think this could just be a feature of $D=4$.

We also show the normalized coefficients in FIG.~\ref{fig:asymptoticoefnorm}, where we divide them with 
$16^n$, which factors out an exponential coefficient.
It is clear that the curve turns up and thus proves the growth is faster than exponential and the factorial
fit is justified.
Such a factorial growth of the coefficients $s_n^C$ then implies a zero radius of convergence of $S_C^{(N)}(\epsal)$, 
defined in~\eqref{SCexpa}, making it an asymptotic series.
We can first check for the optimal truncation order, by studying 
the absolute value of the terms $\epsal^n s_n^C$ for different values of $\epsal$. The value of $n$ at which
$|\epsal^n s_n^C|$ is minimized corresponds to optimal truncation.
\begin{figure}[h!]
    \subfloat{{\includegraphics[width=.46\textwidth]{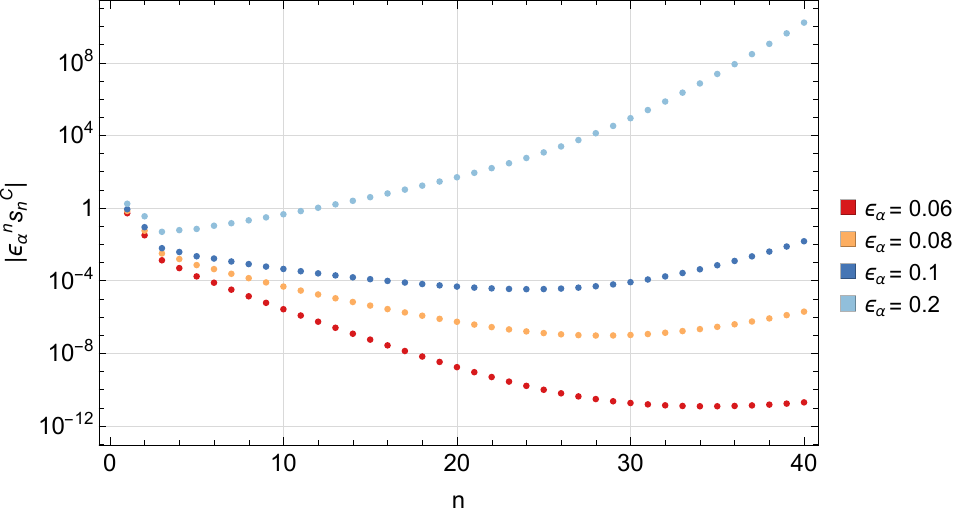} }}%
    \qquad
    \subfloat{{\includegraphics[width=.46\textwidth]{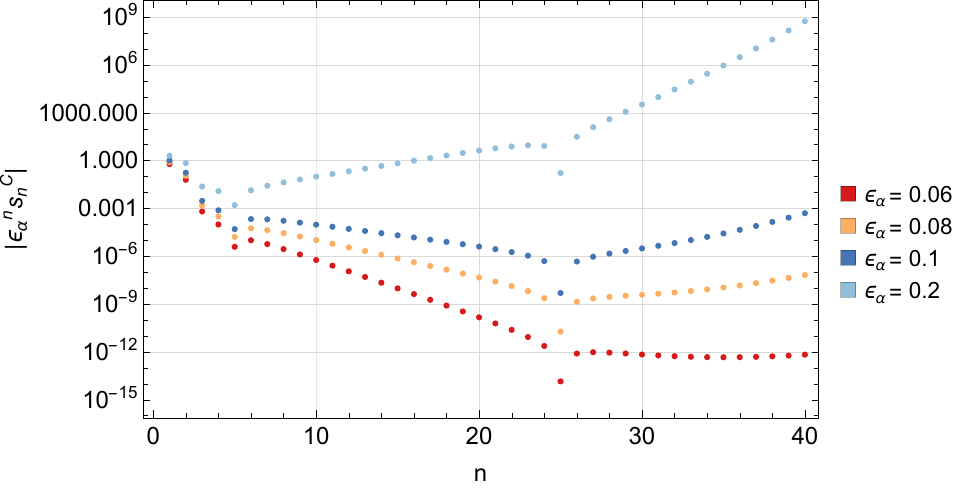} }}%
    \caption{Absolute values of each term of $S^{(10)}_C(\epsal)$ in \eqref{SCexpa} for $\epsal=0.06,0.08,0.1,0.2$. The left panel is for $D=3$, the right for $D=4$.
    The horizontal lines indicate the minimum values. 
    On the right plot we have an outlier at $n=25$, which corresponds to the break from exponential to factorial 
    growth of the $s_n^C$ coefficients (see also Fig.~\ref{fig:asymptoticoefnorm}). The minimum is around that 
    break for $\epsal = 0.08, 0.1$, but is at $N = 36$ for $\epsal = 0.06$.
    }
    \label{fig:truncation-Csmall}
\end{figure}
%
%
%
In FIG.~\ref{fig:truncation-Csmall} we see that for $\epsal$ up to 0.1 the optimal truncation order is at $N$ larger
than 20, both in $D = 3$ and $D = 4$. 
It is roughly at $N \sim A/\epsal$, as one would expect, with $A \sim 2$ as estimated above for the factorial series.
When $\epsal$ gets as big as 0.2 the minimum of $|\epsal^n s_n^C|$ is at much
smaller $N$, of order a few. This implies that for $\epsal \geq 0.2$ talking about optimal truncation is no longer very meaningful.
%
We have performed a preliminary Borel-Pad\'e analysis of the expansion, but our first results do not show the expected convergence
and are inconclusive. 
This requires further study, beyond the scope of the current paper, and we leave it to future work.

%

%
%
\section{Bounce construction in the cubic parametrization} \label{app:cubic}

In this appendix we construct the thin-wall bounce solution, following the same method as 
in~\cite{Ivanov:2022osf}, but starting with the scalar potential in the cubic parametrization:
\begin{equation}
  V_C(\phi) = \frac{1}{2} m^2 \phi^2 + \eta \phi^3 + \frac{\lambda_C}{8} \phi^4 \, .
\end{equation}
Here we take $m^2 > 0$, $\eta > 0$, and $\lambda_C < 4\eta^2 / m^2$. 
With this choice, the minimum at $\phi = 0$ is the false vacuum and has $V(0) = 0$. 
When $0 < \lambda_C < 4 \eta^2 / m^2$ there is another minimum at the negative field value 
$\phi_{\rm TV} = -\frac{3\eta}{\lambda_C}\left( 1 + \sqrt{1 - \frac{2m^2}{9\eta^2}\lambda_C} \right)$, 
the true vacuum, separated from the false vacuum by a potential barrier.
It reaches its maximum at $\phi_{\rm max} = -\frac{3\eta}{\lambda_C}\left( 1 - \sqrt{1 - 
\frac{2m^2}{9\eta^2}\lambda_C} \right)$. 
For a negative quartic, $\lambda_C < 0$, there is still a false vacuum at the origin but two 
possible tunneling directions: one to the left, with a lower barrier, the other to the right, 
with a higher barrier.
Here we restrict our analysis to a positive quartic and take the thin wall limit. 
This corresponds to $\lambda_C \to 4 \eta^2/m^2$, when the two vacua become degenerate.

It is convenient to introduce the dimensionless variables \eqref{dimlessC}.
%
%
Given the dimensions of the quantities involved,
\begin{align}
  [\rho] &= -1 \, , & [m] &= 1 \, , & [\phi] &= \frac{D}{2}-1 \, , &
  \quad [\eta] &= 3 - \frac{D}{2} \, , & [\lambda_C] &= 4 - D \, ,
\end{align}
we see that $\rhotiC, \epsal, \varphi_C$ are dimensionless in any $D$.
We can write the action
\begin{align}
  S & = \Omega \int_0^\infty {\rm d}\rho \ \rho^{D-1} \left( \frac{1}{2} \left( 
  \frac{{\rm d}\phi}{{\rm d} \rho} \right)^2 + V_C(\phi) \right)
  \\
  & = \Omega \frac{m^{6-D}}{4\eta^2} \int_0^\infty {\rm d}\rhotiC \ \rhotiC^{D-1}  \left( 
  \frac{1}{2} \left( \frac{{\rm d}\varphi_C}{{\rm d}\rhotiC} \right)^2 + \tilde V_C(\varphi_C) \right) 
  \\
  & = \Omega \frac{m^{6-D}}{4\eta^2} \int_{-r_C}^\infty {\rm d}z_C \ (z_C+r_C)^{D-1} \left(
  \frac{1}{2} \left( \frac{{\rm d}\varphi_C}{{\rm d}z_C} \right)^2 + \tilde V_C(\varphi_C) \right) \, , \label{Sz}
\end{align}
where
\begin{equation}
  \tilde V_C(\varphi_C) = \frac{1}{2} \varphi_C^2 + \frac{1}{2} \varphi_C^3 + \frac{1-\epsal}{8} \varphi_C^4 \, ,
\end{equation}
is dimensionless and depends only on one parameter, $\epsal$.
In \eqref{Sz} we have introduced
\begin{equation}
  z_C = \rhotiC - r_C \, ,
\end{equation}
where $r_C$ is the dimensionless bubble radius, the physical radius being $r_C/m$.

In this parametrization, the thin wall limit when the vacua become degenerate, corresponds to $\epsal \to 0$.
The false vacuum is fixed at 
\begin{equation}
  \varphi_{C}^{\rm FV} = 0 \, ,
\end{equation}
while the true vacuum is at
\begin{equation}
  \varphi_{C}^{\rm TV} = - \frac{3 + \sqrt{1 + 8\epsal}}{2(1 - \epsal)} 
  = -2 - 4 \epsal + {\cal O}(\epsal^3) \, .
\end{equation}
Note that here the coefficient of the $\epsal^2$ term in the expansion is zero.

We want to find the bounce in the thin wall limit ($\epsal \to 0$). 
From \eqref{Sz} we get the bounce equation
\begin{equation} \label{bounceCeq}
  \frac{{\rm d}^2 \varphi_C}{{\rm d}z_C^2} + \frac{D-1}{z_c+r_C} \frac{{\rm d}\varphi_C}{{\rm d}z_C} - 
  \varphi_C - \frac{3}{2} \varphi_C^2 - \frac{1}{2} (1 - \epsal) \varphi_C^3 = 0 \, ,
\end{equation}
subject to the boundary conditions
\begin{align} \label{bC}
  \varphi_C( z \to \infty)  &= \varphi_{C}^{\rm FV} = 0 \, , 
  & 
  \frac{{\rm d}\varphi_C}{{\rm d}z}( z_C = -r_C \to -\infty) &= 0 \, .
\end{align}
We use the same method as in~\cite{Ivanov:2022osf}. 
We expand the field and the radius in the small parameter $\epsal \ll 1$ as
\begin{align}
  \varphi_C(z) & = \sum_{n = 0} \epsal^n \ \varphi_{Cn}(z) \, , 
  &
  r_C & = \frac{1}{\epsal}  \sum_{n = 0} \epsal^n  \ r_{Cn} \, ;
\end{align}
we substitute into~\eqref{bounceCeq} and solve the bounce equation order by order in $\epsal$, 
with the boundary conditions in~\eqref{bC}. 
We manage to find analytic solutions up to order $\epsal^2$.

We find the following results:
\begin{align}
  \varphi_{C0}(z_C) & = - \frac{2}{1 + e^{z_C}} \, , 
  \\
  \varphi_{C1}(z_C) & = \frac{-8 + e^{z_C} (3 +6z_C)}{2(1 + e^{z_C})^2} \, , 
  \\
  \varphi_{C2}(z_C) & = \frac{e^{-z_C}}{80 (D-1) \left(e^{z_C}+1\right)^3} \left[e^{z_C} \left(e^{2 z_C} \left(15 
  \left(36 z_C^2+64 z_C-115\right) \right. \right. \right.    \nonumber \\
 & \left. \left.\left.  \ \qquad \qquad  \qquad \qquad  -D \left(180 z_C^2+360 z_C+449\right)\right) \right. \right. \nonumber 
  \\
  &   \left. \left. \ \qquad \qquad  \qquad \qquad + e^{z_C} \left(D \left(180 z_C^2+960 z_C+761\right) +
  5 \left(36 z_C^2 - 264 z_C - 635 \right)\right) \right. \right. \nonumber 
  \\
  & \left. \left. \ \qquad \qquad \qquad \qquad + 120 e^{4 z_C} z_C+120 e^{3 z_C} (9 z_C+1)-120\right) 
  \right. \nonumber \\
 &  \ \qquad \qquad \qquad \qquad + 1440 e^{2 z_C} \left(e^{z_C}+1\right) \text{Li}_2\left(-e^{z_C}\right)  \nonumber 
  \\
  &  \left. \ \qquad \qquad \qquad \qquad -120 \left(-7 e^{z_C}+7 e^{2 z_C}+e^{3 z_C}-1\right) 
  \left(e^{z_C}+1\right)^2 \log \left(e^{z_C}+1\right)\right] \, ,
\end{align}
and
\begin{align}
  r_{C0} &= \frac{D-1}{3} \, , &  r_{C1} &= \frac{D-1}{2} \, .
\end{align}
Compared to the solution we found in the linear parametrization in~\cite{Ivanov:2022osf} there are 
two important differences: the functions $\varphi_{Cn}(z_C)$ have no definite parity under $z_C\to -z_C$, and 
the coefficient $r_{C1}$ does not vanish.

The thin-wall solution $\varphi_C(z)$ connects the false vacuum at $z_C = \infty$ ($\rho = \infty$)
to the true vacuum at $z_C = -r_C$ ($\rho = 0$) at every order in $\epsal$ by construction,
as explained in~\cite{Ivanov:2022osf}.

With the solutions above we can compute the bounce action up to second order in $\epsal$.
Up to ${\cal O}(\epsal^3)$, we find
\begin{equation}
  S =  \Omega \frac{m^{6-D}}{4\eta^2} S_C^{(0)} \left( 1 + \epsal \frac{3D+8}{2} + \epsal^2 \frac{9 D^3 - 11 D^2 +138 
  D - 12 D \pi^2 -64}{8(D-1)} \right) \, ,
\end{equation}
with
\begin{equation}
  S_C^{(0)} = \left( \frac{D-1}{3} \right)^{D-1} \frac{2}{3D} \frac{1}{\epsal^{D-1}} \, ,
\end{equation}
being the leading order thin wall bounce action.

\section{The bounce field configuration} \label{app:bounce}
The field configuration corresponding to the bounce is of physical significance as it defines the profile 
of the bubble, which is nucleated in the phase transition.
Let us examine how the bubble wall profiles $\phi_L$ and $\phi_C$ come about for the two different parametrizations, 
linear and cubic.

%
\subsection{Linear parametrization}

In Ref.~\cite{Ivanov:2022osf} we computed the bounce including corrections up to $\Delta^2$.
For computational convenience, we shifted the Euclidean radius $\rho$ by defining the dimensionless variable 
$z_L = \sqrt{\lambda} v \rho - r_L$, with $r_L = (1/\Delta)(r_{L0} + \Delta r_{L1} + \Delta^2 r_{L2} + \cdots)$ the 
dimensionless bubble radius.
Solving the second-order differential equation for the bounce order by order in $\Delta$, we found analytically 
the bounce as a function of $z$, $\varphi_L(z_L) = \varphi_{L0}(z_L) + \Delta \varphi_{L1}(z_L) + \Delta^2 \varphi_{L2}(z_L) + 
\Delta^3 \varphi_{L3}(z_L)$, up to order $\Delta^3$, and we were able to fix the coefficients $r_{L0}, r_{L1}, r_{L2}$ 
using the boundary conditions. 
At each order $n$, the solution $\varphi_{Ln}(z_L)$ contained a term 
\begin{equation}\label{c1ncoeff}
  \varphi_{Ln}(z_L) \supset \frac{c_{1n}}{4\cosh^2(z_L/2)} = \frac{c_{1n}}{2} \frac{{\rm d}\varphi_{L0}}{{\rm d}z_L} \, ,
\end{equation}
proportional to $c_{1n}$.
It turns out, as was pointed out in Ref.~\cite{Ivanov:2022osf}, that the coefficients $c_{1n}$ and $r_{L,n+1}$ are only 
fixed at the order $n + 2$ by the boundary condition $\frac{{\rm d}\varphi}{{\rm d}\rho} = 0$ at $\rho = 0$.
However, they are not independent; what gets fixed is the combination $(1/2)c_{1n} - r_{L,n+1}$. 
This reflects the fact that using the $\Delta$ expansion both for $\varphi_L$ and $r_L$ is redundant, albeit useful 
for the calculation.
We also find that fixing the coefficients in~\eqref{c1ncoeff} of the even orders, $c_{1,2n}$, by requiring that 
$\varphi_{L,2n}(z_L)$ be odd functions of $z_L$, implies that the coefficients $r_{L,2n+1}$ vanish (we were able to 
verify this statement numerically up to high order $n$).

It is useful to rewrite the bounce $\varphi$ as a function of the Euclidean dimensionless radius 
$\rhotiL = z + r_L$ and re-expand the function in small $\Delta$. 
Doing so eliminates the redundancy and fixes the coefficients uniquely at each order in the expansion. 
Up to second order we get the following analytic result:
%
%
%
\begin{equation} \label{bounceuni}
  \varphi^{(2)}_L(\rhotiL) = \varphi_{L0}(\rhotiL) + \left[ \left( \frac{c_{11}}{2} - r_{L2}  \right) 
  \frac{{\rm d}\varphi_{L0}}{{\rm d}\rhotiL} - 1 \right] \Delta + \left[\frac{1}{2} \left(  \frac{c_{11}}{2} - r_{L2} \right)^2  
  \frac{{\rm d}^2\varphi_{L0}}{{\rm d}\rhotiL^2}  + f_2(\rhotiL)\right] \Delta^2 \, ,
\end{equation}
where
%
\begin{align} \label{Colemanb}
  & \varphi_{L0}(\rhotiL)  = \tanh \left[ \frac{1}{2} (\rhotiL - \RtiL) \right] \, , \qquad 
  \RtiL \equiv \frac{r_{L0}}{\Delta} = \frac{D-1}{3\Delta} \, , 
  \\
  & f_2(\rhotiL)  =  \frac{1}{2 (D-1) \left(e^{\rhotiL -\RtiL}+1\right)^2} \left[ 3 e^{\rhotiL -\RtiL} 
  \left(\left(2 D + 1 -3\left( \rhotiL - \RtiL\right) \right)    \left(\RtiL -\rhotiL \right)+\pi ^2\right) \right.  \nn 
  & \qquad \qquad \qquad \left.  
  -3 e^{2 \left(\rhotiL -\RtiL \right)} \left(D - 2 -8 (\rhotiL - \RtiL) \right) +3 e^{3 \left(\rhotiL -\RtiL \right)} 
  \left(\rhotiL -\RtiL \right) \right. \nn
  &\qquad \qquad \qquad \left.   +3 e^{-(\rhotiL -\RtiL)} \left(8 e^{\rhotiL -\RtiL }-8 e^{3 \left(\rhotiL -\RtiL \right)}-e^{4
   \left(\rhotiL -\RtiL \right)}+1\right) \log \left( 1 + e^{\rhotiL -\RtiL }\right) \right. \nn
  &  \qquad \qquad \qquad \left. +36 e^{\rhotiL -\RtiL } \text{Li}_2\left(-e^{\rhotiL -\RtiL }\right)+3 (D-2) \right] \, , 
  \label{f2bounce} \\
  & \frac{c_{11}}{2} - r_{L2}  =   \frac{4 D^2 +D(3\pi^2 -26)  - 6\pi^2 + 40}{3(D-1)} \, .
\end{align}
This result is valid for $D > 1$.

The leading order term in~\eqref{bounceuni} is the one in Coleman's seminal paper~\cite{Coleman:1977py}. 
Then we have corrections up to $\Delta^2$ and the combination of couplings $(c_{11}/2) - r_{L2}$ appears 
both at linear and quadratic order.
At $\Delta^2$ one would also expect terms proportional to $c_{12}$ and $r_{L3}$. 
However, as mentioned above, $c_{12}$ is fixed by requiring that $\varphi_{L2}(z_L)$ be an odd function 
of $z_L$, while $r_{L3} = 0$. 
What is left is the term proportional to the second derivative of $\varphi_{L0}$ plus the function 
$f_2(\rhotiL)$, which was part of $\varphi_{L2}(z_L)$.
Note there are terms in $f_2(\rhotiL)$ proportional to $(\rhotiL - \RtiL)$; as $\RtiL$ is proportional 
to $1/\Delta$ one might worry that they spoil the $\Delta$ power counting. 
They do not, as they are multiplied by exponentials with a positive power of $(\rhotiL - \RtiL)$, which 
preserves the power counting. 

\begin{figure}
 \begin{center}
    \subfloat{{\includegraphics[width=.22\textwidth]{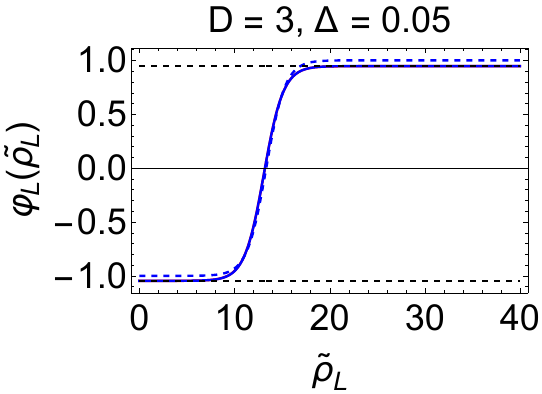} }}%
    \
     \subfloat{{\includegraphics[width=.22\textwidth]{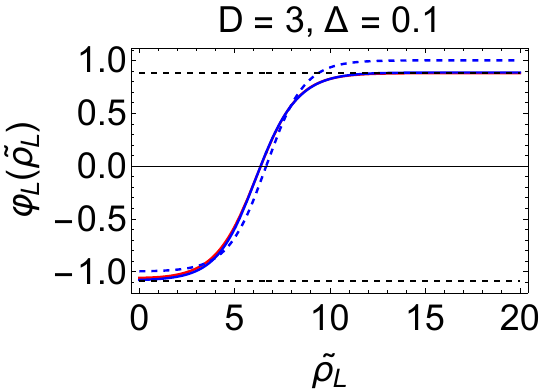} }}%
      \
     \subfloat{{\includegraphics[width=.22\textwidth]{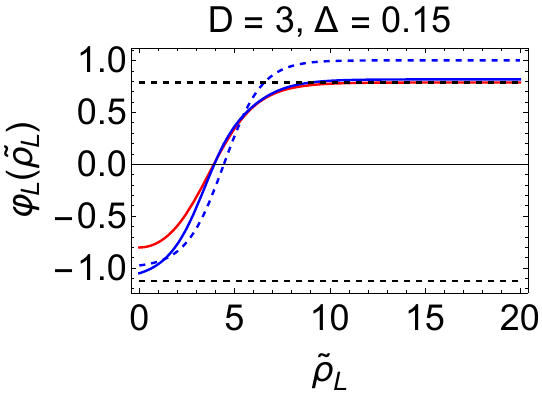} }}%
      \
     \subfloat{{\includegraphics[width=.22\textwidth]{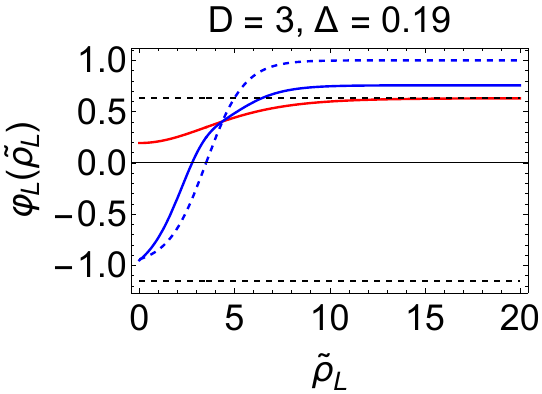} }} \\
     \subfloat{{\includegraphics[width=.22\textwidth]{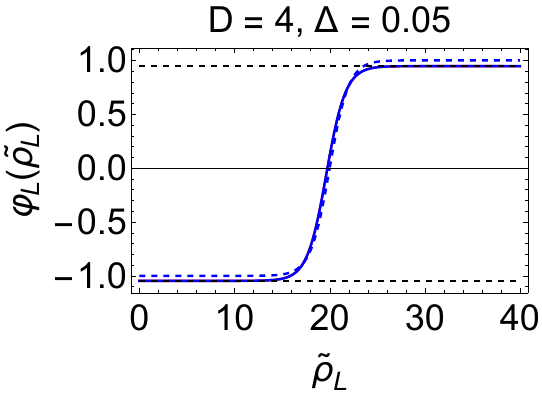} }}%
    \
     \subfloat{{\includegraphics[width=.22\textwidth]{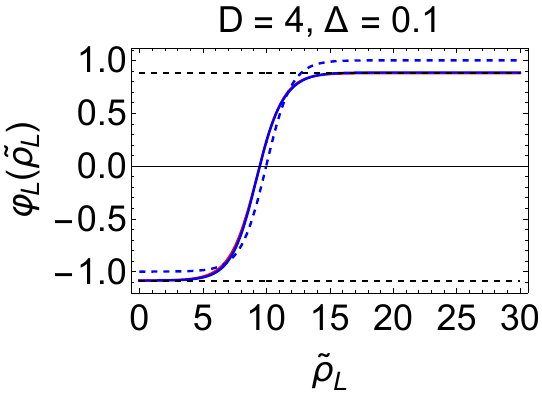} }}%
      \
     \subfloat{{\includegraphics[width=.22\textwidth]{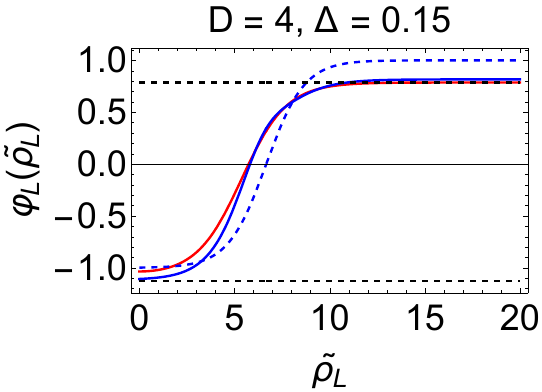} }}%
      \
     \subfloat{{\includegraphics[width=.22\textwidth]{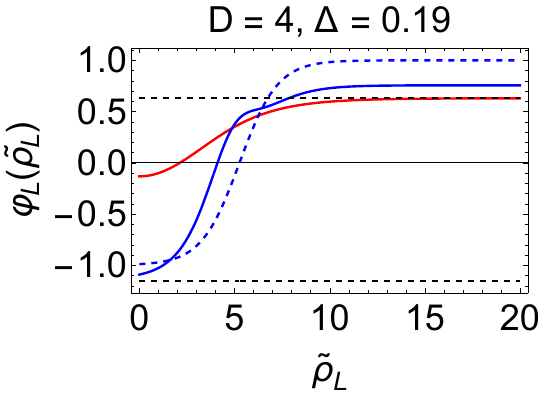} }}
     \end{center}
    \caption{Bounce profile in $D = 3$ (top row) and $D = 4$ (bottom row) for different values of $\Delta$.  
    Small $\Delta$ corresponds to thin wall, while in the last column we are close to 
    $\Delta_{\rm max} = 1/\sqrt{27}$, corresponding to the inflection-point thick wall. 
    The blue dashed line is the thin-wall bounce at the leading order,~\eqref{Colemanb}; 
    the blue plain line is the bounce calculated up to second order in the thin-wall parameter 
    $\Delta$~\eqref{bounceuni}; the red line is the bounce computed numerically with the shooting method. 
    The horizontal black dashed lines mark the values of $\varphi_{\rm FV}$ (top) and $\varphi_{\rm TV}$
    (bottom). 
    The analytic thin-wall bounce of Eq.~\eqref{bounceuni} gives an excellent approximation to 
    the true one up to $\Delta = 0.1$.}
    \label{fig:bounces}
\end{figure}

The result~\eqref{bounceuni} fixes the bounce uniquely up to second order in the thin-wall parameter 
expansion.
%
%
We can see in FIG.~\ref{fig:bounces} how it improves the leading order result of Coleman's, by comparing 
to the bounce calculated numerically (red line).
When we depart from the thin wall regime ($\Delta \to 0$), the analytic result in~\eqref{bounceuni} gives 
an excellent approximation to the true bounce up to $\Delta = 0.1$, and a decent approximation up to 
$\Delta = 0.15$. 
For larger values of $\Delta$ it deviates from the correct profile given by the red line.




%
\subsection{Cubic parametrization}
We can follow the same procedure described in the previous section to construct the 
bounce without redundancy in the cubic parametrization as well. 
Using the expansion in $\epsal$ both for $\varphi_C(z)$ and $r$, we found full analytic 
solutions up to second order, see Appendix~\ref{app:cubic}, and we managed to extract the combination
$ \left( \frac{c_{11}}{2} - r_2 \right)_C$, see \eqref{c2r2ccombo}, from the bounce equation at third order.
A crucial difference compared to the linear case, which makes the calculation in the cubic more involved,
 is that in the cubic we cannot use the $z$-parity 
of the solutions to fix the $c_{1n}$ coefficients, and the odd-order $r_{2n+1}$ do not vanish.

This prevents us from obtaining the final bounce analytically up to order $\epsal^2$,
which would require calculating $c_{12}$ and $r_{C3}$.
Up to order $\epsal$, the physical bounce as a function of the dimensionless radius 
$\rhotiC \equiv m \rho$ is
\begin{equation} \label{bounceunicub}
  \varphi_C(\rhotiC) =  \varphi_{C0}(\rhotiC) + \left[2 \varphi_{C0}(\rhotiC) +\left( 4 + \frac{3}{2}(
  \rhotiC - \RtiC) \right) \frac{{\rm d} \varphi_{C0}}{{\rm d}\rhotiC} + \left( \frac{c_{11}}{2} - r_2 \right)_C 
  \frac{{\rm d} \varphi_{C0}}{{\rm d}\rhotiC}   \right] \epsal \, ,
\end{equation}  
where
\begin{align}
  & \varphi_{C0}(\rhotiC) = -\frac{2}{1+e^{\rhotiC -\RtiC}} \, , \qquad \quad  \RtiC \equiv 
  \frac{r_{C0}}{\epsal} + r_{C1} =   \frac{D-1}{3 \epsal} + \frac{D-1}{2} 
  \\
  & \left( \frac{c_{11}}{2} - r_2 \right)_C = \frac{41D^2 + D(24\pi^2 - 322) - 48\pi^2 + 425}{24(D-1)} \, .
  \label{c2r2ccombo}
\end{align}
Note that $\RtiC$ also contains the $r_{C1}$ term, while the analogous $r_{L1}$ vanished in the linear case.

\begin{figure}
\begin{center}
  \subfloat{{\includegraphics[width=.22\textwidth]{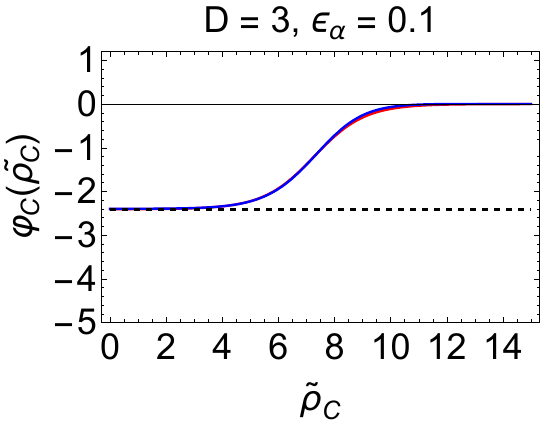} }}%
  \
  \subfloat{{\includegraphics[width=.22\textwidth]{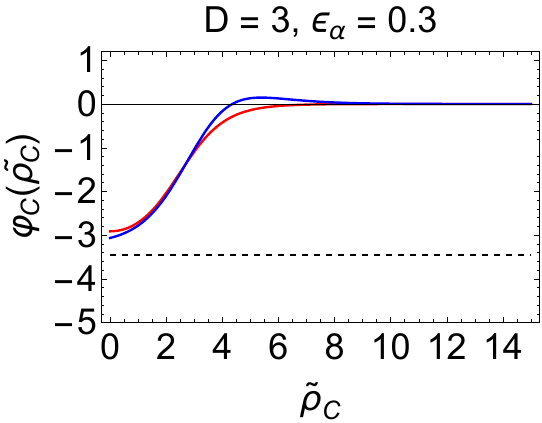} }}%
  \
  \subfloat{{\includegraphics[width=.22\textwidth]{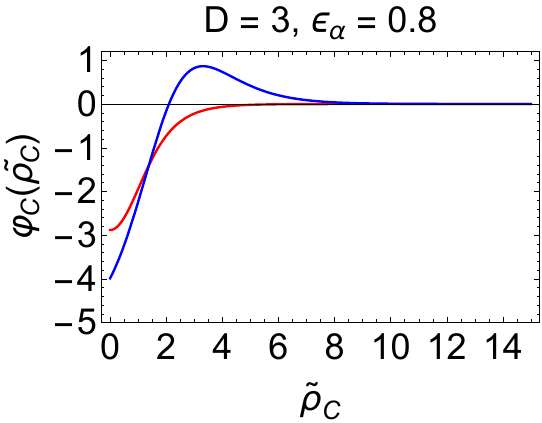} }}%
  \
  \subfloat{{\includegraphics[width=.22\textwidth]{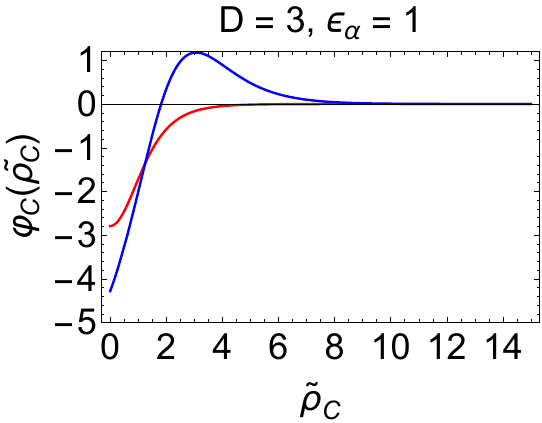} }} 
  \\
  \subfloat{{\includegraphics[width=.22\textwidth]{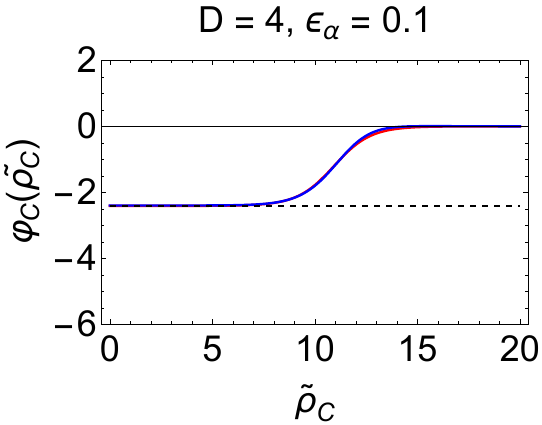} }}%
  \
  \subfloat{{\includegraphics[width=.22\textwidth]{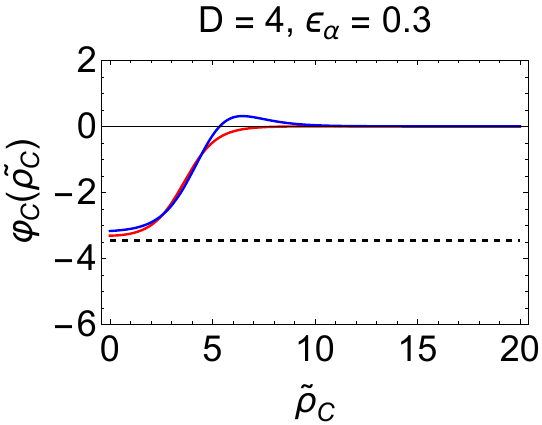} }}%
  \
  \subfloat{{\includegraphics[width=.22\textwidth]{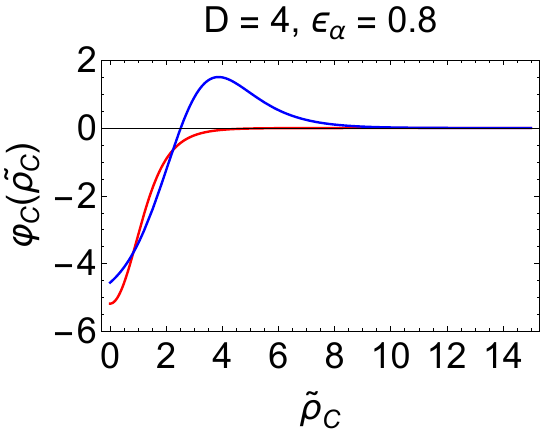} }}%
  \
  \subfloat{{\includegraphics[width=.22\textwidth]{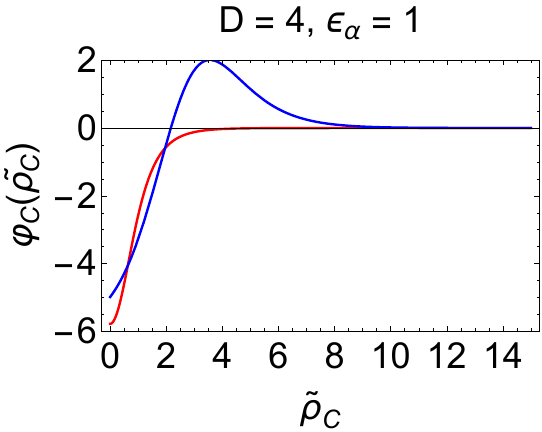} }}
\end{center}
\caption{
  Bounce profile in $D = 3$ (top row) and $D = 4$ (bottom row) for different values of $\epsal$.
  Small $\epsal$ corresponds to thin wall, while in the last column we are in Linde's 
  thick wall regime. 
  The blue plain line is the bounce calculated up to first order in the thin-wall parameter 
  $\epsal$, from~\eqref{bounceunicub}. 
  The red line is the bounce computed numerically with the shooting method. 
  The horizontal black dashed line marks the value of the true vacuum (in the last two columns it is 
  below the scale shown in the vertical axis), while the false vacuum is at $\varphi_C = 0$.
  The analytic thin-wall bounce of Eq.~\eqref{bounceunicub} gives a good approximation to 
  the true one up to $\epsal = 0.3$. 
  }
  \label{fig:bouncescub}
\end{figure}

In FIG.~\ref{fig:bouncescub} we show how the bounce of Eq.~\eqref{bounceunicub} (blue line), which was 
derived in the TW limit, compares to the true one computed numerically (red line) when we depart 
from TW. 
We see that it provides a good approximation up to $\epsal = 0.3$.
For larger $\epsal$ it develops a bump before reaching the FV. 
This gives an indication of the limits of our result, which is not expected to be precise when we approach 
the thick wall. 
It is interesting to note that, apart from the bump, our result at the linear order in $\epsal$ 
approximates decently the value of $\varphi_C(\rhotiC = 0)$ in the whole range $0 \leq \epsal \leq 1$, 
and always reaches the FV at large $\rhotiC$, by construction.

\section{Checks on the functional determinant} \label{app:FunDet}

In this Appendix we perform again the calculations of zero modes removal and of the regularized sums.
Whereas in the main text we carried out the calculations in the cubic parametrization with
\texttt{BubbleDet}, here we use the linear parametrization and employ other numerical methods.
In the end we find excellent agreement among all the different methods.

\subsection{Zero modes}

In the differential equation \eqref{Rlequ}, the term $V^{(2)} - V_{\rm FV}^{(2)}$
vanishes once the field reaches the FV configuration and ensures the finiteness of the solution.
One could worry that adding the offset $\mueps^2$ in 
 \eqref{Rlequoff} for $R_1^\epsilon$ might spoil the good finite behavior and introduce divergences.
By explicitly constructing the solution we did not encounter any divergence at the leading order in $\mueps^2$,
which is all we needed to get $R_1^\prime$.  
It turns out that there are divergences if one goes to higher $\mueps^2$ orders in the calculation.

It is instructive to also perform the calculation with an alternative method:
we can offset not only the second derivative of the potential evaluated on the bounce, 
but also the one evaluated at the FV (this in principle is not needed, as no zero modes appear here), 
to avoid issues with divergences. 
We refer to this method as ``$V_{\rm FV}$ offset'' in FIG.~\ref{fig:zero_removal}. It works as follows.

We first define the second derivative of the shifted false vacuum potential as
\begin{align}
  \hat{V}^{(2)}_\FV \equiv V^{(2)}_\FV + \mueps^2 \, ,
\end{align}
and solve
\begin{align}
  \hat{{\mathcal O}}_{l\FV} \hat{\psi}_{l\FV} \equiv \left(-\frac{\text{d}^2}{
  \text{d} \rho^2} - \frac{D-1}{\rho} \frac{\text{d}}{\text{d} \rho} + 
  \frac{l \left( l + D - 2 \right)}{\rho^2} + \hat{V}^{(2)}_\FV \right)\hat{\psi}_{l\FV}=0 \, ,
\end{align}
for $l = 1$. 
The solution is given by \cite{Ivanov:2022osf}
\begin{align}
  \hat{\psi}_{\nu\FV}(\rho) = c_{\rm FV} \sqrt{\rho} \, 
  I_\nu \left(\rho \sqrt{\hat{V}^{(2)}_\FV} \right) \, ,
\end{align}
where $\nu = D/2$ for $l = 1$. 
In the following we will drop the multipole subscript $\nu$ in $\psi_{\nu\FV}$. 
The ratio of wavefunctions is expanded around $\mueps^2 = 0$
\begin{align}
  \dfrac{1}{\hatpsi}\dfrac{\di\hatpsi }{\di\rho} = \left[\dfrac{1}{\hatpsi}\dfrac{\di\hatpsi }
  {\di\rho}\right]_{\mueps^2=0} +\mueps^2\, \dfrac{\di}{\di\mueps^2}\left[\dfrac{1}
  {\hatpsi}\dfrac{\di\hatpsi }{\di\rho}\right]_{\mueps^2=0} + \cdots \, .
\end{align}
The first term is just
\begin{align}
  \left[\dfrac{1}{\hatpsi}\dfrac{\di\hatpsi }{\di\rho}\right]_{\mueps^2=0} = 
  \dfrac{1}{\psi_\FV}\dfrac{\di\psi_\FV }{\di\rho} \, ,
\end{align}
that appears in the differential equation for generic multipoles, and we define
\begin{align}
    \hat{X}\equiv\dfrac{\di}{\di\mueps^2}\left[\dfrac{1}{\hatpsi}\dfrac{\di\hatpsi }{\di\rho}\right]_{\mueps^2=0} \, .
\end{align}

The differential equation for $R_1^\epsilon$ is therefore
\begin{align}
  \left(\dfrac{\partial^2}{\partial\rho^2}+2\frac{\dot{\psi}_\FV}{\psi_\FV}\dfrac{\partial}
  {\partial\rho}-(V^{(2)}-V^{(2)}_\FV)+2\mueps^2\hat{X}\dfrac{\partial}
  {\partial\rho}\right)\left(R_1+\mueps^2\delta R_1\right)=0 \, .
\end{align}
The first contribution is the usual differential equation for $R_1$
\begin{align}
  \left(\dfrac{\partial^2}{\partial\rho^2}+2\frac{\dot{\psi}_\FV}{\psi_\FV}
  \dfrac{\partial}{\partial\rho}-(V^{(2)}-V^{(2)}_\FV)\right)R_1=0 \, ,
\end{align}
which will be one of the two coupled differential equations. 
To obtain the equation for $\delta R_1$, we truncate at order $\mueps^2$
\begin{align}
  \left(\dfrac{\partial^2}{\partial\rho^2}+2\frac{\dot{\psi}_\FV}{\psi_\FV}
  \dfrac{\partial}{\partial\rho}-(V^{(2)}-V^{(2)}_\FV)\right)\delta R_1=-2\hat{X}
  \dfrac{\partial R_1}{\partial\rho} \, .
\end{align}
Our strategy is then to fix a value of the parameter $\epsal$, numerically solve the 
differential equation for $R_1$ first, plug in the result into the differential 
equation for $\delta R_1$ and solve for $\delta R_1$.
The result, denoted ``FindBounce, $V_{\rm FV}$ offset'',
 is in perfect agreement with that of \eqref{delR1eq}, as shown in FIG.~\ref{fig:zero_removal}.

\begin{figure}[t]
    \centering
    \includegraphics[width=0.9\textwidth]{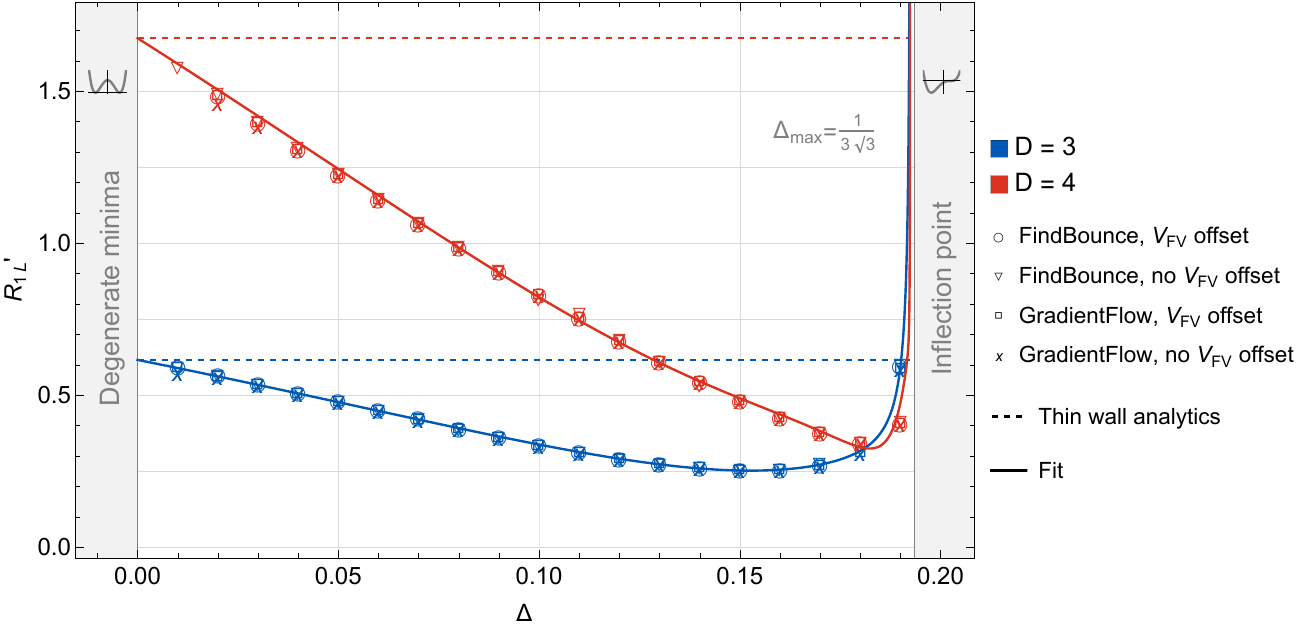}
    \caption{\footnotesize{
   We plot $R_1^\prime$ as a function of $\Delta$.
   We use a few different methods for the numerical evaluation, as described in the text. We also
   show the fits given by \eqref{fitzerolin} as solid lines, and 
   the analytic result \eqref{R1primethin} obtained in \cite{Ivanov:2022osf} 
   in the thin-wall limit ($\Delta \to 0$) as dashed lines.
   }
    }
    \label{fig:zero_removal_linear}
\end{figure}

As a further check, we also perform the numerical calculation in the linear parametrization using a couple of numerical methods, and show
the result in FIG.~\ref{fig:zero_removal_linear}. The solid lines in the figure are obtained starting from the fits \eqref{fitzero3}
and \eqref{fitzero4}, where $R_1^\prime(\epsal)$ was given in the cubic parametrization, and translating them into the linear parametrization,
\begin{equation} \label{fitzerolin}
R_{1L}^\prime(\Delta) = \frac{\lambda v^2}{m^2(\lambda,v,\Delta)} R_1^\prime (\epsal(\Delta)) = \frac{1}{f_{m^2}(\Delta)} R_1^\prime(\epsal(\Delta)) \, .
\end{equation}
Here we used the map from Appendix~\ref{sec:map}. We find very good agreement between these translated fits
and the numerical points calculated directly in the linear parametrization. Also note the trend of increasing $R_{1L}^\prime$ as we 
approach $\Delta_{\rm max}$. That is well understood from \eqref{fitzerolin}: the function $f_{m^2}(\Delta)$ goes to zero as 
$\Delta \to \Delta_{\rm max}$, so $R_{1L}^\prime$ goes to infinity. This makes the vacuum decay rate go to zero. In this limit we reach the inflection point, and
we go from quantum tunneling to classical rolling, so we expect indeed the vacuum decay rate to vanish. 

\subsection{Regularized sums}

In this section we compute numerically the regularized sums defined in \eqref{rensum3} and \eqref{rensumD4} in the 
linear parametrization, using our own code and implementing a couple of methods.
 We then translate the results obtained in the cubic parametrization and find 
very good agreement among all the different methods.

 For the regularization we use two different procedures: (i) involves the $\zeta$-function~\cite{Dunne:2006ct} and is equivalent
 to that used by \texttt{BubbleDet},
(ii) involves Feynman diagrams~\cite{Baacke:2003uw}. We confirm numerically, as argued formally in~\cite{Dunne:2006ct}, 
that they lead to the same result.

Let us describe the Feynman diagrammatic regularization~\cite{Baacke:2003uw}, where we calculate the subtraction of the sum in
a slightly different way compared to \eqref{rensum3} and \eqref{rensumD4}.
We define a perturbative expansion of $R_\nu^{(n)}$ in orders of insertions of $\left(\ddV - \ddV_\text{FV} \right)$,
which solves the fluctuation operator
\begin{align} \label{RFeyn}
  & \frac{\mathrm{d}^2 R_\nu^{(n)}}{\mathrm{d}\rho^2} + 2 \frac{\mathrm{d} R_\nu^{(n)}}{\mathrm{d}\rho} 
  \left( \frac{\mathrm{d} \psi_{\nu\text{FV}} / \mathrm{d}\rho}{\psi_{\nu\text{FV}}} \right) -
  \left(\ddV - \ddV_\text{FV} \right) R_{\nu}^{(n-1)}  = 0  \, .
\end{align}
At the leading order there is no insertion of interactions, meaning no term with $\left(\ddV - \ddV_\text{FV} \right)$ in the equation,
and the solution which satisfies the boundary conditions is trivially given by $R_{\nu}^{(0)} = 1$. 
We plug this into \eqref{RFeyn} with $n=1$, and solve for $R_\nu^{(1)}$. Then we have to plug the $R_\nu^{(1)}$ solution
into the $n=2$ equation, and proceed this way iteratively. 
Note that $n = 1$ can be mapped into a one-loop Feynman diagram with one insertion of the operator $\left(\ddV - \ddV_\text{FV} \right)$,
$n = 2$ into a diagram with two insertions, and so on. Thus, $n=1$ corresponds to a linear (quadratic) divergence in $D=3$ ($D=4$),
$n = 2$ to a logarithmic divergence in $D = 4$, and so on.
We solve for $R_\nu^{(1)}$ and $R_\nu^{(2)}$ numerically. 

The sum over multipoles regularized in the $\overline{\rm MS}$ scheme is given by
\begin{equation} \label{rensum3F}
  \Sigma_3^{\rm Feyn} = \sum_{\nu = 1/2} 2\nu \left(\ln R_\nu - R_\nu^{(1)} \right) + \overline T^{(1)}_3 \, ,
\end{equation}
in $D=3$, where the renormalized part is given by
\begin{align}
  \overline T^{(1)}_3 &= -\frac{\sqrt{V^{(2)}_{\rm FV}}}{4\pi}\overline V^{(2)}(0) \, ,
\end{align}
and $\overline V^{(2)}$ denotes the Fourier transform of the second derivative of the potential, 
\begin{align}
  \overline V^{(2)}(|k|) &= \int \text{d}^D x \, e^{-i k x} \left(V^{(2)}(|x|) - V^{(2)}_{\rm FV} \right) \, .
\end{align}
This Fourier transform is defined in generic $D$ spacetime dimensions. 

In $D = 4$ we have
\begin{equation}\label{rensum4F}
  \overline  \Sigma_4^{\rm Feyn} = \sum_{\nu = 1} \nu^2 \left(\ln R_\nu - R_\nu^{(1)} - R_\nu^{(2)} + 
    \frac12 R_\nu^{(1)2} \right) + \overline T^{(1)}_4 - \frac12\overline T^{(2)}_4\, ,
\end{equation}
where
\begin{align} \label{T41}
  \overline T^{(1)}_4 &= -\frac{V^{(2)}_{\rm FV}}{16\pi^2} \left(1+\ln\frac{\mu^2}{V^{(2)}_{\rm FV}}\right)\overline V^{(2)}(0) \, ,
  \\
  \overline T^{(2)}_4 &= \int_0^\infty \frac{{\rm d}k \, k^3}{128\pi^4}\left(2+\ln\frac{\mu^2}{V^{(2)}_{\rm FV}}-
  \frac{\sqrt{k^2+4V^{(2)}_{\rm FV}}}{2k}\ln\frac{k^2+2V^{(2)}_{\rm FV}+k\sqrt{k^2+4V^{(2)}_{\rm FV}}}{k^2+2V^{(2)}_{\rm FV}-
  k \sqrt{k^2 + 4 V^{(2)}_{\rm FV}}}\right) [\overline V^{(2)}(k)]^2 \, .
  \label{T42}
\end{align}
It is easy to check, by switching to the dimensionless variables, that $ \overline T^{(1)}_3$, $ \overline T^{(1)}_4 $,  $\overline T^{(2)}_4$
are functions only of $\Delta$ or of $\epsal$, apart from the terms with $\ln\frac{\mu^2}{V^{(2)}_{\rm FV}}$, 
which contain the dimensionful parameters $\lambda v^2$ or $m^2$ in the argument of the logarithm.
Here, again, the $\mu$ dependence
will cancel out against the analogous contributions from the renormalized bounce action.   

Working with the linear parametrization, 
we split the sum at a multipole $\nu_{max}$, which is chosen manually for each value of $\Delta$.
For a given $\nu$, we solve for $R_\nu$ and find that this quantity initially grows as a function of $\nu$, has a 
peak at a value $\nu_{peak}$ and then decays, because of the subtraction of high-$\nu$ divergence.
We choose a value of $\nu_{max} > \nu_{peak}$ and perform the sum numerically up to this value.
We then interpolate the high-$\nu$ part with inverse powers of $\nu$ and perform the sum from $\nu_{max}+1$ to 
$\infty$ analytically in terms of Riemann $\zeta$ functions.

\begin{figure}
    \centering
    \includegraphics[width=0.94\textwidth]{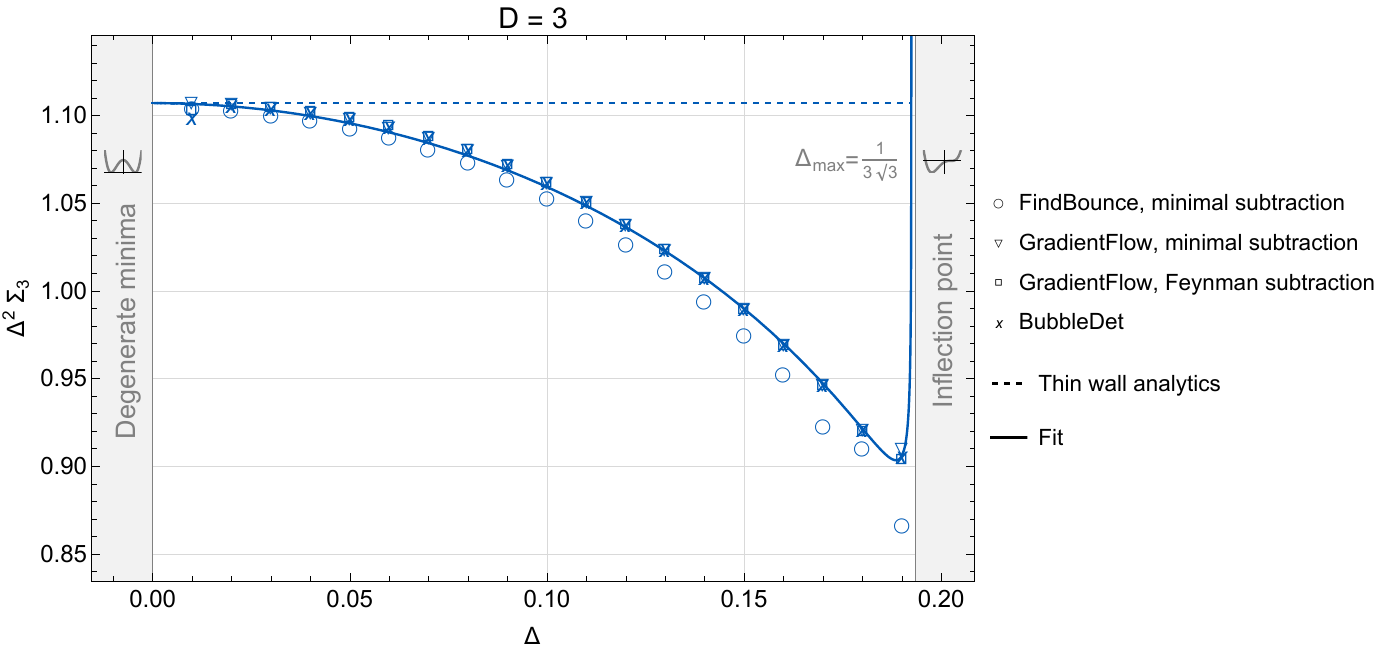}
    \caption{\footnotesize{ 
 We plot $\Delta^2 \Sigma_3$ as a function of $\Delta$.
 To compute the renormalized sum $\Sigma_3$ we use both the recipe \eqref{rensum3} (minimal subtraction) and \eqref{rensum3F}
 (Feynman subtraction). The calculation is performed in the linear parametrization
  using a couple of different numerical methods, as labeled in the legend. The horizontal
 dashed line corresponds to the analytic result $\Sigma_3 =\frac{1}{\Delta^2} \frac{20 + 9\ln 3}{27}$ obtained in the thin wall limit~\cite{Ivanov:2022osf},
 $\Delta \to 0$. The solid line corresponds to the fit \eqref{fitC3} translated into the linear parametrization as explained in the text.  
     }}
    \label{fig:RenSum3D}
\end{figure}
\begin{figure}
    \centering
    \includegraphics[width=0.94\textwidth]{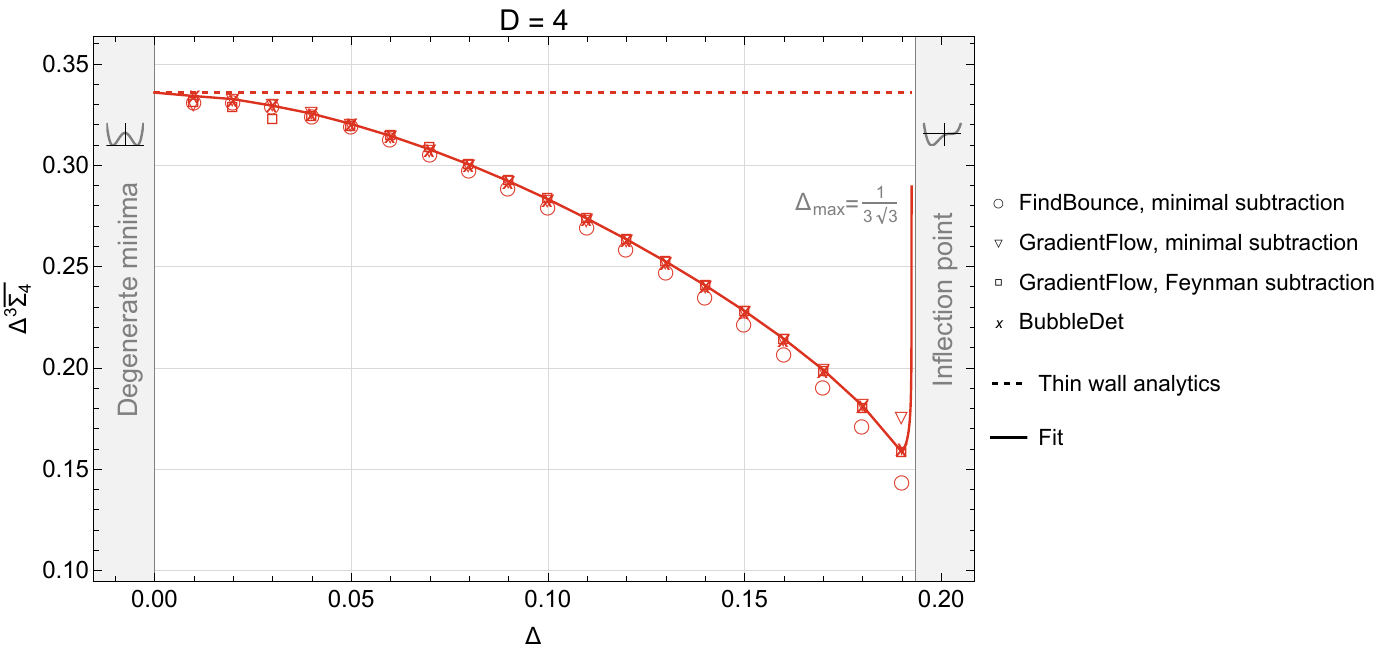}
    \caption{\footnotesize{We plot 
    $\Delta^3 \overline \Sigma_4 $
    as a function of $\Delta$.
 To compute the renormalized sum $\overline \Sigma_4$ we use both the recipe \eqref{rensumD4} (minimal subtraction) and \eqref{rensum4F}
 (Feynman subtraction). The calculation is performed in the linear parametrization using a few different numerical methods, as labeled in the legend. 
 Here we set $\mu = \sqrt{\lambda}v$ in \eqref{T41}, \eqref{T42}.
 The horizontal
 dashed line corresponds to the analytic result $\overline \Sigma_4 =\frac{1}{\Delta^3} \frac{27 - 2\pi \sqrt{3}}{48}$ obtained in the thin wall limit~\cite{Ivanov:2022osf},
 $\Delta \to 0$. The solid line corresponds to the fit \eqref{fitC4} translated into the linear parametrization as explained in the text.}}
    \label{fig:RenSum4D}
\end{figure}

In FIG.~\ref{fig:RenSum3D} and \ref{fig:RenSum4D} we plot the regularized sums, separately for $D = 3$ 
and $D = 4$, calculated with a few different numerical methods.
The bounce is obtained using either \texttt{FindBounce} or the gradient flow method, and we use the two 
regularization schemes discussed above, for comparison.
We compare our results against those computed with \texttt{BubbleDet}~\cite{Ekstedt:2023sqc} as well.
We find very good agreement among all the different numerical methods in both $D = 3$ and $D = 4$. Our results also provide a numerical
check that using either the $\zeta$-function subtraction scheme [see \eqref{rensum3} and \eqref{rensumD4}], which is the minimal subtraction,
or the $\overline{\rm MS}$ scheme with the diagrammatic approach [see \eqref{rensum3F} and \eqref{rensum4F}], the renormalized
sums are the same in the end. This was already shown for $D=4$ in \cite{Dunne:2006ct}, here we see that it holds true in $D=3$ as well.

We also check that taking the regularized sums discussed in the main text, computed in the cubic parametrization,
and translating them into the linear parametrization using the map in Appendix~\ref{sec:map},
 we find agreement with the numerics discussed in this section.  
The translation requires some care. 
First, one has to keep in mind that the sums $\Sigma_3$ and $\overline\Sigma_4$ in \eqref{rensum3}
and \eqref{rensumD4} contain $\ln R_1^\prime$ at the multipole $\nu =3/2$ ($\nu = 2$) in $D=3$ ($D=4$).
Using \eqref{fitzerolin}, we see that going from $R_1^\prime (\epsal)$ to $R_{1L}^\prime(\Delta)$ we have a factor
of $f_{m^2}(\Delta)$. 
In $D=3$ this implies that $\Sigma_3(\epsal) \to \Sigma_3(\epsal(\Delta)) - 3 \ln(f_{m^2}(\Delta))$.
Similarly, in $D=4$ this observation requires subtracting $-4\ln(f_{m^2}(\Delta))$ from $\overline\Sigma_4$. 
Moreover, in $D=4$ we have the term $\ln(\mu / m)$ entering the sum $\overline\Sigma_4$ through $\tilde I_2$ defined
 in \eqref{I2tilde}. 
Translating that, we get $\ln(\mu / m) \to \ln(\mu/(\sqrt{\lambda}v)) - 1/2 \ln(f_{m^2}(\Delta))$. 
In the numerics done in the linear parametrization and
shown in FIGs.~\ref{fig:RenSum3D} and \ref{fig:RenSum4D} we set $\mu = \sqrt{\lambda} v$.
To compare properly then, we have to subtract $- 1/2 \ln(f_{m^2}(\Delta)) I_2(\epsal(\Delta))$ from
 $\tilde I_2$ computed in the cubic. 
 Here, $I_2(\epsal(\Delta))$ is computed numerically from \eqref{Isubtract} as a function of $\epsal$, then translated 
 into a function of $\Delta$.
 At the end of the day, taking the fits \eqref{fitC3} and \eqref{fitC4}, and correcting with the proper additions of 
 logs of $f_{m^2}(\Delta)$ as just explained, we get the solid lines in FIGs.~\ref{fig:RenSum3D} and \ref{fig:RenSum4D}.
 They provide a fit in excellent agreement with the numerical points. Again we see that the regularized sums
 blow up as we approach $\Delta_{\rm max}$ due to the $\ln(f_{m^2}(\Delta))$, 
 which has the consequence of making the decay rate vanish at the inflection point, as expected.
  
Note that the \texttt{FindBounce} points in FIGs.~\ref{fig:RenSum3D} and \ref{fig:RenSum4D}  
are slightly lower compared to the others, especially at larger values of $\Delta$. 
We found the reason in the extension of the bounce solution in $\rho$ space: compared to gradient flow, where 
the maximum radius is set by hand, in \texttt{FindBounce} this is automatically set when the routine is solving for 
the bounce. 
This in turn reflects in a different evaluation of integrals in $\rho$, such as the minimal subtraction
integrals of $\ddV$. 
One can try to manually increase the maximum radius, but the bounce solution will still slightly 
differ from the gradient flow solution. 
However, the discrepancies are at the \% level and will not significantly affect the final result for the decay rate.

\bibliography{References}

\end{document}